\RequirePackage{fix-cm} 
\documentclass[twocolumn]{svjour3}

\makeatletter
\def\cl@chapter{\@elt {theorem}}
\makeatother

\usepackage{times}

\usepackage{balance}

\usepackage[hyphens]{xurl}
\usepackage[utf8]{inputenc}
\usepackage[misc,geometry]{ifsym} 
\usepackage[T1]{fontenc} 
\usepackage{esvect}
\usepackage{url}            
\usepackage{booktabs}    
\usepackage{array}
\usepackage{hyperref}
\hypersetup{colorlinks,allcolors=black}
\usepackage{comment}
\usepackage{multirow}
\usepackage{xcolor}
\usepackage{colortbl}
\usepackage[export]{adjustbox}
\usepackage{subcaption}
\usepackage{listings}
\usepackage{pifont}
 \usepackage{amsmath}
\usepackage{rdfmacros}
\usepackage{inconsolata}
\usepackage{amssymb}

\usepackage[hyphens]{url}
\usepackage{hyperref}
\usepackage[hyphenbreaks]{breakurl}

\usepackage[english]{babel}  
\addto\extrasenglish{  
    
}

\newif\ifarxiv
\arxivtrue
\newcommand{\ja}[2]{\ifarxiv #2\else #1\fi}

\newcolumntype{a}{ >{\columncolor[HTML]{baffba}} c }
\newcolumntype{b}{ >{\columncolor[HTML]{ffdaff}} c }
\newcolumntype{d}{ >{\columncolor[HTML]{eeee99}} c }
\newcolumntype{e}{ >{\columncolor[HTML]{cceaea}} c }
\newcolumntype{f}{ >{\columncolor[HTML]{ffdada}} l }
\newcolumntype{g}{ >{\columncolor[HTML]{ffdada}} c }
\newcolumntype{n}{ >{\columncolor[HTML]{f3f3f3}} l }
\newcolumntype{y}{ >{\columncolor[HTML]{f3f3f3}} c }

\newcommand{\cmark}{\ding{51}}%
\newcommand{\pmark}{\ensuremath{\sim}}%

\definecolor{Gray}{gray}{0.9} 
\definecolor{Cyan}{rgb}{0.88,1,1}
\definecolor{Blue}{rgb}{0.92,0.72,0.47}

\hypersetup{
	pdftitle={A Survey of RDF Stores \& SPARQL Engines for Querying Knowledge Graphs},    
	pdfauthor={Waqas Ali, Muhammad Saleem, Bin Yao, Aidan Hogan, Axel-Cyrille Ngonga Ngomo},     
	pdfsubject={VLDBJ 2021},   
	pdfkeywords={sparql engine;} {query;} {rdf store;} {knowledge graph}, 
}

\begin{document}

\title{A Survey of RDF Stores \& SPARQL Engines\\for Querying Knowledge Graphs}

\titlerunning{A Survey of RDF Stores \& SPARQL Engines for Querying Knowledge Graphs}  


\author{
 Waqas Ali \and
 Muhammad Saleem \and
 Bin Yao \and\\ 
 Aidan Hogan \and
 Axel-Cyrille Ngonga Ngomo
}

\authorrunning{Ali et al.} 

\institute{W. Ali \at
              SEIEE, 
              Shanghai Jiao Tong University, 
              Shanghai, China. \\
              \email{waqasali@sjtu.edu.cn}
           \and
           M. Saleem \at
              AKSW, 
              University of Leipzig, 
              Leipzig, Germany. \\
              \email{saleem@informatik.uni-leipzig.de}
           \and
          \Letter \hspace{0.1cm} B. Yao (Hangzhou Qianjiang Distinguished Expert) \at
            SEIEE, 
              Shanghai Jiao Tong University, 
              Shanghai, China.
              \newline 
              Hangzhou Institute of Advanced Technology, Hangzhou, China.\\
              \email{yaobin@cs.sjtu.edu.cn}
           \and       
           A. Hogan \at
              DCC, University of Chile \& IMFD, 
              Santiago, Chile.\\
              \email{ahogan@dcc.uchile.cl} 
           \and
           A.-C. Ngonga Ngomo  \at
              DICE, Paderborn University, 
              Paderborn, Germany.\\
              \email{axel.ngonga@upb.de}
}

\date{Received: date / Accepted: date}

\maketitle

\begin{abstract}
RDF has seen increased adoption in recent years, prompting the standardization of the SPARQL query language for RDF, and the development of local and distributed engines for processing SPARQL queries. This survey paper provides a comprehensive review of techniques and systems for querying RDF knowledge graphs. While other reviews on this topic tend to focus on the distributed setting, the main focus of the work is on providing a comprehensive survey of state-of-the-art storage, indexing and query processing techniques for efficiently evaluating SPARQL queries in a local setting (on one machine). To keep the survey self-contained, we also provide a short discussion on graph partitioning techniques used in the distributed setting. We conclude by discussing contemporary research challenges for further improving SPARQL query engines. \ja{An}{This} extended version also provides a survey of over one hundred SPARQL query engines and the techniques they use, along with twelve benchmarks and their features. 


\end{abstract}

\section{Introduction}

The Resource Description Framework (RDF) is a graph-based data model where triples of the form $(s,p,o)$ denote directed labeled edges $s \xrightarrow{p} o$ in a graph. RDF has gained significant adoption in the past years, particularly on the Web. As of 2019, over 5 million websites publish RDF data embedded in their webpages~\cite{webdatacommons}. RDF has also become a popular format for publishing knowledge graphs on the Web, the largest of which -- including 
Bio2RDF,
DBpedia,
PubChemRDF,
UniProt,
and 
Wikidata -- contain billions of triples. 
These de\-ve\-lop\-ments have brought about the need for optimized techniques and engines for querying large RDF graphs. We refer to engines that allow for storing, indexing and processing joins over RDF as \textit{RDF stores}.

While various query languages have historically been proposed for RDF, the SPARQL Protocol and RDF Query Language (SPARQL) has become the standard~\cite{sparql11}. The first version of SPARQL was standardized in 2008, while SPARQL~1.1 was released in 2013~\cite{sparql11}. SPARQL is an expressive language that supports not only joins, but also variants of the broader relational algebra (projection, selection, union, difference, etc.). Various new features were added in SPARQL~1.1, such as \textit{property paths} for matching arbitrary-length paths in the RDF graph. Hundreds of SPARQL query services, called \textit{endpoints}, have emerged on the Web~\cite{sparqlinfra2013}, with the most popular endpoints receiving millions of queries per day~\cite{lsq2015,MalyshevKGGB18}. We refer to engines that support storing, indexing and processing SPARQL (1.1) queries over RDF as \textit{SPARQL engines}. Since SPARQL supports joins, we consider any SPARQL engine to also be an RDF store.

Efficient data storage, indexing and join processing are key to RDF stores (and thus, to SPARQL engines):

\begin{itemize}
    \item \textit{Storage.} Different engines store RDF data using different structures (tables, graphs, etc.), encodings (integer IDs, string compression, etc.) and media (main memory, disk, etc.). Which storage to use may depend on the scale of the data, the types of query features supported, etc. 
    \item \textit{Indexing.} Indexes are used in RDF stores for fast lookups and query execution. Different index types can support different operations with varying time--space trade-offs. 
    
    \item \textit{Join Processing.} At the core of evaluating queries lie efficient methods for processing joins. Aside from traditional pairwise joins, recent years have seen the emergence of novel techniques, such as multiway and worst-case optimal joins, as well as GPU-based join processing. Optimizing the order of evaluation of joins can also be important to ensure efficient processing.
   
 \end{itemize}
 
Beyond processing joins, SPARQL engines must offer efficient support for more expressive query features:

\begin{itemize}
    \item \textit{Query Processing.} SPARQL is an expressive language containing a variety of query features beyond joins that need to be supported efficiently, such as filter expressions, optionals, path queries, etc.
\end{itemize}

RDF stores can further be divided into two categories: (1) \textit{local stores} (also called single-node stores) that manage RDF data on one machine and (2) \textit{distributed stores} that partition RDF data over multiple machines. While local stores are more lightweight, the resources of one machine limit scalability~\cite{Wylot2018RDFDS,Pan2018,huang2011scalable}. Various kinds of distributed RDF stores have thus been proposed \cite{Hammoud:2015:DDR:2735703.2735705,huang2011scalable,Schtzle2014SempalaIS,Schtzle2015S2RDFRQ} that typically run on clusters of shared-nothing machines.   

In this survey, we describe storage, indexing, join processing and query processing techniques employed by local RDF stores, as well as high-level strategies for partitioning RDF graphs as needed for distributed storage. \ja{An extended online version of this survey \cite{online}}{An appendix in this extended version} further compares 135 local and distributed RDF engines in terms of the techniques they use, as well as 12 benchmarks in terms of the types of data and queries they contain. The goal of this survey is to give a succinct introduction of the different techniques used by RDF query engines, and also to help users to choose the appropriate engine or benchmark for a given use case.

The rest of the paper is structured as follows. \autoref{sec:review} discusses and contrasts this survey with related literature. \autoref{sec:preliminaries} provides preliminaries for RDF and SPARQL. Sections~\ref{sec:storage}, \ref{sec:index}, \ref{sec:join} and \ref{sec:query} review techniques for storage, indexing, join processing and query processing, respectively.
\autoref{sec:partition} explains different graph partitioning techniques for distributing storage over multiple machines. \autoref{sec:systems} introduces additional content available in the \ja{online extended version~\cite{online}}{appendix of this extended version}, which surveys 135 local and distributed RDF engines, along with 12 SPARQL benchmarks. \autoref{sec:conclusion} concludes the paper with subsections for current trends and research challenges regarding efficient RDF-based data management and query processing. 

\section{Literature Review}
\label{sec:review}

We first discuss related studies. More specifically, we summarize peer-reviewed tertiary literature (surveys in journals, short surveys in proceedings, book chapters, surveys with empirical comparisons, etc.) from the last 10 years collating techniques, engines and/or benchmarks for querying RDF. We summarize the topics covered by these works in \autoref{tab:surveys}. We use \cmark, \pmark\ and blank cells to denote detailed, partial or little/no discussion, respectively, when compared with the current survey (the bottom row). We also present the number of engines and benchmarks included in the
extended version of this survey. If the respective publication does not formally list all systems/benchmarks (e.g., as a table), we may write $n+$ as an estimate for the number discussed in the text.

\begin{table*}
\centering
\caption{Prior tertiary literature on RDF query engines; the abbreviations are: \textbf{Sto.}/Storage, \textbf{Ind.}/Indexing, \textbf{J.Pr.}/Join Processing, \textbf{Q.Pr.}/Query Processing, \textbf{Dis.}/Distribution, \textbf{Eng.}/Engines, \textbf{Ben.}/Benchmarks} \ja{*Refers to engines and benchmarks in an extended version of the paper~\cite{online}.}{}
 \begin{tabular}{lclccccr@{}lr@{}l}
 \toprule
\multirow{2}{*}{\textbf{Study}} & \multirow{2}{*}{\textbf{Year}} & 
\multicolumn{5}{c}{\textbf{Techniques}} & \multicolumn{2}{c}{\multirow{2}{*}{\textbf{Eng.}}} & \multirow{2}{*}{\textbf{Bench.}}  \\  
 & & \hspace{.5cm} \textbf{Sto.} & \textbf{Ind.} & \textbf{J.Pr.} & \textbf{Q.Pr.}  & \textbf{Dis.} & & & & \\
 \midrule
 
 Sakr et al. \cite{sakr2010relational} & 2010  & \hspace{.5cm} \cmark & \pmark & & & & 10&+ & & \\
 
 Svoboda et al. \cite{svoboda2011linked} & 2011  & \hspace{.5cm} \pmark & \pmark & & & \pmark & 14 & & 6 & \\

 Faye et al. \cite{Faye2012ASO} & 2012  & \hspace{.5cm} \cmark & \pmark & \pmark & & & 13 & & & \\
  
 Luo et al. \cite{LuoPFHV12} & 2012  & \hspace{.5cm} \cmark & \cmark & & & & 20&+ & & \\
 
 Kaoudi et al. \cite{kaoudi2015rdf} & 2015  & \hspace{.5cm} \cmark & \pmark & \pmark & \pmark & \cmark & 17 & & & \\

 Ma et al. \cite{Ma2016StoringMR} & 2016  & \hspace{.5cm} \cmark & \pmark & & & \pmark & 17 & & 6 & \\
 
 Özsu \cite{zsu2016ASO} & 2016  & \hspace{.5cm} \cmark &  & \pmark & \pmark & \pmark & 35&+ & & \\
 
 Abdelaziz et al. \cite{AbdelazizHKK17} & 2017  & \hspace{.5cm} \cmark & \pmark & \pmark & \pmark & \cmark & 21 & & 4 & \\
 
 Elzein et al. \cite{ELZEIN2018375}  & 2018  & \hspace{.5cm} \cmark &  & \pmark & \pmark & \pmark & 15&+ & & \\ 

 Janke \& Staab \cite{janke2018storing} & 2018  & \hspace{.5cm} \pmark & \pmark & \pmark & \pmark & \cmark & 50&+ & 9 & \\

 Pan et al. \cite{Pan2018} & 2018  & \hspace{.5cm} \cmark & \pmark & \pmark & & \pmark & 25&+ & 4 & \\ 

 Wylot et. al \cite{Wylot2018RDFDS} & 2018  & \hspace{.5cm} \cmark & \pmark & \pmark & \pmark & \cmark & 24 & & 8 & \\ 
 
 Yasin et al. \cite{yasin2018comprehensive} & 2018  & \hspace{.5cm} &  & \pmark & \cmark & \pmark & 14 & & & \\ 
 
 Alaoui \cite{10.1145/3368756.3369047} & 2019  & \hspace{.5cm} \cmark & & & & \pmark & 30&+ & & \\ 

 Chawla et al. \cite{article11} & 2020  & \hspace{.5cm} \cmark & \pmark & \pmark &  & \pmark & 46 & & 9 & \\
 
 Zambom et al. \cite{10.1145/3341105.3375753} & 2020 & \hspace{.5cm} \pmark & & \pmark & & \cmark & 24 & & & \\

\midrule

 Ali et al.  &   & \hspace{.5cm} \cmark & \cmark & \cmark & \cmark & \cmark & 135 & \ja{*}{} & 12 & \ja{*}{} \\
  \hline
   \end{tabular}
    \label{tab:surveys}
  \end{table*}

Sakr et al.~\cite{sakr2010relational} present three schemes for storing RDF data in relational databases, surveying works that use the different schemes. Svoboda et al. \cite{svoboda2011linked} provide a brief survey on indexing schemes for RDF divided into three categories: local, distributed and global. Faye et al.~\cite{Faye2012ASO} focus on both storage and indexing schemes for local RDF engines, divided into native and non-native storage schemes. Luo et al.~\cite{LuoPFHV12} also focus on RDF storage and indexing schemes under the relational-, entity-, and graph-based perspectives in local RDF engines. Compared to these works, we present join processing, query processing and partitioning techniques; furthermore, these works predate the standardization of SPARQL 1.1, and thus our discussion includes more recent storage and indexing techniques, as well as support for new features such as property paths.

Later surveys began to focus on distributed RDF stores. Kaoudi et al.~\cite{kaoudi2015rdf} present a survey of RDF stores explicitly designed for a cloud-based environment. Ma et al.~\cite{Ma2016StoringMR} provide an overview of RDF storage in relational and NoSQL databases. Özsu~\cite{zsu2016ASO} presents a survey that focuses on storage techniques for RDF within local and distributed stores, with a brief overview of query processing techniques in distributed and decentralized (Linked Data) settings. Abdelaziz et al.~\cite{AbdelazizHKK17} survey 22 distributed RDF stores, and compare 12 experimentally in terms of pre-processing cost, query performance, scalability, and workload adaptability. Elzein et al.~\cite{ELZEIN2018375} present a survey on the storage and query processing techniques used by RDF stores on the cloud. Janke \& Staab~\cite{janke2018storing} present lecture notes discussing RDF graph partitioning, indexing, and query processing techniques, with a focus on distributed and cloud-based RDF engines. Pan et al.~\cite{Pan2018} provide an overview of local and distributed storage schemes for RDF. Yasin et al.~\cite{yasin2018comprehensive} discussed SPARQL (1.1) query processing in the context of distributed RDF stores. Wylot et al.~\cite{Wylot2018RDFDS} present a comprehensive survey of storage and indexing techniques for local (centralized) and distributed RDF stores, along with a discussion of benchmarks; most of their survey is dedicated to distributed and federated stores. Alaoui~\cite{10.1145/3368756.3369047} proposes a categorization scheme for RDF engines, including memory-, cloud-, graph- and binary-bases stores. The survey by Chawla et al.~\cite{article11} reviews distributed RDF engines in terms of storage, partitioning, indexing, and retrieval. The short survey by Zambom \& dos Santos~\cite{10.1145/3341105.3375753} discusses mapping RDF data into NoSQL databases. All of these works focus on techniques for storing RDF, particularly in distributed settings, where our survey is more detailed in terms of join and query processing techniques, particularly in local settings.

Local RDF stores are those most commonly found in practice~\cite{sparqlinfra2013}. To the best of our knowledge, our survey provides the most comprehensive discussion thus far on storage, indexing, join processing and querying processing techniques for SPARQL in a local setting, where, for example, we discuss novel techniques for established features -- such as novel indexing techniques based on compact data structures, worst-case optimal and matrix-based join processing techniques, multi-query optimization, etc. -- as well as techniques for novel features in SPARQL 1.1 -- such as indexing and query processing techniques for evaluating property paths -- that are not well-represented in the existing literature. To keep our survey self-contained, we also present partitioning techniques for RDF graphs, and include distributed stores and benchmarks in our survey. Per \autoref{tab:surveys}, the survey of engines and benchmarks found in the online version is more comprehensive than seen in previous works~\cite{online}. Conversely, some of the aforementioned works are more detailed in certain aspects, particularly distributed stores; we refer to this literature for further details as appropriate.

\section{Preliminaries}
\label{sec:preliminaries}

Before beginning the core of the survey, we first introduce some preliminaries regarding RDF and SPARQL.

\subsection{RDF}

The RDF data model~\cite{rdf11} uses \textit{RDF terms} from three pairwise disjoint sets: the set $\I$ of \textit{Internationalized Resource Identifiers} (\textit{IRIs})~\cite{iris} used to identify resources; the set $\L$ of \textit{literals} used for (language-tagged or plain) strings and datatype values; and the set $\B$ of \textit{blank nodes}, interpreted as existential variables. An \textit{RDF triple} $(s,p,o) \in \I\B \times \I \times \I\B\L$ contains a subject $s$, a predicate $p$ and an object $o$.\footnote{In this paper, we abbreviate the union of sets $M_1 \cup \ldots \cup M_n$ with $M_1 \ldots M_n$. Hence, $\I\B\L$ stands for $\I \cup \B \cup \L$.} A set of RDF terms is called an \textit{RDF graph} $G$, where each triple $(s,p,o) \in G$ represents a directed labeled edge $s \xrightarrow{p} o$. The sets $\subjs{G}$, $\preds{G}$ and $\objs{G}$ stand for the set of subjects, predicates and objects in $G$, respectively. We further denote the set of \textit{nodes} in $G$ by $\nodes{G} \da \subjs{G} \cup \objs{G}$.

\begin{figure*}
\setlength{\hgap}{1.9cm}
\setlength{\vgap}{1cm}
\begin{tikzpicture}[baseline]
\node[iri,anchor=mid] (fp) {foaf:Person};

\node[iri,anchor=mid,left=\hgap of fp] (a) {:Alice}
  edge[arrout] node[lab] {rdf:type} (fp);

\node[lit,anchor=mid,above=1.3\vgap of a] (aage) {"26"\dt xsd:int}
  edge[arrin] node[lab,pos=0.6] {foaf:age} (a);
  
\node[iri,anchor=mid,right=\hgap of fp] (b) {:Bob}
  edge[arrout,bend left=14] node[lab] {foaf:knows} (a)
  edge[arrin,bend right=14] node[lab] {foaf:knows} (a)
  edge[arrout] node[lab] {rdf:type} (fp);

\node[lit,anchor=mid,above=1.3\vgap of b] (bage) {"21"\dt xsd:int}
  edge[arrin] node[lab,pos=0.6] {foaf:age} (b);
  
\node[iri,anchor=mid,above=1.3\vgap of fp] (p) {\_:p}
  edge[arrin,bend left=6] node[lab,pos=0.4] {foaf:pastProject} (b)
  edge[arrin,bend right=6] node[lab,pos=0.4] {foaf:currentProject} (a);

\node[iri,anchor=mid,above=\vgap of p] (pr) {foaf:Project}
  edge[arrin] node[lab] {rdf:type} (p);

\node[lit,anchor=mid,above=\vgap of aage] (pnen) {"RDF Engine"@en}
  edge[arrin] node[lab] {rdfs:label} (p);

\node[lit,anchor=mid,above=\vgap of bage] (pnes) {"Motor RDF"@es}
  edge[arrin] node[lab] {rdfs:label} (p);
  
\node[iri,anchor=mid,below=\vgap of a] (sw) {:SW}
  edge[arrin] node[lab] {foaf:topic\_interest} (a);
  
\node[iri,anchor=mid,below=\vgap of b] (db) {:DB}
  edge[arrin] node[lab] {foaf:topic\_interest} (b)
  edge[arrin] node[lab] {skos:related} (sw)
  edge[arrin,bend left=6] node[lab] {foaf:topic\_interest} (a);

\node[iri,anchor=mid,below=\vgap of sw] (web) {:Web}
  edge[arrin] node[lab] {skos:broader} (sw);

\node[iri,anchor=mid,below=\vgap of db] (cs) {:CS}
  edge[arrin] node[lab] {skos:broader} (db)
  edge[arrin] node[lab] {skos:broader} (web);

\end{tikzpicture}
\hfill
\begin{tabular}{lll} 
\toprule
\textbf{Subject} & \textbf{Predicate} & \textbf{Object} \\ [0.4ex] 
 \midrule
 
\texttt{:Alice} & \texttt{rdf:type}  & \texttt{foaf:Person} \\
\texttt{:Alice} & \texttt{foaf:age}  & \texttt{"26"\dt xsd:int} \\ 
\texttt{:Alice} & \texttt{foaf:topic\_interest}  & \texttt{:DB} \\
\texttt{:Alice} & \texttt{foaf:topic\_interest}  & \texttt{:SW} \\ 
\texttt{:Alice} & \texttt{foaf:knows}  & \texttt{:Bob} \\
\texttt{:Alice} & \texttt{foaf:currentProject}  & \texttt{\_:p} \\

\texttt{:Bob} & \texttt{rdf:type}  & \texttt{foaf:Person} \\
\texttt{:Bob} & \texttt{foaf:age}  & \texttt{"21"\dt xsd:int} \\ 
\texttt{:Bob} & \texttt{foaf:topic\_interest}  & \texttt{:DB} \\ 
\texttt{:Bob} & \texttt{foaf:knows}  & \texttt{:Alice} \\ 
\texttt{:Bob} & \texttt{foaf:pastProject}  & \texttt{\_:p} \\ 
 
\texttt{\_:p} & \texttt{rdf:type}  & \texttt{foaf:Project} \\ 
\texttt{\_:p} & \texttt{rdfs:label}  & \texttt{"RDF Engine"@en} \\  
\texttt{\_:p} & \texttt{rdfs:label}  & \texttt{"Motor RDF"@es} \\ 
 
\texttt{:SW} & \texttt{skos:broader} & \texttt{:Web} \\
\texttt{:SW} & \texttt{skos:related} & \texttt{:DB} \\

\texttt{:Web} & \texttt{skos:broader} & \texttt{:CS} \\

\texttt{:DB} & \texttt{skos:broader} & \texttt{:CS} \\

\bottomrule
 \end{tabular}
 
 \caption{Graphical (left) and triple-based representation (right) of an example RDF graph \label{fig:rdfg}}
\end{figure*}
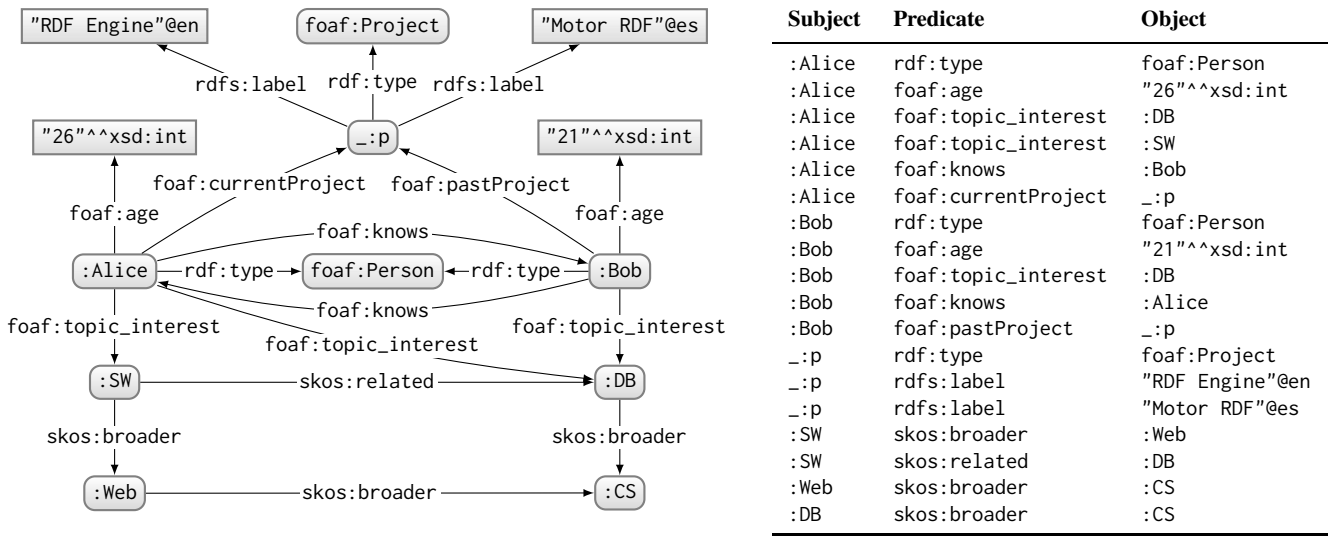

An example RDF graph, representing information about two university students, is shown in \autoref{fig:rdfg}. We include both a graphical representation and a triple-based representation. RDF terms such as \texttt{:DB}, \texttt{foaf:age}, etc., denote prefixed IRIs.\footnote{We use the blank prefix (e.g., \texttt{:DB}) as an arbitrary example. Other prefixes used can be retrieved at \url{http://prefix.cc/}.} For example, \texttt{foaf:age} stands for the full IRI  \url{http://xmlns.com/foaf/0.1/age} if we define the prefix \texttt{foaf} as \url{http://xmlns.com/foaf/0.1/}. Terms such as \texttt{"Motor RDF"@es} denote strings with (optional) language tags, and terms such as \texttt{"21"\dt xsd:int} denote datatype values. Finally we denote blank nodes with the underscore prefix, where \texttt{\_:p} refers to the existence of a project shared by Alice and Bob. Terms used in the predicate position (e.g., \texttt{foaf:age}, \texttt{skos:broader}) are known as \textit{properties}. RDF defines the special property \texttt{rdf:type}, which indicates the \textit{class} (e.g., \texttt{foaf:Person}, \texttt{foaf:Project}) of a resource.

The semantics of RDF can be defined using RDF Schema (RDFS)~\cite{rdfs11}, covering class and property hierarchies, property domains and ranges, etc. Further semantics can be captured with the Web Ontology Language (OWL)~\cite{owl2}, such as class and property equivalence; inverse, transitive, symmetric and reflexive properties; set- and restriction-based class definitions; and more besides. Since our focus is on querying RDF graphs, we do not discuss these standards in detail.

\subsection{SPARQL}

Various query languages for RDF have been proposed down through the years, such as RQL~\cite{KarvounarakisMACPST03}, SeRQL~\cite{StuckenschmidtVBH05}, etc. We focus our discussion on SPARQL~\cite{sparql11}, which is now the standard language for querying RDF, and refer to the work by Haase et al.~\cite{HaaseBEV04} for information on its predecessors. 

We define the core of SPARQL in terms of \textit{basic graph patterns} that express the core pattern matched against an RDF graph; \textit{navigational graph patterns} that match arbitrary-length paths; \textit{complex graph patterns} that introduce various language features, such as \texttt{OPTIONAL}, \texttt{UNION}, \texttt{MINUS}, etc.~\cite{AnglesABHRV17}; and \textit{query types} that specify what result to return.

\paragraph{Basic Graph Patterns (BGPs)} At the core of SPARQL lie \textit{triple patterns}, which are RDF triples that allow variables from the set $\V$ (disjoint with $\I\B\L$) in any position. A \textit{basic graph pattern} (\textit{BGP}) is a set of triple patterns. Since blank nodes in BGPs act as variables, we assume they have been replaced with variables. We use $\vars{B}$ to denote the set of variables in the BGP $B$. Given an RDF graph $G$, the \textit{evaluation} of a BGP $B$, denoted $B(G)$, returns a set of \textit{solution mappings}. A solution mapping $\mu$ is a partial mapping from the set $\V$ of variables to the set of RDF terms $\I\B\L$. We write $\dom{\mu}$ to denote the set of variables for which $\mu$ is defined. Given a triple pattern $t$, we use $\mu(t)$ to refer to the image of $t$ under $\mu$, i.e., the result of replacing any variable $v \in \dom{\mu}$ appearing in $t$ with $\mu(v)$. $\mu(B)$ stands for the image of the BGP $B$ under $\mu$; i.e., $\mu(B) \da \{ \mu(t) \mid t \in B\}$. The evaluation of a BGP $B$ on an RDF graph $G$ is then given as
$B(G) \da \{ \mu \mid \mu(B) \subseteq G\text{ and }\dom{\mu} = \vars{B} \}$. In the case of a singleton BGP $\{t\}$, we may write $\{t\}(G)$ as $t(G)$.

In \autoref{fig:bgp}, we provide an example of a BGP along with its evaluation. Each row of the results refers to a solution mapping. Some solutions map different variables to the same term; each such solution is thus a homomorphism from the BGP to the RDF graph. 

\begin{figure}
\centering
\setlength{\hgap}{1.7cm}
\setlength{\vgap}{1cm}
\begin{lstlisting}[style=sparqld]
SELECT * WHERE {
  ?a a foaf:Person ; foaf:knows ?b ; foaf:topic_interest ?ia .
  ?b a foaf:Person ; foaf:knows ?a ; foaf:topic_interest ?ib .
}
\end{lstlisting}

\vspace{0.2cm}

\begin{tikzpicture}[baseline]
\node[iri,anchor=mid] (fp) {foaf:Person};

\node[var,anchor=mid,left=\hgap of fp] (a) {?a}
  edge[arrout] node[lab] {rdf:type} (fp);
  
\node[var,anchor=mid,right=\hgap of fp] (b) {?b}
  edge[arrout,bend left=20] node[lab] {foaf:knows} (a)
  edge[arrin,bend right=20] node[lab] {foaf:knows} (a)
  edge[arrout] node[lab] {rdf:type} (fp);

\node[var,anchor=mid,below=\vgap of a] (ia) {?ia}
  edge[arrin] node[lab,xshift=0.2cm] {foaf:topic\_interest} (a);

\node[var,anchor=mid,below=\vgap of b] (ib) {?ib}
  edge[arrin] node[lab,xshift=-0.2cm] {foaf:topic\_interest} (b);
\end{tikzpicture}

\vspace{0.2cm}

\begin{tabular}{llll} 
\toprule
\texttt{?a} & \texttt{?b} & \texttt{?ia} & \texttt{?ib} \\ [0.4ex] 
 \midrule
 
\texttt{:Alice} & \texttt{:Bob}  & \texttt{:DB} & \texttt{:DB} \\
\texttt{:Alice} & \texttt{:Bob}  & \texttt{:SW} & \texttt{:DB} \\
\texttt{:Bob} & \texttt{:Alice}  & \texttt{:DB} & \texttt{:DB} \\
\texttt{:Bob} & \texttt{:Alice}  & \texttt{:DB} & \texttt{:SW} \\
\bottomrule
 \end{tabular}
 
 \caption{A BGP in SPARQL syntax and as a graph (above), with its evaluation over the graph of \autoref{fig:rdfg} (below) \label{fig:bgp}}
\end{figure} 

\paragraph{Navigational Graph Patterns (NGPs)} A key feature of graph query languages is the ability to match paths of arbitrary length~\cite{AnglesABHRV17}. In SPARQL~(1.1), this ability is captured by \textit{property paths}~\cite{sparql11}, which are regular expressions $\E$ that paths should match, defined recursively as follows:
\begin{itemize}
\item if $p$ is an IRI, then $p$ is a path expression (\textit{property});
\item if $e$ is a path expression, then $\mcir e$ (\textit{inverse}), $e\texttt{*}$ (\textit{zero-or-more}, aka.\ \textit{Kleene star}), $e\texttt{+}$ (\textit{one-or-more}), and $e\texttt{?}$ (\textit{zero-or-one}) are path expressions.
\item If $e_1,e_2$ are path expressions, then $e_1 / e_2$ (\textit{concatenation}) and $e_1 | e_2$ (\textit{disjunction}) are path expressions.
\item if $P$ is a set of IRIs, then $!P$ and $!\mcir P$ are path expressions (\textit{negated property set});\footnote{SPARQL uses the syntax \texttt{!($p_1$|$\ldots$|$p_k$|$p_{k+1}$|$\ldots$|$p_n$)} which can be written as $!P|!\mcir P'$, where $P = \{p_1,\ldots,p_k\}$ and $P' = \{p_{k+1},\ldots,p_n \}$~\cite{sparql11,KostylevR0V15}.}
\end{itemize}
%

The evaluation of path expressions on an RDF graph $G$ returns pairs of nodes in $G$ connected by paths that match the expression, as defined in \autoref{tab:path}. These path expressions are akin to \textit{2-way regular path queries} (2RPQs) extended with negated property sets~\cite{KostylevR0V15,AnglesABHRV17}.

\begin{table}
\caption{Evaluation of path expressions \label{tab:path}}
\footnotesize
\begin{tabular}{r@{}l@{}l}
\toprule
$\ev{p}{G}$ & $\da$ & $\{ (s,o) \mid (s,p,o) \in G \}$ \\
\midrule
$\ev{\texttt{\mcir}e}{G}$ & $\da$ & $\{ (s,o) \mid (o,s) \in \ev{e}{G} \}$ \\
$\ev{e\texttt{+}}{G}$ & $\da$ & $\{ (y_1,y_n) \mid \text{ for }1 \leq i < n : \exists (y_i,y_{i+1}) \in \ev{e}{G} \}$ \\
$\ev{e\texttt{?}}{G}$ & $\da$ & $\ev{e}{G} \cup \{ (x,x) \mid x \in \so{G} \}$ \\
$\ev{e\texttt{*}}{G}$ & $\da$ & $\ev{(e\texttt{+})\texttt{?}}{G}$ \\
\midrule
$\ev{e_1 \texttt{/} e_2}{G}$ & $\da$ & $\{ (x,z) \mid \exists y : (x,y) \in \ev{e_1}{G} \wedge (y,z) \in \ev{e_2}{G} \}$ \\
$\ev{e_1 \texttt{|} e_2}{G}$ & $\da$ & $\ev{e_1}{G} \cup \ev{e_2}{G}$ \\
\midrule
$\ev{!P}{G}$ & $\da$ & $\{ (s,o) \mid (s,p,o) \in G \wedge p \notin P\}$ \\
$\ev{!\mcir P}{G}$ & $\da$ & $\{ (s,o) \mid (o,p,s) \in G \wedge p \notin P\}$ \\
\bottomrule
\end{tabular}
\end{table}

We call a triple pattern $(s,e,o)$ that further allows a path expression as the predicate (i.e., $e \in \E\V$) a \textit{path pattern}. A \textit{navigational graph pattern} (NGP) is then a set of path patterns. Given a navigational graph pattern $N$, let $\paths{N} \da \preds{N} \cap \E$ denote the set of path expressions used in $N$. Given an RDF graph $G$ and a set of path expressions $E \subseteq \E$, we denote by $G_E \da G \cup (\bigcup_{e \in E} \{ (s,e,o) \mid (s,o) \in \ev{e}{G}\})$ the result of materializing all paths matching $E$ in $G$. The evaluation of the navigational graph pattern $N$ on $G$ is then\\
$N(G) \!\da\! \{ \mu \mid \mu(N) \subseteq G_{\paths{N}}\text{ and }\dom{\mu} = \vars{N} \}$.

We provide an example of a navigational graph pattern and its evaluation in \autoref{fig:ngp}.

\begin{figure}
\centering
\setlength{\hgap}{1.9cm}
\setlength{\vgap}{1cm}
\begin{lstlisting}[style=sparqld]
SELECT * WHERE {
  ?a a foaf:Person ; foaf:knows ?b ;
    foaf:topic_interest/skos:related*/foaf:topic_interest ?b .
  ?b a foaf:Person ; foaf:knows ?a .
}
\end{lstlisting}

\vspace{0.2cm}

\begin{tikzpicture}[baseline]
\node[iri,anchor=mid] (fp) {foaf:Person};

\node[var,anchor=mid,left=\hgap of fp] (a) {?a}
  edge[arrout] node[lab] {rdf:type} (fp);
  
\node[var,anchor=mid,right=\hgap of fp] (b) {?b}
  edge[arrout,bend left=20] node[lab] {foaf:knows} (a)
  edge[arrin,bend right=20] node[lab] {foaf:knows} (a)
  edge[arrin,bend left=40] node[lab] {foaf:topic\_interest/skos:related*/\mcir foaf:topic\_interest} (a)
  edge[arrout] node[lab] {rdf:type} (fp);
\end{tikzpicture}

\vspace{0.4cm}

\begin{tabular}{ll} 
\toprule
\texttt{?a} & \texttt{?b} \\ [0.4ex] 
 \midrule
 
\texttt{:Alice} & \texttt{:Alice} \\
\texttt{:Alice} & \texttt{:Bob}  \\
\texttt{:Bob} & \texttt{:Alice} \\
\texttt{:Bob} & \texttt{:Bob} \\
\bottomrule
 \end{tabular}
 
 \caption{Example NGP (above) and its evaluation over the graph of \autoref{fig:rdfg} (below) \label{fig:ngp}}
\end{figure} 

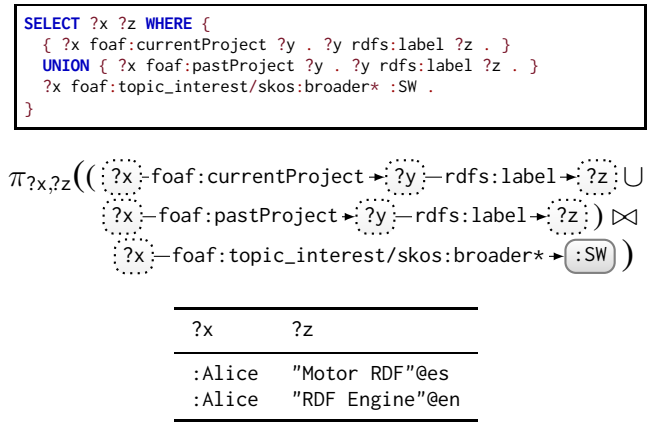
\begin{figure}
\centering
\setlength{\vgap}{1cm}
\begin{lstlisting}[style=sparqld]
SELECT ?x ?z WHERE { 
  { ?x foaf:currentProject ?y . ?y rdfs:label ?z . } 
  UNION { ?x foaf:pastProject ?y . ?y rdfs:label ?z . }
  ?x foaf:topic_interest/skos:broader* :SW .
}
\end{lstlisting}

\vspace{0.2cm}

\begin{tabular}{@{}l@{}l@{}}
{\large $\pi_{\texttt{?x},\!\texttt{?z}}$}{\Large(} & {\large(} \begin{tikzpicture}[baseline=-3pt]
\node[var,anchor=mid] (x1) {?x};

\node[var,anchor=mid,right=1.6\hgap of x1] (y1) {?y}
  edge[arrin] node[lab,pos=0.52] {foaf:currentProject} (x1);
  
\node[var,anchor=mid,right=\hgap of y1] (z1) {?z}
  edge[arrin] node[lab] {rdfs:label} (y1);
\end{tikzpicture} {\large$\cup$} \\[2pt]

 &  ~~ \begin{tikzpicture}[baseline=-3pt]
\node[var,anchor=mid] (x1) {?x};

\node[var,anchor=mid,right=1.4\hgap of x1] (y1) {?y}
  edge[arrin] node[lab] {foaf:pastProject} (x1);
  
\node[var,anchor=mid,right=\hgap of y1] (z1) {?z}
  edge[arrin] node[lab] {rdfs:label} (y1);
\end{tikzpicture} {\large) $\bowtie$} \\[2pt]

& ~~~~ \begin{tikzpicture}[baseline=-3pt]
\node[var,anchor=mid] (x1) {?x};

\node[iri,anchor=mid,right=2.7\hgap of x1] (sw) {:SW}
  edge[arrin] node[lab] {foaf:topic\_interest/skos:broader*} (x1);
\end{tikzpicture} {\Large)}
\end{tabular}

\vspace{0.4cm}

\begin{tabular}{ll} 
\toprule
\texttt{?x} & \texttt{?z} \\ [0.4ex] 
 \midrule

\texttt{:Alice} & \texttt{"Motor RDF"@es} \\
\texttt{:Alice} & \texttt{"RDF Engine"@en} \\
\bottomrule
 \end{tabular}
 
 \caption{Example CGP (above) and its evaluation over the graph of \autoref{fig:rdfg} (below) \label{fig:cgp}}
\end{figure}

\paragraph{Complex Graph Patterns (CGPs)} Complex graph patterns (CGPs) introduce additional language features that can combine and transform the results of one or more graph patterns. More specifically, evaluating BGPs and NGPs returns solution mappings that can be viewed as relations, (i.e., tables), where variables are attributes (i.e., column names) and tuples (i.e., rows) contain the RDF terms bound by each solution mapping (see Figures~\ref{fig:bgp}--\ref{fig:cgp}). CGPs support combining and transforming the results of BGPs/NGPs with language features that include \texttt{FILTER} (selection: $\sigma$), \texttt{SELECT} (projection: $\pi$), \texttt{UNION} (union: $\cup$), \texttt{EXISTS} (semi-join: $\ltimes$), \texttt{MINUS} (anti-join: $\vartriangleright$\footnote{The definition of \texttt{MINUS} is slightly different from anti-join in that mappings with no overlapping variables on the right are ignored.}) and \texttt{OPTIONAL} (left-join: $\loj$). These language features correspond to the relational algebra defined in \autoref{tab:ra}. The default operator is a natural inner join ($\bowtie$).
\autoref{fig:cgp} provides an example of a CGP combining two BGPs and an NGP using union, join and projection.

\begin{table}
\caption{Core relational algebra of SPARQL \label{tab:ra}}
\footnotesize
\begin{tabular}{r@{~}l@{~}l}
\toprule
$\sigma_R(M)$ & $\da$ & $\{\ \mu \in M \mid R(\mu) \}$ \\
$\pi_V(M)$ & $\da$ & $\{\ \mu' \mid \exists \mu \in M\!:\!\com{\mu}{\mu'} \wedge\!\dom{\mu'}\!=\!V\!\cap\!\dom{\mu} \}$ \\
$M_1 \bowtie M_2$ & $\da$ & $\{\ \mu_1 \cup \mu_2 \mid \mu_1 \in M_1 \wedge \mu_2 \in M_2 \wedge \com{\mu_1}{\mu_2} \}$ \\
$M_1 \cup M_2$ & $\da$ & $\{\ \mu \mid \mu \in M_1 \vee \mu \in M_2 \}$ \\
$M_1 \ltimes M_2$ & $\da$ & $\{ \mu_1 \in M_1 \mid \exists \mu_2 \in M_2 : \com{\mu_1}{\mu_2} \}$ \\
$M_1 \vartriangleright M_2$ & $\da$ & $\{\ \mu_1 \in M_1 \mid \nexists \mu_2 \in M_2 : \com{\mu_1}{\mu_2} \}$ \\
$M_1 \loj M_2$ & $\da$ & $(M_1 \bowtie M_2) \cup (M_1 \vartriangleright M_2)$ \\
\bottomrule
\end{tabular}
\end{table}

\paragraph{Named graphs} SPARQL allows for querying multiple RDF graphs through the notion of a \textit{SPARQL dataset}, defined as $D \da \{ G, (n_1,G_1), \ldots, (n_k,G_k))\}$ where $G, G_1 \ldots, G_n$ are RDF graphs; $n_1, \ldots, n_k$ are pairwise distinct IRIs; $G$ is known as the \textit{default graph}; and each pair $(n_1,G_1)$ (for $1 \leq i \leq n$) is known as a \textit{named graph}. Letting $N', N''$ denote sets of IRIs, $n',n''$ IRIs and $v$ a variable, SPARQL then provides a number of features for querying different graphs:

\begin{itemize}
\item \texttt{FROM} $N'$ \texttt{FROM NAMED} $N''$: activates a dataset with a default graph composed of the merge of all graphs $G'$ such that $(n',G') \in D$ and $n' \in N'$, and the set of all named graphs $(n'',G'') \in D$ such that $n'' \in N''$;
\item \texttt{GRAPH} $n'$: evaluates a graph pattern on the graph $G'$ if the named graph $(n',G')$ is active;
\item \texttt{GRAPH} $v$: takes the union of the evaluation of a graph pattern over each $G'$ such that $(n',G')$ is active, binding $v$ to $n'$ for each solution generated from $G'$;
\end{itemize}

\noindent
Without \texttt{FROM} or \texttt{FROM NAMED}, the active dataset is the indexed dataset $D$. Without \texttt{GRAPH}, graph patterns are evaluated on the active default graph. \textit{Quad stores} disallow empty named graphs, such that $D \da \{ G, (n_1,G_1), \ldots, (n_k,G_k))\}$ is viewed as $D = G \times \{\star \} \cup (\bigcup_{(n_i,G_i) \in D} G_i \times \{ n_i \})$, i.e., a set of quads using $\star \not\in \I\B\L$ as a special symbol for the default graph. In this case, a quad $(s,p,o,n)$ denotes a triple $(s,p,o)$ in the default graph if $n = \star$, or a triple in the named graph $G'$ such that $(n,G') \in D$ if $n \in \I$. We can define CGPs involving quad patterns analogously.

\paragraph{Other SPARQL features} SPARQL supports features beyond CGPs, which include aggregation (group-by with count, sum, etc.), solution modifiers (ordering and slicing solutions), bag semantics (preserving result multiplicity), federation (fetching solutions from remote services), entailment and more besides. SPARQL also supports different query types, such as \texttt{SELECT}, which returns a sequence of solution mappings; \texttt{CONSTRUCT}, which returns an RDF graph based on the solution mappings; \texttt{DESCRIBE}, which returns an RDF graph describing indicated RDF terms; and \texttt{ASK}, which returns true if some solution mapping is found, or false otherwise. 

\section{Storage}
\label{sec:storage}

Data storage refers to how data are represented in memory. Different storage mechanisms store different elements of data contiguously in memory, offering trade-offs in terms of compression and efficient data access. This section reviews various categories of RDF storage.

\subsection{Triple table}

A \textit{triple table} stores an RDF graph $G$ as a single ternary relation. \autoref{fig:rdfg} shows an RDF graph with its triple table on the right-hand side. One complication when storing triple tables in relational databases is that such systems assume a column to have a single type, which may not be true for RDF objects in particular; a workaround is to store a string encoding of the terms, though this may complicate their ordering.

Rather than storing full RDF terms in the triple table, stores may apply dictionary encoding, where RDF terms are mapped one-to-one with numeric object identifiers (OIDs), with OIDs being stored in the table and decoded using the dictionary as needed. Since OIDs consume less memory and are faster to process than strings, such an approach works better for queries that involve many intermediate results but generate few final results; on the other hand, such an approach suffers when queries are simple and return many results, or when selective filters are specified that require decoding the term before filtering. To find a better trade-off, some RDF engines (e.g., Jena 2 \cite{jena2}) only use OIDs for
strings with lengths above a threshold.
 
The most obvious physical storage is to store triples contiguously (row-wise). This allows for quickly retrieving the full triples that match (e.g.) a given triple pattern. However, some RDF engines based on relational storage  (e.g., Virtuoso~\cite{Erling2010}) rather use (or provide an option for) column-wise storage, where the values along a column are stored contiguously, often following a particular order. Such column-wise storage allows for better compression, and for quickly reading many values from a single column.

Triple tables can be straightforwardly extended to quad tables in order to support SPARQL datasets~\cite{Erling2010,harris4store}.

\subsection{Vertical partitioning}\label{ssec:vp}

The \textit{vertical partitioning} approach~\cite{swstore} uses a binary relation for each property $p \in \preds{G}$ whose tuples encode subject--object pairs for that property. In \autoref{fig:vp} we exemplify two such binary relations. Physical storage can again use OIDs, row-based or column-based storage, etc.

When compared with triple tables, vertical partitioning generates relations with fewer rows, and more specific domains for columns (e.g., the object column for \texttt{foaf:age} can be defined as an integer type). However, triple patterns with variable predicates may require applying a union on all relations. Also, RDF graphs may have thousands of properties~\cite{wikidata}, which may lead to a schema with many relations.

Vertical partitioning can be used to store quads by adding a \textbf{Graph} column to each table~\cite{Erling2010,harris4store}.

\begin{figure}[t]
  \footnotesize
  \begin{tabular}{ll} 
     \multicolumn{2}{c}{\texttt{rdf:type}}\\
     \midrule
     \textbf{Subject} & \textbf{Object} \\ 
     \midrule
     \texttt{:Alice} & \texttt{foaf:Person}    \\
     \texttt{:Bob} & \texttt{foaf:Person}   \\
     \texttt{\_:p} & \texttt{foaf:Project}   \\
    \bottomrule  
 \end{tabular}
 \hfill
 \begin{tabular}{ll} 
     \multicolumn{2}{c}{\texttt{foaf:age}}\\
     \midrule
     \textbf{Subject} & \textbf{Object} \\ 
     \midrule
     \texttt{:Alice} & 26    \\
     \texttt{:Bob} & 21   \\
    \bottomrule 
 \end{tabular}
  \caption{Vertical partitioning for two properties in \autoref{fig:rdfg} \label{fig:vp}}
\end{figure}

\subsection{Extended vertical partitioning}\label{sec:evp}

S2RDF~\cite{Schtzle2015S2RDFRQ} uses \textit{extended vertical partitioning} based on \textit{semi-join reductions} (we recall from Table~\ref{tab:ra} that a semi-join $M_1 \ltimes M_2$, aka.\ \texttt{FILTER EXISTS}, returns the tuples in $M_1$ that are ``joinable''with $M_2$). Letting $\vbx, \vby, \vbz$ denote variables and $\cnp, \cnq$ denote RDF terms, then for each property pair $(\cnp,\cnq) \in \preds{G} \times \preds{G}$ such that $\cnp \neq \cnq$, extended vertical partitioning stores three semi-join reductions: 
\begin{enumerate}
\item $(\vbx,\cnp,\vby)(G) \ltimes (\vby,\cnq,\vbz)(G)$ (\textsc{o}--\textsc{s}), 
\item $(\vbx,\cnp,\vby)(G) \ltimes (\vbx,\cnq,\vbz)(G)$ (\textsc{s}--\textsc{s}),
\item $(\vbx,\cnp,\vby)(G) \ltimes (\vbz,\cnq,\vbx)(G)$ (\textsc{s}--\textsc{o}). 
\end{enumerate}
The semi-join $(\vbx,\cnp,\vby)(G) \ltimes (\vbz,\cnq,\vby)(G)$ (\textsc{o}--\textsc{o}) is not stored as most \textsc{o}--\textsc{o} joins have the same predicate, and thus would occur in the same relation. In Figure~\ref{fig:sj} we give an example of a semi-join reduction for two predicates from the running example; empty semi-joins are omitted. 

In comparison with vertical partitioning, observing that $(M_1 \ltimes M_2) \Join (M_2 \ltimes M_1) \equiv M_1 \Join M_2$, we can apply joins over the corresponding semi-join reductions knowing that each tuple read from each side will contribute to the join, thus reducing I/O. The cost involves storing (and updating) each tuple in up to $3(|\preds{G}|-1)$ additional relations; omitting empty semi-joins can help to mitigate this issue~\cite{Schtzle2015S2RDFRQ}. Extended vertical partitioning also presents complications for variable predicates, graphs with many properties, etc.

\begin{figure}[tb]
\centering
\footnotesize 

\begin{tabular}{ll} 
\multicolumn{2}{c}{\texttt{skos:broader}} \\
\multicolumn{2}{c}{$\ltimes_{\textsc{s--s}}\, \texttt{skos:related}$} \\
\midrule
\textbf{Subject} & \textbf{Object} \\
\midrule
\texttt{:SW} & \texttt{:Web}\\
\bottomrule
\end{tabular}
\hspace{0.2cm}
\begin{tabular}{ll} 
\multicolumn{2}{c}{\texttt{skos:broader}} \\
\multicolumn{2}{c}{$\ltimes_{\textsc{s--o}}\, \texttt{skos:related}$} \\
\midrule
\textbf{Subject} & \textbf{Object} \\
\midrule
\texttt{:DB} & \texttt{:CS}\\
\bottomrule
\end{tabular}

\vspace{0.4cm}

\begin{tabular}{ll} 
\multicolumn{2}{c}{\texttt{skos:related}} \\
\multicolumn{2}{c}{$\ltimes_{\textsc{o--s}}\, \texttt{skos:broader}$} \\
\midrule
\textbf{Subject} & \textbf{Object} \\
\midrule
\texttt{:SW} & \texttt{:DB}\\
\bottomrule
\end{tabular}
\hspace{0.2cm}
\begin{tabular}{ll} 

\multicolumn{2}{c}{\texttt{skos:related}} \\
\multicolumn{2}{c}{$\ltimes_{\textsc{s--s}}\, \texttt{skos:broader}$} \\
\midrule
\textbf{Subject} & \textbf{Object} \\
\midrule
\texttt{:SW} & \texttt{:DB}\\
\bottomrule
\end{tabular}
 \caption{Example semi-join reduction for two properties \label{fig:sj}}
\end{figure}

 \subsection{Property table}
 
\textit{Property tables} aim to emulate the $n$-ary relations typical of relational databases. A property table usually contains one subject column, and $n$ further columns to store objects for the corresponding properties of the given subject. The subject column then forms a primary key for the table. The tables to define can be based on classes, clustering~\cite{PhamB16}, coloring~\cite{10.1145/2463676.2463718}, etc., to group subjects with common properties. We provide an example of a property table based on the class \texttt{foaf:Person} in \autoref{fig:pt} for the RDF graph of \autoref{fig:rdfg}.

Property tables can store and retrieve multiple triples with a given subject as one tuple (e.g., to find people with $\textbf{age}<30$ and $\textbf{interest}=\texttt{:SW}$) without needing joins. Property tables often store terms of the same type in the same column, enabling better compression. Complications arise for multi-valued (\ldots-to-many) or optional (zero-to-\ldots) properties. In the example of \autoref{fig:rdfg}, Alice is also interested in SW, which does not fit in the cell. Furthermore, Alice has no past project, and Bob has no current project, leading to nulls. 
Changes to the graph may also require re-normalization; for example, even though each person currently has only one value for \textbf{knows}, adding that Alice knows another person would require renormalizing the tables. Complications also arise when considering variable predicates, RDF graphs with many properties or classes, quads, etc.

\begin{figure}[tb]
  \footnotesize
  \centering
  \begin{tabular}{lrllll} 
     \multicolumn{6}{c}{\texttt{foaf:Person}}\\
     \midrule
     \textbf{Subject} & \textbf{age} & \textbf{topic} & \textbf{knows} & \textbf{cProj}  & \textbf{pProj} \\ 
     \midrule
     \texttt{:Alice} & \texttt{26} & \cellcolor{red!30!white}\texttt{:DB} & \texttt{:Bob} & \texttt{\_:p} & \textsc{null}   \\
     \texttt{:Bob} & \texttt{21} & \texttt{:DB} & \texttt{:Alice} & \textsc{null}  & \texttt{\_:p}   \\
    \bottomrule  
 \end{tabular}
 \caption{Example property table for people \label{fig:pt}}
\end{figure}

\subsection{Graph-based storage}\label{sec:gstore}

While the previous three storage mechanisms rely on relational storage, \textit{graph-based storage} is adapted specifically for the graph-based model of RDF. Key characteristics of such models that can be exploited for storage include the \textit{adjacency} of nodes, the fixed \textit{arity} of graphs, etc. 

Graphs have bounded arity (3 for triples, 4 for quads), which can be exploited for specialized storage. Engines like 4store~\cite{harris4store} and YARS2~\cite{10.5555/1785162.1785179} build \textit{native triple/quad tables}, which differ from relational triple/quad tables in that they have fixed arity, fixed attributes (\textsc{s},\textsc{p},\textsc{o}(,\textsc{g})), and more general domains (e.g., the \textsc{o} column can contain any RDF term). 

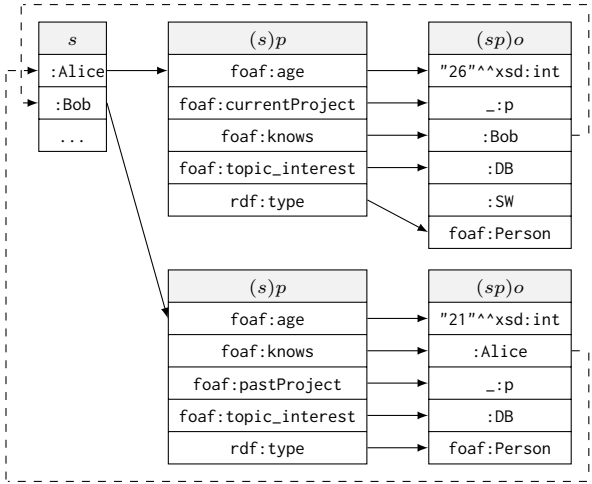
\begin{figure}[t]
\setlength{\hgap}{0.8cm}
\setlength{\vgap}{0.7cm}
\newcommand{\pd}{{\color{white}.}}
\def\mystrut{\vrule height 8pt depth 3pt width 0pt} 
\centering
\begin{tikzpicture}
\tikzset{every node/.style={inner sep=0.4pt}}
\node [map,rectangle split,rectangle split parts=4] (sa)  {\nodepart{one}\mystrut $s$   \nodepart{two}\mystrut \texttt{~:Alice~} \nodepart{three}\mystrut \texttt{:Bob} \nodepart{four}\mystrut \texttt{... }};

\node [map,rectangle split,rectangle split parts=6,right=\hgap of sa.north east,anchor=north west] (pa)  {\nodepart{one}\mystrut $(s)p$  \nodepart{two}\mystrut \texttt{foaf:age} \nodepart{three}\mystrut \texttt{foaf:currentProject} \nodepart{four}\mystrut \texttt{foaf:knows} \nodepart{five}\mystrut \texttt{\pd foaf:topic\_interest\pd}  \nodepart{six}\mystrut \texttt{rdf:type}};

\node [map,rectangle split,rectangle split parts=7,right=\hgap of pa.north east,anchor=north west] (oa)  {\nodepart{one}\mystrut $(sp)o$  \nodepart{two}\mystrut \texttt{\pd"26"\dt xsd:int\pd} \nodepart{three}\mystrut \texttt{\_:p} \nodepart{four}\mystrut \texttt{:Bob}  \nodepart{five}\mystrut \texttt{:DB}  \nodepart{six}\mystrut \texttt{:SW}  \nodepart{seven}\mystrut  \texttt{foaf:Person}};

\draw  (sa.two east) edge[arrout] (pa.two west);
\draw  (pa.two east) edge[arrout] (oa.two west);
\draw  (pa.three east) edge[arrout] (oa.three west);
\draw  (pa.four east) edge[arrout] (oa.four west);
\draw  (pa.five east) edge[arrout] (oa.five west);
\draw  (pa.six east) edge[arrout] (oa.seven west);

\node [map,rectangle split,rectangle split parts=6,below=\vgap of pa.south,anchor=north] (pb)  {\nodepart{one}\mystrut $(s)p$  \nodepart{two}\mystrut \texttt{foaf:age} \nodepart{three}\mystrut \texttt{foaf:knows} \nodepart{four}\mystrut \texttt{foaf:pastProject} \nodepart{five}\mystrut \texttt{\pd foaf:topic\_interest\pd}  \nodepart{six}\mystrut \texttt{rdf:type}};

\node [map,rectangle split,rectangle split parts=6,right=\hgap of pb.north east,anchor=north west] (ob)  {\nodepart{one}\mystrut $(sp)o$  \nodepart{two}\mystrut  \texttt{\pd"21"\dt xsd:int\pd}   \nodepart{three}\mystrut \texttt{:Alice} \nodepart{four}\mystrut \texttt{\_:p}  \nodepart{five}\mystrut \texttt{:DB}  \nodepart{six}\mystrut \texttt{foaf:Person}};

\draw  (sa.three east) edge[arrout] (pb.two west);
\draw  (pb.two east) edge[arrout] (ob.two west);
\draw  (pb.three east) edge[arrout] (ob.three west);
\draw  (pb.four east) edge[arrout] (ob.four west);
\draw  (pb.five east) edge[arrout] (ob.five west);
\draw  (pb.six east) edge[arrout] (ob.six west);

\node [right=0.2cm of oa.four east] (b1) {};
\node [above=0.2cm of oa.one north] (x) {};
\node at (x-|b1) (b2) {};
\node [left=0.2cm of sa.three west] (b4) {};
\node at (x-|b4) (b3) {};

\draw  (oa.four east) edge[dashed] (b1); 
\draw  (b1) edge[dashed] (b2);
\draw  (b2) edge[dashed] (b3);
\draw  (b3) edge[dashed] (b4);
\draw  (b4) edge[arrout,dashed] (sa.three west);

\node [right=0.2cm of ob.three east] (a1) {};
\node [below=0.2cm of ob.six south] (y) {};
\node at (y-|a1) (a2) {};
\node [left=0.4cm of sa.two west] (a4) {};
\node at (y-|a4) (a3) {};

\draw  (ob.three east) edge[dashed] (a1); 
\draw  (a1) edge[dashed] (a2);
\draw  (a2) edge[dashed] (a3);
\draw  (a3) edge[dashed] (a4);
\draw  (a4) edge[arrout,dashed] (sa.two west);
\end{tikzpicture}

\caption{Example adjacency list for two subjects with dashed links indicating index-free adjacency pointers} \label{fig:adj}
\end{figure}

Graphs often feature local repetitions that are compressible with \textit{adjacency lists} (e.g., Hexastore~\cite{Weiss2008HexastoreSI}, gStore~\cite{Zou2011}, SpiderStore~\cite{BinnaGZPS11}, Trinity.RDF~\cite{Zeng:2013:DGE:2488329.2488333}, GRaSS~\cite{LyuWLFW15}). These lists are akin to tries, where subject or subject--predicate prefixes are followed by the rest of the triple. Such tries can be stored row-wise in blocks of triples; or column-wise, where blocks elements from one column point to blocks of elements from the next column. \textit{Index-free adjacency} can enable efficient navigation, where terms in the suffix directly point to the location on disk of their associated prefix. We refer to \autoref{fig:adj} for an example. Such structures can also include inverse edges (e.g., Trinity.RDF~\cite{Zeng:2013:DGE:2488329.2488333}, GRaSS~\cite{LyuWLFW15}). 

An alternative is to decompose an RDF graph into its constituent components for storage. AMBER~\cite{doi:10.1002/9781119528227.ch5} uses a \textit{multigraph representation} where an RDF graph $G$ is decomposed into a set of (non-literal) nodes $V \da \nodes{G} \cap \I\B$, 
a set of edges $E \da \{ (s,o) \in V\times V \mid \exists p: (s,p,o) \in G \}$, an edge-labeling function of the form $L_E : V \rightarrow 2^\I$ such that $L_E(s,o) \da \{ p \mid (s,p,o) \in G\}$, and an attribute-labeling function of the form $L_V : \I\B \rightarrow 2^{\I \times \L}$ such that $L_V(s) \da \{ (p,o) \mid (s,p,o) \in G \wedge o \in \L\}$, as seen in \autoref{fig:mgr} (in practice, AMBER uses dictionary-encoding).

\begin{figure}[tb]
\centering
\footnotesize 
$L_E$:~~\begin{tabular}{ll} 
\toprule
\textbf{Edge}  & \textbf{Label}  \\ \midrule
(\texttt{:Alice},\texttt{:Bob}) & $\{\texttt{foaf:knows}\}$ \\
(\texttt{:Alice},\texttt{:DB}) & $\{\texttt{foaf:topic\_interest}\}$ \\
(\texttt{:Bob},\texttt{:Alice}) & $\{\texttt{foaf:knows}\}$ \\
\ldots & \ldots \\
\bottomrule
\end{tabular}
\vspace{0.2cm}

$L_V$:~~\begin{tabular}{ll} 
\toprule
\textbf{Node}  & \textbf{Attributes}  \\ \midrule
\texttt{:Alice} & $\{ (\texttt{foaf:age},\texttt{"26"\dt xsd:int})\}$ \\
\texttt{:Bob} & $\{ (\texttt{foaf:age},\texttt{"21"\dt xsd:int})\}$ \\
\texttt{:DB} & $\{ \}$ \\
\ldots & \ldots \\
\bottomrule
 \end{tabular}
\caption{Example of the multi-graph representation \label{fig:mgr}}
\end{figure}

\subsection{Tensor-based storage}\label{sec:tensor}

Another type of native graph storage uses tensors, viewing a dictionary-encoded RDF graph $G$ with $m = |\so{G}|$ nodes and $n = |\preds{G}|$ predicates as an $m \times n \times m$ 3-order tensor $\mathfrak{T}$ of bits such that $\mathfrak{T}_{i,j,k} = 1$ if the $i$\textsuperscript{th} node links to the $k$\textsuperscript{th} node with the $j$\textsuperscript{th} property, or $\mathfrak{T}_{i,j,k} = 0$ otherwise. A popular variant uses an adjacency matrix per property (e.g., BitMat~\cite{atrebitmat}, BMatrix~\cite{BrisaboaCBF20}, QDags~\cite{NavarroRR20}), akin to vertical partitioning, as seen in \autoref{fig:bitmat}. A third option (considered, e.g., by MAGiQ~\cite{JamourACK19}) is to encode the full graph as an adjacency matrix where each cell indicates the property id connecting the two nodes; this matrix cannot directly represent pairs of nodes connected by more than one property.

While abstract tensor-based representations may lead to highly-sparse matrices or tensors, compact data structures offer compressed representations that support efficient operations~\cite{atrebitmat,JamourACK19,BrisaboaCBF20,NavarroRR20}. Often such matrices/tensors are stored in memory, or loaded into memory when needed. Such representations may also enable query processing techniques that leverage hardware acceleration, e.g., for processing joins on GPUs (as we will discuss in Section~\ref{sec:la}).

\begin{figure}[tb]
\centering
\footnotesize 
\begin{tabular}{rcccc} 
\multicolumn{5}{c}{\texttt{skos:broader}} \\
\midrule
 \cellcolor{gray!15} & \texttt{\cellcolor{gray!15}:CS}  & \texttt{\cellcolor{gray!15}:DB} & \texttt{\cellcolor{gray!15}:SW} & \texttt{\cellcolor{gray!15}:Web} \\ 
\texttt{\cellcolor{gray!15}:CS} & 0 & 0 & 0 & 0 \\
\texttt{\cellcolor{gray!15}:DB} & 1 & 0 & 0 & 0 \\
\texttt{\cellcolor{gray!15}:SW} & 0 & 0 & 0 & 1 \\
\texttt{\cellcolor{gray!15}:Web} & 1 & 0 & 0 & 0 \\
\bottomrule
 \end{tabular}
 
 \caption{Example bit matrix for \texttt{skos:broader} \label{fig:bitmat}}
\end{figure}

\subsection{Miscellaneous storage}

Aside from relational-based and graph-based storage, other engines have proposed to leverage other forms of storage as implemented by existing systems. A common example is the use of NoSQL key-value, tabular or document stores for distributed storage (see~\cite{janke2018storing,10.1145/3341105.3375753,Wylot2018RDFDS} for more details).

\subsection{Discussion}

Early works on storing RDF tended to rely on relational storage, which had been subject to decades of developments and optimizations before the advent of RDF (e.g.,~\cite{jena2,swstore,Erling2010}). Though such an approach still has broad adoption~\cite{Erling2010}, more recent storage techniques aim to exploit the graph-based characteristics of RDF -- and SPARQL -- in order to develop dedicated storage techniques (e.g.,~\cite{atrebitmat,Weiss2008HexastoreSI,Zou2011}), including those based on tensors/matrices~\cite{atrebitmat,JamourACK19,BrisaboaCBF20,NavarroRR20}. A recent trend is to leverage NoSQL storage (e.g.,~\cite{Ladwig_cumulusrdf:linked,Papailiou2013H2RDFHD,rdfmongo}) in order to distribute the management of RDF data.

\section{Indexing} 
\label{sec:index}

Indexing enables efficient lookup operations on RDF graphs (i.e., $O(1)$ or $O(\log|G|)$ time to return the first result or an empty result). The most common such operation is to find triples that match a given triple pattern. However, indexes can also be used to match non-singleton BGPs (with more than one triple pattern), to match path expressions, etc. We now discuss indexing techniques proposed for RDF graphs.

\subsection{Triple indexes}

The goal of \textit{triple indexes} is to efficiently find triples matching a triple pattern. Letting $\cns,\cnp,\cno$ denote RDF terms and $\vbs,\vbp,\vbo$ variables, there are $2^3 = 8$ abstract patterns: $(\vbs,\vbp,\vbo)$, $(\vbs,\vbp,\cno)$, $(\vbs,\cnp,\vbo)$, $(\cns,\vbp,\vbo)$, $(\vbs,\cnp,\cno)$, $(\cns,\vbp,\cno)$, $(\cns,\cnp,\vbo)$ and $(\cns,\cnp,\cno)$. Unlike relational data\-bases, where often only the primary key of a relation will be indexed by default and further indexes must be manually specified, most RDF stores aim to have a complete index by default, covering all eight possible triple patterns. However, depending on the type of storage chosen, this might not always be feasible.

\begin{figure}
\setlength{\hgap}{0.08cm}
\setlength{\vgap}{0.4cm}
\newcommand{\frc}{0.4}
\newcommand{\sfrc}{0.4}
$\begin{pmatrix}
\begin{tabular}{|c|c|cc|}
\hline
0 & 0 & 0 & 0 \\\cline{1-2}
1 & 0 & 0 & 0 \\\hline
0 & 0 & \multicolumn{1}{c|}{0} & 1 \\\hline
1 & 0 & \multicolumn{1}{c|}{0} & 0 \\\hline
\end{tabular}
\end{pmatrix}$
\hfill
\begin{tikzpicture}[baseline=14pt]
\node[rectangle,draw,fill=white] (b11) {};
\node[rectangle,draw,fill=white,right=\hgap of b11] (b12) {};
\node[rectangle,draw,fill=black,right=\hgap of b12] (b13) {};
\node[rectangle,draw,fill=white,right=\hgap of b13] (b14) {};

\node (b1) at ($(b12)!0.5!(b13)$) {};

\node[rectangle,fill=white,right=\hgap of b14] (b21) {};
\node[rectangle,fill=white,right=\hgap of b21] (b22) {};
\node[rectangle,fill=white,right=\hgap of b22] (b23) {};
\node[rectangle,fill=white,right=\hgap of b23] (b24) {};

\node (b2) at ($(b22)!0.5!(b23)$) {};

\node[rectangle,draw,fill=white,right=\hgap of b24] (b31) {};
\node[rectangle,draw,fill=white,right=\hgap of b31] (b32) {};
\node[rectangle,draw,fill=black,right=\hgap of b32] (b33) {};
\node[rectangle,draw,fill=white,right=\hgap of b33] (b34) {};

\node (b3) at ($(b32)!0.5!(b33)$) {};

\node[rectangle,draw,fill=white,right=\hgap of b34] (b41) {};
\node[rectangle,draw,fill=black,right=\hgap of b41] (b42) {};
\node[rectangle,draw,fill=white,right=\hgap of b42] (b43) {};
\node[rectangle,draw,fill=white,right=\hgap of b43] (b44) {};

\node (b4) at ($(b42)!0.5!(b43)$) {};

\node[rectangle,draw,fill=gray,above=\vgap of b1] (a1) {};
\node[rectangle,draw,fill=white,above=\vgap of b2] (a2) {};
\node[rectangle,draw,fill=gray,above=\vgap of b3] (a3) {};
\node[rectangle,draw,fill=gray,above=\vgap of b4] (a4) {};

\node (a) at ($(a2)!0.5!(a3)$) {};

\node[rectangle,draw,fill=gray,above=\vgap of a] (r) {};

\draw (r) edge[arrout] (a1);
\draw (r) edge[arrout] (a2);
\draw (r) edge[arrout] (a3);
\draw (r) edge[arrout] (a4);

\draw (a1) edge[arrout] (b11);
\draw (a1) edge[arrout] (b12);
\draw (a1) edge[arrout] (b13);
\draw (a1) edge[arrout] (b14);

\draw (a3) edge[arrout] (b31);
\draw (a3) edge[arrout] (b32);
\draw (a3) edge[arrout] (b33);
\draw (a3) edge[arrout] (b34);

\draw (a4) edge[arrout] (b41);
\draw (a4) edge[arrout] (b42);
\draw (a4) edge[arrout] (b43);
\draw (a4) edge[arrout] (b44);
\end{tikzpicture}

\caption{A quadtree index based on the bit matrix of Figure~\ref{fig:bitmat}; the root represents the full matrix, while children denote four sub-matrices of the parent; a node is colored black if it contains only 1's, white if it contains only 0's, and gray if it contains both; only gray nodes require children \label{fig:qtree}}
\end{figure}
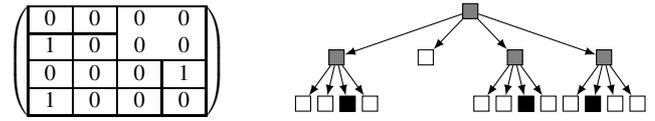

When a storage scheme such as vertical partitioning is used~\cite{swstore}, only the five patterns where the predicate is constant can be efficiently supported (by indexing the subject and object columns). If the RDF graph is stored as a (binary) adjacency matrix for each property~\cite{atrebitmat,NavarroRR20}, again only constant-predicate patterns can be efficiently supported. Specialized indexes can be used to quickly evaluate such patterns, where QDags~\cite{NavarroRR20} uses \textit{quadtrees}: a hierarchical index structure that recursively divides the matrix into four sub-matrices; we provide an example quadtree in Figure~\ref{fig:qtree}. A similar structure, namely a $k^2$\textit{-tree}, is used by BMatrix~\cite{BrisaboaCBF20}.

Otherwise, in triple tables, or similar forms of graph-based storage, all triple patterns can be efficiently supported with \textit{triple permutations}. \autoref{fig:adj} illustrates a single \textsc{spo} permutation. A total of $3! = 6$ permutations are possible and suffice to cover all eight abstract triple patterns if the index structure permits prefix lookups; for example, in an \textsc{spo} permutation we can efficiently support four abstract triple patterns $(\vbs,\vbp,\vbo)$, $(\cns,\vbp,\vbo)$, $(\cns,\cnp,\vbo)$ and $(\cns,\cnp,\cno)$ as we require the leftmost terms of the permutation to be filled. In fact, with only ${3 \choose \lfloor 3/2 \rfloor} = 3$ permutations -- e.g., \textsc{spo}, \textsc{pos} and \textsc{osp} -- we can cover all eight abstract triple patterns.
Such index permutations can be implemented using standard data structures such as ISAM files~\cite{10.5555/1785162.1785179}, B(+)Trees~\cite{Neumann2010}, AVL trees~\cite{kowari}, as well as compact data structures, such as adjacency lists~\cite{Weiss2008HexastoreSI} (see \autoref{fig:adj}) and tries~\cite{8959165}, etc.

Recent works use compact data structures to reduce redundancy for index permutations, and thus the space required for triple indexing. Perego et al.~\cite{8959165} use tries to index multiple permutations, over which they apply \textit{cross-compression}, whereby the order of the triples given by one permutation is used to compress another permutation. Other approaches remove the need for multiple permutations.  RDFCSA~\cite{BrisaboaCFN15} and Ring~\cite{ArroyueloHNRRS21} use a \textit{compact suffix-array} (\textit{CSA}) such that one permutation suffices to efficiently support all triple patterns. Intuitively speaking, triples can be indexed cyclically in a CSA, such that in an \textsc{spo} permutation, one can continue from \textsc{o} back to \textsc{s}, thus covering \textsc{spo}, \textsc{pos} and \textsc{osp} permutations in one CSA index~\cite{BrisaboaCFN15}. The Ring indexing scheme is also bidirectional, where in an \textsc{spo} permutation, one can move from \textsc{o} forwards to \textsc{s} or backwards to \textsc{p}. 

\subsection{Entity-based indexes}\label{sec:estore}

\textit{Entity-based indexes} optimize graph patterns that ``center on'' a particular entity. BGPs can be reduced to joins over their triple patterns; for example, $\{(\vbx,\cnp,\vby), (\vby,\cnq,\vbz)\}(G) \!=\! \{ (\vbx,\cnp,\vby) \} (G) \!\Join\! \{ (\vby,\cnp,\vbz) \}(G)$. \textit{Star joins} are frequently found in BGPs, defined to be a join on a common subject, e.g., $\{(\vbw,\cnp,\vbx), (\vbw,\cnq,\vby), (\vbw,\cnr,\cnz)\}$. Star joins may sometimes also include \textsc{s}--\textsc{o} joins on the common variable, e.g., $\{(\vbw,\cnp,\vbx), (\vbw,\cnq,\vby), (\cnz,\cnr,\vbw)\}$~\cite{LyuWLFW15}. Star joins retrieve data surrounding a particular entity (in this case $\vbw$). Entity-based indexes permit efficient evaluation of such joins.

Property tables can enable efficient star joins so long as the relevant tables can be found efficiently and there are indexes on the relevant columns (e.g., for $\cnp$, $\cnq$ and/or $\cnr$). 

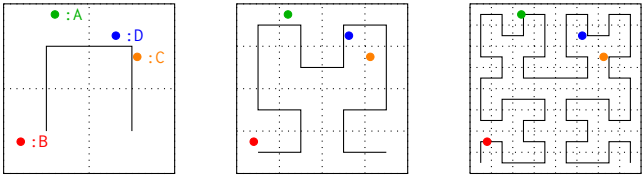
\begin{figure}
\newlength{\ul}
\setlength{\ul}{4pt}
\newlength{\dl}
\setlength{\dl}{1.4pt}
\begin{tikzpicture}[x=\ul,y=\ul] 
\draw (0,0) -- (0,16) -- (16,16) -- (16,0) -- (0,0);
\draw[step=8\ul,dotted] (0,0) grid (16,16);
\draw (4,4) -- (4,12) -- (12,12) -- (12,4);

\filldraw[color=green!70!black] (4.8,15) circle[radius=\dl];
\node[right=0.3pt of {(4.8,15)}] {\scriptsize\color{green!70!black}\texttt{:A}};
\filldraw[color=red] (1.6,3) circle[radius=\dl];
\node[right=0.3pt of {(1.6,3)}] {\scriptsize\color{red}\texttt{:B}};
\filldraw[color=orange] (12.5,11) circle[radius=\dl];
\node[right=0.3pt of {(12.5,11)}] {\scriptsize\color{orange}\texttt{:C}};
\filldraw[color=blue] (10.5,13) circle[radius=\dl];
\node[right=0.3pt of {(10.5,13)}] {\scriptsize\color{blue}\texttt{:D}};
\end{tikzpicture}
\hfill
\begin{tikzpicture}[x=\ul,y=\ul] 
\draw (0,0) -- (0,16) -- (16,16) -- (16,0) -- (0,0);
\draw[step=4\ul,dotted] (0,0) grid (16,16);
\draw (2,2) -- (6,2) -- (6,6) -- (2,6) -- (2,14) -- (6,14) -- (6,10) -- (10,10) -- (10,14) -- (14,14) -- (14,6) -- (10,6) -- (10,2) -- (14,2);

\filldraw[color=green!70!black] (4.8,15) circle[radius=\dl];
\filldraw[color=red] (1.6,3) circle[radius=\dl];
\filldraw[color=orange] (12.5,11) circle[radius=\dl];
\filldraw[color=blue] (10.5,13) circle[radius=\dl];
\end{tikzpicture}
\hfill
\begin{tikzpicture}[x=\ul,y=\ul] 
\draw (0,0) -- (0,16) -- (16,16) -- (16,0) -- (0,0);
\draw[step=2\ul,dotted] (0,0) grid (16,16);
\draw (1,1) -- (1,3) -- (3,3) -- (3,1) -- (7,1) -- (7,3) -- (5,3) -- (5,5) -- (7,5) -- (7,7) -- (3,7) -- (3,5) -- (1,5) -- (1,9) -- (3,9) -- (3,11) -- (1,11) -- (1,15) -- (3,15) -- (3,13) -- (5,13) -- (5,15) -- (7,15) -- (7,11) -- (5,11)  -- (5,9)  -- (11,9) -- (11,11) -- (9,11) -- (9,15) -- (11,15) -- (11,13) -- (13,13) -- (13,15) -- (15,15) -- (15,11) -- (13,11) -- (13,9) -- (15,9) -- (15,5) -- (13,5) -- (13,7) -- (9,7) -- (9,5)  -- (11,5) -- (11,3) -- (9,3) -- (9,1) -- (13,1) -- (13,3) -- (15,3) -- (15,1);

\filldraw[color=green!70!black] (4.8,15) circle[radius=\dl];
\filldraw[color=red] (1.6,3) circle[radius=\dl];
\filldraw[color=orange] (12.5,11) circle[radius=\dl];
\filldraw[color=blue] (10.5,13) circle[radius=\dl];
\end{tikzpicture}
\caption{Space-filling indexing with a Hilbert curve \label{fig:hilbert}}
\end{figure}

The EAGRE system~\cite{6544856} uses an index for property tables where entities with $n$ properties are encoded in $n$-dimensional space. A space-filling curve (e.g., a Z-order or Hilbert curve) is then used for indexing. \autoref{fig:hilbert} illustrates the idea, where four entities are indexed (abbreviating \texttt{:Alice}, \texttt{:Bob}, \texttt{:Carol}, \texttt{:Dave}) with respect to two dimensions (say \texttt{foaf:age} for $x$ and integer-encoded values of \texttt{foaf:knows} for $y$). We show the first-, second- and third-order Hilbert curves from left to right. Letting $d$ denote the number of dimensions, the $n$\textsuperscript{th}-order Hilbert curve assigns an ordinal to $2^{dn}$ regions of the space based on the order in which it visits the region; e.g., starting with region 1 on the bottom left and following the curve, $\texttt{:A}$ is in the region of ordinal 2, 7 and 26, respectively. The space-filling curve thus ``flattens'' multidimensional data into one dimension (the ordinal), which can be indexed sequentially.

Property tables are complicated by multi-valued properties, missing values, etc. A more flexible approach is to index \textit{signatures} of entities, which are bit-vectors encoding the property--value pairs of the entity. One such example is the \textit{vertex signature tree} of gStore~\cite{Zou2011}, which encodes all outgoing $(p,o)$ pairs for a given entity $s$ into a bit vector akin to a Bloom filter, and indexes these bit vectors hierarchically allowing for fast, approximate containment checks that quickly find candidate entities for a subset of such pairs. GRaSS~\cite{LyuWLFW15} further optimizes for star subgraphs that include both outcoming and incoming edges on entities, where a custom \textit{FDD-index} allows for efficient retrieval of the subgraphs containing a triple that matches a triple pattern.

\subsection{Property-based indexes}\label{sec:propind}

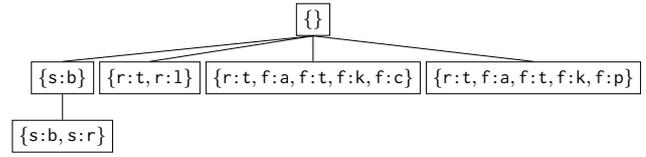
\begin{figure}
\setlength{\hgap}{0.08cm}
\setlength{\vgap}{0.4cm}
\resizebox{\linewidth}{!}{
\begin{tikzpicture}
\node[rectangle,draw,fill=white] (root) {$\{\}$};

\node[rectangle,draw,below=\vgap of root] (al) {$\{ \texttt{r:t}, \texttt{f:a}, \texttt{f:t}, \texttt{f:k}, \texttt{f:c} \}$};

\node[rectangle,draw,right=\hgap of al] (bo) {$\{ \texttt{r:t}, \texttt{f:a}, \texttt{f:t}, \texttt{f:k}, \texttt{f:p} \}$};

\node[rectangle,draw,left=\hgap of al] (pr) {$\{ \texttt{r:t}, \texttt{r:l} \}$};

\node[rectangle,draw,left=\hgap of pr] (to1) {$\{ \texttt{s:b} \}$};

\draw (root.south) -- (al.north);
\draw (root.south) -- (bo.north);
\draw (root.south) -- (pr.north);
\draw (root.south) -- (to1.north);

\node[rectangle,draw,below=\vgap of to1] (to2) {$\{ \texttt{s:b}, \texttt{s:r} \}$}
   edge (to1);
\end{tikzpicture}
}

\caption{Lattice of node signatures with abbreviated terms (e.g., \texttt{s:b} denotes \texttt{skos:broader}) \label{fig:lattice}}
\end{figure}

Returning to the star join $\{(\vbw,\cnp,\vbx),(\vbw,\cnq,\vby),(\vbw,\cnr,\cnz)\}$, another way to quickly return candidate bindings for the variable \vbw\ is to index nodes according to their adjacent properties; then we can find nodes that have at least the adjacent properties $\cnp, \cnq, \cnr$. Such an approach is used by RDFBroker~\cite{SintekK06}, which defines the signature of a node $s$ as $\Sigma(s) = \{ p \mid \exists o: (s,p,o) \in G\}$; for example, the signature of \texttt{:SW} in Figure~\ref{fig:rdfg} is $\Sigma(\texttt{:SW}) = \{ \texttt{skos:broader}, \texttt{skos:related} \}$ (analogous to \textit{characteristic sets} proposed later~\cite{NeumannM11}). A property table is then created for each signature. At query time, property tables whose signatures subsume $\{ \cnp, \cnq, \cnr \}$ are found using a lattice of signatures. We provide an example in Figure~\ref{fig:lattice} with respect to the RDF graph of Figure~\ref{fig:rdfg}, where children subsume the signatures of their parent.

AxonDB~\cite{MeimarisPMA17} uses \textit{extended characteristic sets} where each triple $(s,p,o)$ in the RDF graph is indexed with the signatures (i.e., characteristic sets) of its subject and object; i.e., $(\Sigma(s),\Sigma(o))$. Thus the triple $(\texttt{:SW}, \texttt{skos:related},\texttt{:DB})$ of Figure~\ref{fig:rdfg} would be indexed with the extended characteristic set $(\{\texttt{skos:broader}, \texttt{skos:related}\}\,,\, \{\texttt{skos:broader}\})$.\newline The index then allows for efficiently identifying two star joins that are connected by a given property $p$.

\subsection{Path indexes}\label{sec:pathind}

A path join involves successive \textsc{s}--\textsc{o} joins between triple patterns; e.g., $\{(\vbw,\cnp,\vbx),(\vbx,\cnq,\vby),(\vby,\cnr,\cnz)\}$, where the start and end nodes ($\vbw,\cnz$) may be variables or constants. While path joins have fixed length, navigational graph patterns may further match arbitrary length paths. A number of indexing approaches have been proposed to speed up querying paths. 

A path can be seen as a string of arbitrary length; e.g., a path $\{(\cnw,\cnp,\cnx),(\cnx,\cnq,\cny),(\cny,\cnr,\cnz)\}$ can be seen as a string $\cnw\cnp\cnx\cnq\cny\cnr\cnz\$$, where \$ indicates the end of the string; alternatively, if intermediate nodes are not of importance, the path could be represented as the string $\cnw\cnp\cnq\cnr\cnz\$$. The Yaanii system~\cite{CappellariVR12} builds an index of paths of the form $\cnw\cnp\cnx\cnq\cny\cnr\cnz\$$ that are clustered according to their template of the form $\cnw\cnp\cnq\cnr\cnz\$$. Paths are then indexed in B+trees, which are partitioned by template. Fletcher et al.~\cite{FletcherPP16} also index paths in B+trees, but rather than partition paths, they apply a maximum length of at most $k$ for the paths included. Text indexing techniques can also be applied for paths (viewed as strings). Maharjan et al.~\cite{MaharjanLL09} and the HPRD system~\cite{LiuH10} both leverage suffix arrays -- a common indexing technique for text -- to index paths. The downside of path indexing approaches is that they may index an exponential number of paths; in the case of HPRD, for example, users are thus expected to specify which paths to index~\cite{LiuH10}.

\begin{figure}
\setlength{\hgap}{1cm}
\setlength{\vgap}{0.4cm}
\scriptsize
\centering
\begin{tikzpicture}
\node[rectangle,draw,fill=white] (cs) {$(1,7)$ \texttt{:CS}};

\node[rectangle,draw,below=\vgap of cs] (db) {$(5,4)$ \texttt{:DB}};
\node[rectangle,draw,left=\hgap of db] (ai) {$(2,3)$ \texttt{:AI}};
\node[rectangle,draw,right=\hgap of db] (web) {$(6,6)$ \texttt{:Web}};

\node[rectangle,draw,below=\vgap of ai,xshift=-1.1cm] (ml) {$(3,1)$ \texttt{:ML}};
\node[rectangle,draw,below=\vgap of ai,xshift=1.1cm] (kr) {$(4,2)$ \texttt{:KR}};
\node[rectangle,draw,below=\vgap of web] (sw) {$(7,5)$ \texttt{:SW}};

\draw (cs.south) -- (db.north);
\draw (cs.south) -- (ai.north);
\draw (cs.south) -- (web.north);
\draw (web.south) -- (sw.north);
\draw (ai.south) -- (ml.north);
\draw (ai.south) -- (kr.north);

\end{tikzpicture}
\caption{Preorder and postorder on a \texttt{skos:narrower} tree; e.g., \texttt{:CS} has preorder $1$ and postorder $7$ \label{fig:prepost}}
\end{figure}
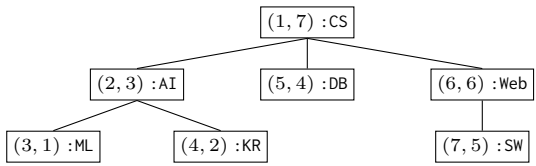

Other path indexes are inspired by prior works for path queries over trees (e.g., for XPath). Barto{\v{n}}~\cite{Barton04} proposes a tree-based index based on \textit{preorder} and \textit{postorder traversal}. A preorder traversal starts at the root and traverses children in a depth-first manner from left to right. A postorder traversal starts at the leftmost leaf and traverses all children, from left to right, before moving to the parent. We provide an example preorder and postorder traversal in Figure~\ref{fig:prepost}. Given two nodes $m$ and $n$ in the tree, a key property is that $m$ is a descendant of $n$ if and only if $m$ is greater than $n$ for preorder and less than $n$ for postorder. Barto{\v{n}}~\cite{Barton04} uses this property to generate an index on ascending preorder so as to linearize the tree and quickly find descendants based on postorder. To support graphs, Barto{\v{n}} uses a decomposition of the graph into a forest of trees that are then indexed~\cite{Barton04}.

\begin{figure}
\setlength{\hgap}{1cm}
\setlength{\vgap}{0.4cm}
\scriptsize
\centering
\begin{tikzpicture}
\node[rectangle,draw,fill=white] (cs) {$(2,1,2)$ \texttt{:CS}};

\node[rectangle,draw,below=\vgap of cs] (db) {$(5,2,10)$ \texttt{:DB}};
\node[rectangle,draw,left=\hgap of db] (ai) {$(3,2,6)$ \texttt{:AI}};
\node[rectangle,draw,right=\hgap of db] (web) {$(7,2,14)$ \texttt{:Web}};

\node[rectangle,draw,below=\vgap of ai] (ml) {$(11,3,66)$ \texttt{:ML}};
\node[rectangle,draw,below=\vgap of db] (dm) {$(13,15,390)$ \texttt{:DM}};
\node[rectangle,draw,below=\vgap of web] (sw) {$(17,35,1190)$ \texttt{:SW}};

\draw (cs.south) -- (db.north);
\draw (cs.south) -- (ai.north);
\draw (cs.south) -- (web.north);
\draw (web.south) -- (sw.north);
\draw (db.south) -- (sw.north);
\draw (db.south) -- (dm.north);
\draw (ai.south) -- (dm.north);
\draw (ai.south) -- (ml.north);

\end{tikzpicture}
\caption{PLSD index on an example \texttt{skos:narrower} hierarchy; terms (e.g., \texttt{:CS}) are indexed externally \label{fig:plsd}}
\end{figure}
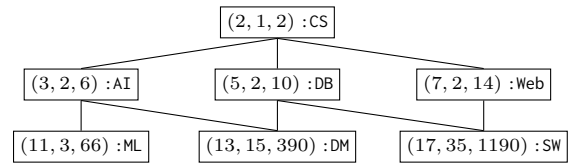

Another type of path index, called \textit{PLSD}, is used in System $\Pi$~\cite{WuLHW09} for indexing the transitivity of a single property, optimizing for path queries of the form $(\vbs,\cnp*,\vbo)$, or $(\cns,\cnp*,\vbo)$, etc. For a given property $p$, each incident (subject or object) node $x$ is assigned a triple of numbers $(i,j,k) \in \mathbb{N}^3$, where $i$ is a unique prime number that identifies the node $x$, $j$ is the least common multiple of the $i$-values of $x$'s parents (i.e., nodes $y$ such that $(y,p,x) \in G$), and $k$ is the least common multiple of the $k$-values of $x$'s parents and the $i$-value of $x$. We provide an example in Figure~\ref{fig:plsd}. PLSD can further handle cycles by multiplying the $k$-value of all nodes by the $i$ value of all nodes in its strongly-connected component. Given the $i$-value of a node, the $i$-values of its parents and ancestors can be retrieved by factorizing $j$ and $k/i$ respectively. However, multiplication may give rise to large numbers, where no polynomial time algorithm is known for the factorization of binary numbers.

Gubichev et al.~\cite{GubichevBS13} use a path index of directed graphs, called FERRARI~\cite{SeufertABW13}, for each property in an RDF graph. First, a \textit{condensed graph} is computed by merging nodes of strongly connected components into one ``supernode''; adding an artificial root node (if one does not exist), the result is a directed acyclic graph (DAG) that preserves reachability. A \textit{spanning tree} -- a subgraph that includes all nodes and is a tree -- of the DAG is computed and labeled with its postorder. All subtrees thus have contiguous identifiers, where the maximum identifies the root; e.g., in Figure~\ref{fig:prepost}, the subtree at \iri{AI} has the interval $[1,3]$, where $3$ identifies the root. Then there exists a (directed) path from $x$ to $y$ if and only if $y$ is in the subtree interval for $x$. Nodes in a DAG may, however, be reachable through paths not in the spanning tree. Hence each node is assigned a \textit{set} of intervals for nodes that can be reached from it, where overlapping and adjacent intervals are merged; we must now check that $y$ is in one of the intervals of $x$. To improve time and space at the cost of precision, approximate intervals are proposed that merge non-overlapping intervals; e.g., $[4,6], [8,9]$ is merged to $[4,9]$, which can reject reachability for nodes with id less than 2 or greater than 9, but has a $\frac{1}{6}$ chance of a false positive for nodes in $[4,9]$, which must be verified separately. 

\subsection{Join indexes}

The results of joins can also be indexed. Groppe et al.~\cite{GroppeGL07} proposed to construct $6 \times 2^4 = 96$ indexes for $6$ types of non-symmetric joins between two triple patterns (\textsc{s}--\textsc{s}, \textsc{s}--\textsc{p},  \textsc{s}--\textsc{o},  \textsc{p}--\textsc{p},  \textsc{p}--\textsc{o}, \textsc{o}--\textsc{o}). Hash maps are used to cover the $2^4$ permutations of the remaining elements (not considering the join variable). Given the high space cost, only frequently encountered joins are sometimes indexed~\cite{10.5555/1083592.1083734,rdfjoin}.

\subsection{Structural indexes}\label{sec:structure}

Another family of indexes -- known as \textit{structural indexes}~\cite{LuoPFHV12} --  rely on a high-level summary of the RDF graph.

\begin{figure}
\setlength{\hgap}{1cm}
\setlength{\vgap}{0.4cm}
\scriptsize
\centering
\begin{tikzpicture}
\node[rectangle,draw,fill=white] (r) {$(\texttt{:Alice},2)$};

\node[below=\vgap of r] (r0) {};
\node[rectangle,draw,left=\hgap of r0] (r1) {$(\texttt{\_:p},2)$};
\node[rectangle,draw,right=\hgap of r0] (r2) {$(\texttt{:CS},1)$};

\node[below=\vgap of r1] (r10) {};
\node[rectangle,draw,left=0.7\hgap of r10] (r11) {$\ldots$};
\node[rectangle,draw,right=0.7\hgap of r10] (r12) {$\ldots$};

\node[below=\vgap of r2] (r20) {};
\node[rectangle,draw,left=0.7\hgap of r20] (r21) {$\ldots$};
\node[rectangle,draw,right=0.7\hgap of r20] (r22) {$\ldots$};

\draw (r.south) -- (r1.north);
\draw (r.south) -- (r2.north);

\draw (r1.south) -- (r11.north);
\draw (r1.south) -- (r12.north);

\draw (r2.south) -- (r21.north);
\draw (r2.south) -- (r22.north);

\end{tikzpicture}
\caption{Distance-based indexing (GRIN) \label{fig:grin}}
\end{figure}
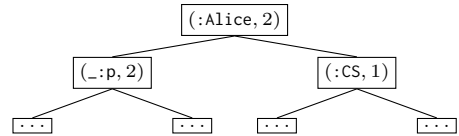

Some structural indexes are based on distance measures. GRIN~\cite{UdreaPS07} divides the graph hierarchically into regions based on the distance of its nodes to selected centroids. These regions form a tree, where the non-leaf elements indicate a node $x$ and a distance $d$ referring to all nodes at most $d$ steps from $x$. The root element chooses a node and distance such that all nodes of the graph are covered. Each non-leaf element has two children that capture all nodes of their parent. Each leaf node contains a set of nodes $N$, which induces a subgraph of triples between the nodes of $N$; the leaves can then be seen as partitioning the RDF graph. We provide an example in Figure~\ref{fig:grin} for the RDF graph of Figure~\ref{fig:rdfg}, where all nodes are within distance two of \texttt{:Alice}, which are then divided into two regions: one of distance at most two from \texttt{\_:p}, and another of distance at most one from \texttt{:CS}. The index can continue dividing the graph into regions, and can then be used to find subgraphs within a particular distance from a given node (e.g., a node given in a BGP).

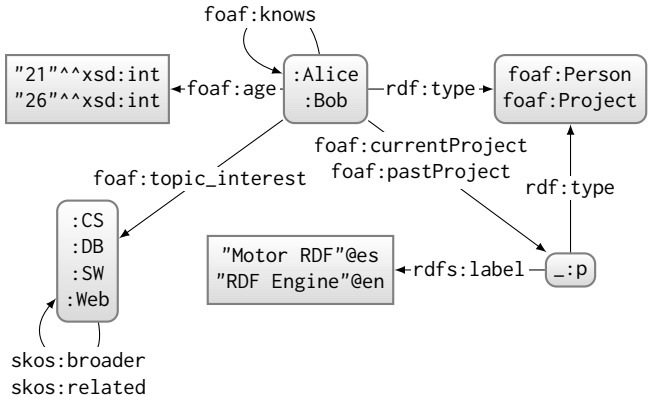
\begin{figure}
\setlength{\hgap}{1.65cm}
\setlength{\vgap}{1cm}
\begin{tikzpicture}[baseline]
\node[iri,anchor=mid] (fp) {\begin{tabular}{@{}c@{}}foaf:Person\\foaf:Project\end{tabular}};
  
\node[iri,anchor=mid,left=\hgap of fp] (ab) {\begin{tabular}{@{}c@{}}:Alice\\:Bob\end{tabular}}
  edge[in=160,out=100,looseness=6,arrout] node[lab] {foaf:knows} (ab)
  edge[arrout] node[lab] {rdf:type} (fp);
  
\node[lit,anchor=mid,left=0.9\hgap of ab] (age) {\begin{tabular}{@{}c@{}}"21"\dt xsd:int\\"26"\dt xsd:int\end{tabular}}
  edge[arrin,pos=0.55] node[lab] {foaf:age} (ab);
  
\node[iri,anchor=mid,below=1.7\vgap of fp] (p) {\_:p}
  edge[arrin] node[lab,pos=0.7] {\begin{tabular}{@{}c@{}}foaf:currentProject\\foaf:pastProject\end{tabular}} (ab)
  edge[arrout] node[lab] {rdf:type} (fp);

\node[lit,anchor=mid,left=1.2\hgap of p] (names) {\begin{tabular}{@{}c@{}}"Motor RDF"@es\\"RDF Engine"@en\end{tabular}}
  edge[arrin] node[lab] {rdfs:label} (p);
  
\node[iri,anchor=mid,below=\vgap of age] (temas) {\begin{tabular}{@{}c@{}}:CS\\:DB\\:SW\\:Web\\\end{tabular}}
  edge[arrin] node[lab] {foaf:topic\_interest} (ab)
  edge[in=230,out=280,looseness=4,arrout] node[lab,yshift=-0.2cm,xshift=0.2cm] {\begin{tabular}{@{}c@{}}skos:broader\\skos:related\end{tabular}} (temas);
\end{tikzpicture}

\caption{Quotient graph with six supernodes \label{fig:quotient}}
\end{figure}

Another type of structural index relies on some notion of a \textit{quotient graph}~\cite{CebiricGKKMTZ19}, where the nodes of a graph $\nodes{G}$ are partitioned into $\{ X_1, \ldots, X_n \}$ pairwise-disjoint sets such that $\bigcup_{i=1}^n X_i = \nodes{G}$. Then edges of the form $(X_i,p,X_j)$ are added if and only if there exists $(x_i,p,x_j) \in G$ such that $x_i \in X_i$ and $x_j \in X_j$. Intuitively, a quotient graph merges nodes from the input graph into ``supernodes'' while maintaining the input (labeled) edges between the supernodes. We provide an example of a quotient graph in Figure~\ref{fig:quotient} featuring six supernodes. Any partitioning of nodes can form a quotient graph, ranging from a single supernode with all nodes $\nodes{G}$ and loops for all properties in $\preds{G}$, to the graph itself replacing each node $x \in \nodes{G}$ with the singleton $\{ x \}$. If the input graph yields solutions for a BGP, then the quotient graph will also yield solutions (with variables now matching supernodes). For example, taking the BGP of Figure~\ref{fig:bgp}, matching \texttt{foaf:Person} to the supernode containing \texttt{foaf:Person} in Figure~\ref{fig:quotient}, then the variables \texttt{?a} and \texttt{?b} will match the supernode containing \texttt{:Alice} and \texttt{:Bob}, while \texttt{?ia} and \texttt{?ib} will match to the supernode containing \texttt{:CS}, \texttt{:DB}, \texttt{:SW} and \texttt{:Web}; while we do not know the exact solutions for the input graph, we know they must correspond to elements of the supernodes matched in the quotient graph.

DOGMA~\cite{dogma2009} partitions an RDF graph into subgraphs, from which a balanced binary tree is computed, where each parent node contains a quotient-like graph of both its children. The (O)SQP approach~\cite{TranLR13} creates an in-memory index graph, which is a quotient graph whose partition is defined according to various notions of bisimulation. 

SAINT-DB~\cite{PicalausaLFHV12} adopts a similar approach, where supernodes are defined directly as a partition of the triples of the RDF graph, and edges between supernodes are labeled with the type of join (\textsc{s}--\textsc{s}, \textsc{p}--\textsc{o}, etc.) between them.

\subsection{Quad indexes}

Most quad indexes follow the triple index scheme~\cite{kowari,10.5555/1785162.1785179,harris4store,Erling2010}, extending it to add another element. The number of permutations then grows to $2^4 = 16$ abstract index patterns, $4! = 24$ potential permutations, and ${4 \choose \lfloor 4/2 \rfloor} = 6$ flat (ISAM/B+Tree/AVL tree/trie) permutations or $2$ circular (CSA) permutations to efficiently support all abstract quad patterns. A practical compromise is to maintain a selection of permutations that cover the most common patterns~\cite{Erling2010}; for example, a pattern $(\cns,\vbp,\cno,\cng)$ may be uncommon in practice, and could be supported reasonably well by evaluating (e.g.) $(\cns,\vbp,\cno,\vbg)$ and filtering on $\vbg = \cng$.

The RIQ system~\cite{katib2016riq} proposes a custom index for quads called a PV-index for finding (named) graphs that match a BGP. Each graph is indexed by hashing all seven abstract patterns on triples with some constant, generating seven \textit{pattern vectors} for each graph. For example, a triple $(s,p,o)$ in a graph named $g$ will be hashed as $(s,p,o)$, $(s,p,?)$, $(s,?,o)$, $(?,p,o)$, $(s,?,?)$, $(?,p,?)$, $(?,?,o)$, where $?$ is an arbitrary fixed token, and each result will be added to one of seven pattern vectors for $g$ for that abstract pattern. Basic graph patterns can be encoded likewise, where locality sensitive hashing is then used to group and retrieve similar pattern vectors for a given basic graph pattern.

\subsection{Miscellaneous Indexing}

RDF stores may use legacy systems, such as NoSQL stores, for indexing. Since such approaches are not tailored to RDF, and often correspond conceptually to one of the indexing schemes already discussed, we refer to more dedicated surveys of such topics for further details~\cite{janke2018storing,10.1145/3341105.3375753,Wylot2018RDFDS}. Other stores provide specialized indexes for particular types of values such as spatial or temporal data~\cite{10.1145/3297280.3299732,10.1007/978-3-642-35176-1_19}; we do not discuss such specialized indexes in detail.

\subsection{Discussion}

While indexing triples or quads is conceptually the most straightforward approach, a number of systems have shown positive results with entity- and property-based indexes that optimize the evaluation of star joins, path indexes that optimize the evaluation of path joins, or structural indexes that allow for identifying query-relevant regions of the graph. Different indexing schemes often have different time--space trade-offs: more comprehensive indexes enable faster queries at the cost of space and more costly updates. 

\section{Join Processing}
\label{sec:join}

RDF stores employ diverse query processing strategies, but all require translating logical operators that represent the query, into ``physical operators'' that implement algorithms for efficient evaluation of the operation. The most important such operators -- as we now discuss -- are natural joins.

\subsection{Pairwise join algorithms}\label{sec:pairwise}

We recall that the evaluation of a BGP $\{ t_1, \ldots t_n \}(G)$ can be rewritten as $t_1(G) \bowtie \ldots \bowtie t_n(G)$, where the evaluation of each triple pattern $t_i$ ($1 \leq i \leq n$) produces a relation of arity $|\vars{t_i}|$. Thus the evaluation of a BGP $B$ produces a relation of arity $|\vars{B}|$. The relational algebra -- including joins -- can then be used to combine or transform the results of one or more BGPs, giving rise to CGPs. The core of evaluating graph patterns is thus analogous to processing relational joins. The simplest and most well-known such algorithms perform \textit{pairwise joins}; for example, a pairwise strategy for computing $\{ t_1, \ldots t_n \}(G)$ may evaluate $((t_1(G) \bowtie t_2(G)) \bowtie \ldots) \bowtie t_n(G)$. 

Without loss of generality, we assume a join of two graph patterns $P_1(G) \bowtie P_2(G)$, where the join variables are denoted by $V = \{ v_1, \ldots, v_n \} = \vars{P_1} \cap \vars{P_2}$. Well-known algorithms for performing pairwise joins include \textit{(index) nested-loop joins}, where $P_1(G) \bowtie P_2(G)$ is reduced to evaluating $\bigcup_{\mu \in P_1(G)} \{ \mu \} \bowtie \mu(P_2)(G)$; \textit{hash joins}, where each solution $\mu \in P_1(G)$ is indexed by hashing on the key $(\mu(v_1),\ldots,\mu(v_n))$ and thereafter a key is computed likewise for each solution in $P_2(G)$ to probe the index with; and \textit{(sort-)merge joins}, where $P_1(G)$ and $P_2(G)$ are (sorted if necessary and) read in the same order with respect to $V$, allowing the join to be reduced to a merge sort. Index nested-loop joins tend to perform well when $|P_1(G)| \ll |P_2(G)|$ (assuming that $\mu(P_2)(G)$ can use indexes) since it does not require reading all of $P_2(G)$. Otherwise hash or merge joins can perform well~\cite{Neumann2010}. Pairwise join algorithms are then used in many RDF stores (e.g.,~\cite{Harth:2005:OIS:1114687.1114857,Erling2010,Neumann2010}).

Techniques to optimize pairwise join algorithms include \textit{sideways information passing}~\cite{BeeriR91}, which passes data across different parts of the query, often to filter intermediate results. Neumann and Weikum~\cite{NeumannW09} propose \textit{ubiquitous sideways information passing} (\textit{U-SIP}) for computing joins over RDF, which shares global ranges of values for a given query variable. U-SIP is implemented differently for different join types. For merge joins, where data are read in order, a maximum value for a variable can be shared across pairwise joins, allowing individual operators to skip ahead to the current maximum. For hash joins, a global \textit{domain filter} is employed -- consisting of a maximum value, a minimum value, and Bloom filters -- for filtering the results of each variable.

\subsection{Multiway joins}\label{sec:multiway}

\textit{Multiway join algorithms} exploit the commutativity and associativity of joins to evaluate two \textit{or more} operands at once. For example, in order to compute $\{ t_1, \ldots t_n \}(G)$, a multiway join algorithm may evaluate $(t_1(G) \bowtie \ldots \bowtie t_k(G)) \bowtie (t_{k+1}(G) \bowtie \ldots \bowtie t_n(G))$ where $k \geq 2$, or it may even simply evaluate everything at once as $(t_{1}(G) \bowtie \ldots \bowtie t_n(G))$.

Some of the previous storage and indexing schemes we have seen lend themselves naturally to processing certain types of multiway joins in an efficient manner. Entity-based indexes allow for processing star joins efficiently, while path indexes allow for processing path joins efficiently (see Section~\ref{sec:index}). A BGP can be decomposed into sub-BGPs that can be evaluated per the corresponding multiway join, with pairwise joins being applied across the sub-BGPs; for example: $\{(\vbw,\cnp,\vbx),(\vbw,\cnq,\vby),(\vbw,\cnr,\cnz),(\vbx,\cnq,\vby),(\vbx,\cnr,\cnz)\}$ may be divided into the sub-BGPs $\{(\vbw,\cnp,\vbx),\!(\vbw,\cnq,\vby),\!(\vbw,\cnr,\cnz) \}$ and $\{ (\vbx,\cnq,\vby), (\vbx,\cnr,\cnz)\}$, which are evaluated separately as multiway joins before being themselves joined. Even in the case of (sorted) triple/quad tables, multiway joins can be applied taking advantage of the locality of processing, where, for example, in an \textsc{spo} index permutation, triples with the same subject will be grouped together. Similar locality can be exploited in distributed settings (see, e.g., SMJoin~\cite{GalkinEACVA17}).

\subsection{Worst case optimal joins}\label{sec:wcojoins}

A new family of join algorithms have arisen due to the \textit{AGM bound}~\cite{AtseriasGM13}, which puts an upper bound on the number of solutions that can be returned from relational join queries. The result can be adapted straightforwardly to the case of BGPs. Let $B = \{ t_1, \ldots, t_n \}$ denote a BGP with $\vars{B} = V$. Now define a \textit{fractional edge cover} as a mapping $\lambda : B \rightarrow \mathbb{R}_{[0,1]}$ that assigns a real value in the interval $[0,1]$ to each triple pattern of $B$ such that for all $v \in V$, it holds that $\sum_{t \in B_v} \lambda(t) \geq 1$, where $B_v$ denotes the set of triple patterns in $B$ that mention $v$. The AGM bound tells us that if $B$ has the fractional edge cover $\lambda$, then for any RDF graph it holds that $|B(G)| \leq \prod_{i=1}^n |t_i(G)|^{\lambda(t_i)}$; this bound is ``tight''.

To illustrate the AGM bound, consider the BGP $B = \{ t_1, t_2, t_3 \}$ from Figure~\ref{fig:wco}. There exists a fractional edge cover $\lambda$ of $B$ such that $\lambda(t_1) = \lambda(t_2) = \lambda(t_3) = \frac{1}{2}$; taking $\texttt{?a}$, we have that $B_\texttt{?a} = \{ t_1,t_3\}$, $\lambda(t_1) + \lambda(t_3) = 1$, and thus $\texttt{?a}$ is ``covered'', and we can verify the same for $\texttt{?b}$ and $\texttt{?c}$. Then the AGM bound is given as the inequality $|B(G)| \leq \prod_{i=1}^n|t_i(G)|^{\lambda(t_i)}$. For $G$ the graph in Figure~\ref{fig:wco}, $|t_1(G)| = |t_2(G)| = |t_3(G)| = 5$, and hence $|B(G)| \leq 5^\frac{3}{2}$. In reality, for this graph, $|B(G)| = 5$, thus satisfying the inequality, but there exists a graph where $B = \prod_{i=1}^n|t_i(G)|^{\lambda(t_i)}$. 

\begin{figure}
\setlength{\hgap}{2.1cm}
\setlength{\vgap}{1cm}
\centering
\begin{tikzpicture}[baseline]
\node[iri,anchor=mid] (cs) {:CS};

\node[iri,anchor=mid,above=\vgap of cs] (sw) {:SW}
  edge[arrout,bend left=30] node[lab] {s:b} (cs)
  edge[arrin,bend right=30] node[lab] {s:n} (cs);

\node[iri,anchor=mid,left=\hgap of sw] (db) {:DB}
  edge[arrout,bend left=11] node[lab] {s:b} (cs)
  edge[arrin,bend right=11] node[lab] {s:n} (cs)
  edge[arrin] node[lab] {s:r} (sw);

\node[iri,anchor=mid,right=\hgap of sw] (w) {:Web}
  edge[arrout,bend left=11] node[lab] {s:b} (cs)
  edge[arrin,bend right=11] node[lab] {s:n} (cs)
  edge[arrin,bend left=11] node[lab] {s:r} (sw)
  edge[arrout,bend right=11] node[lab] {s:r} (sw);
  
\node[iri,anchor=mid] (ir) at (cs-|w) {:IR}
  edge[arrout,bend left=11] node[lab] {s:b} (cs)
  edge[arrin,bend right=11] node[lab] {s:n} (cs)
  edge[arrin] node[lab] {s:r} (w);

\node[iri,anchor=mid] (ir) at (cs-|db) {:AI}
  edge[arrout,bend left=11] node[lab] {s:b} (cs)
  edge[arrin,bend right=11] node[lab] {s:n} (cs)
  edge[arrout] node[lab] (ml) {s:r} (db);
  
\node[left=0.3cm of ml] {$G:$};
\end{tikzpicture}
\medskip

\setlength{\hgap}{0.7cm}
\begin{tikzpicture}
\node[var,anchor=mid] (a) {?a};
  
\node[var,anchor=mid,right=\hgap of fp] (c) {?c}
  edge[arrout] node[lab] (m) {s:r} (a);

\node[below=0.07cm of m] {$(t_3)$};
  
\node[var,anchor=mid,above=\vgap of m] (b) {?b}
  edge[arrin] node[lab] (ml) {s:b} (a)
  edge[arrout] node[lab] (mr) {s:n} (c);
  
\node[above right=0.01cm of mr] {$(t_2)$};
\node[above left=0.01cm of ml] {$(t_1)$};    

\node[left=0.6cm of ml] {$B:$};

\node[right=1cm of mr] {$B(G):$};

\node[right=2cm of mr]{
\begin{tabular}{lll}
\toprule
\texttt{?a} & \texttt{?b} & \texttt{?c} \\
\midrule
\texttt{:DB} & \texttt{:CS} & \texttt{:AI}  \\
\texttt{:DB} & \texttt{:CS} & \texttt{:SW} \\
\texttt{:IR} & \texttt{:CS} & \texttt{:Web} \\
\texttt{:SW} & \texttt{:CS} & \texttt{:Web} \\
\texttt{:Web} & \texttt{:CS} & \texttt{:SW} \\
\bottomrule
\end{tabular}
};
\end{tikzpicture}

\caption{Example RDF graph $G$, BGP $B$ and its evaluation $B(G)$; the IRIs \texttt{s:b}, \texttt{s:n} and \texttt{s:r} abbreviate \texttt{skos:broader}, \texttt{skos:narrower} and \texttt{skos:related}, resp. \label{fig:wco}}
\end{figure}

Recently, join algorithms have been proposed that can enumerate the results for a BGP $B$ over a graph $G$ in time $O(agm(B,G))$, where $agm(B,G)$ denotes the AGM bound of $B$ over $G$. Since such an algorithm must at least spend $O(agm(B,G))$ time writing the results in the worst case, such algorithms are deemed \textit{worst-case optimal} (\textit{wco})~\cite{NgoPRR18}. Though such algorithms were initially proposed in a relational setting~\cite{NgoPRR18,Veldhuizen14}, they have recently been adapted for processing joins over RDF graphs ~\cite{abs-1811-10955,HRRSiswc19,NavarroRR20,ArroyueloHNRRS21}. Note that traditional pairwise join algorithms are not wco. If we try to evaluate $\{ t_1, t_2 \}(G)$ by pairwise join, for example, in order to later join it with $t_3(G)$, the AGM bound becomes quadratic as $\lambda(t_1) = \lambda(t_2) = 1$, and thus we have the bound $|t_1(G)| \cdot |t_2(G)|$, which exceeds the AGM bound for $B$. This holds for any pairwise join in $B$. Note that $\{ t_1, t_2 \}(G)$ will indeed produce (25) quadratic results, mapping \var{a} to \iri{CS} and \var{b} and \var{c} to $\{ \iri{AI}, \iri{DB}, \iri{IR}, \iri{SW}, \iri{Web} \}^2$.

Wco join algorithms -- including \textit{Leapfrog Triejoin} (LTJ) ~\cite{Veldhuizen14} -- perform a multiway join that resolves a BGP $B$ variable-by-variable rather than pattern-by-pattern. First an ordered sequence of variables is selected; say $(\var{a}, \var{b}, \var{c})$. Then the set of partial solutions $M_{\{\var{a}\}} = \{ \mu \mid \dom{\mu} = \{ \var{a} \}\text{ and }\mu(B_\var{a})(G) \neq \emptyset \}$ are computed for the first variable \var{a} such that each image of $B_\var{a}$ under $\mu \in M_{\{\var{a}\}}$ has some solutions for $G$; e.g., $M_{\{\var{a}\}} = \{ \{ \var{a}/\iri{DB} \}, \{ \var{a}/\iri{IR} \}, \newline \{ \var{a}/\iri{SW} \}, \{ \var{a}/\iri{Web} \} \}$ in Figure~\ref{fig:wco}, since replacing $\var{a}$ in $B_\var{a}$ with $\iri{DB}$, $\iri{IR}$, $\iri{SW}$ or $\iri{Web}$ yields a BGP with solutions over $G$. Next we compute $M_{\{\var{a},\var{b}\}} = \{ \mu \cup \mu' \mid \mu \in M_{\{\var{a}\}}, \dom{\mu'} = \{ \var{b} \}\text{ and } \mu'(\mu(B_\var{b}))(G) \neq \emptyset \}$, ``eliminating'' the next variable \var{b}. In the example of Figure~\ref{fig:wco}, $M_{\{\var{a},\var{b}\}} = \{ \{ \var{a}/\iri{DB}, \var{b}/\iri{CS} \}, \ldots, \{ \var{a}/\iri{Web}, \var{b}/\iri{CS} \} \}$, where each solution $\mu \in M_{\{\var{a}\}}$ is extended with $\{ \var{b} / \iri{CS} \}$. Finally, $M_{\{\var{a},\var{b},\var{c}\}}$ is computed analogously, eliminating the last variable, and yielding the five results seen in Figure~\ref{fig:wco}.

To be wco-compliant, the algorithm must always be able to efficiently compute $M_{\{v\}}$, i.e., solutions $\mu$ with $\dom{\mu} = \{v\}$, such that $\mu(B_v)(G) \neq \emptyset$. To compute $M_{\{\var{a}\}}$ in the running example, we need to efficiently intersect all nodes with an outgoing \texttt{s:b} edge and an incoming \texttt{s:r} edge. This is typically addressed by being able to read the results of a triple pattern, in sorted order, for any variable, which enables efficient intersection by allowing to seek ahead to the maximum current value of all triple patterns involving a given variable. Jena-LTJ~\cite{HRRSiswc19}, which implements an LTJ-style join algorithm for SPARQL, enables this by maintaining all six index permutations over triples, while Ring~\cite{ArroyueloHNRRS21} requires only one permutation. Wco algorithms often outperform traditional join algorithms for complex BGPs~\cite{abs-1811-10955,HRRSiswc19}.

\subsection{Translations to linear algebra}\label{sec:la}

Per Section~\ref{sec:tensor}, dictionary-encoded RDF graphs are sometimes represented as a bit tensor, or as a bit matrix for each property (see Figure~\ref{fig:bitmat}), etc. Viewed in this light, some query algebra can then be reduced to linear algebra~\cite{MetzlerM15a}; for example, joins become matrix/tensor multiplication. To illustrate, we can multiply the bit (adjacency) matrix from Figure~\ref{fig:bitmat} for \texttt{skos:broader} by itself:
\begin{align*}
\setlength\arraycolsep{5pt}
\begin{pmatrix}
    0 & 0 & 0 & 0 \\
    1 & 0 & 0 & 0 \\
    0 & 0 & 0 & 1 \\
    1 & 0 & 0 & 0 
\end{pmatrix}
\begin{pmatrix}
    0 & 0 & 0 & 0 \\
    1 & 0 & 0 & 0 \\
    0 & 0 & 0 & 1 \\
    1 & 0 & 0 & 0 
\end{pmatrix} 
=&~
\setlength\arraycolsep{5pt}
\begin{pmatrix}
    0 & 0 & 0 & 0 \\
    0 & 0 & 0 & 0 \\
    1 & 0 & 0 & 0 \\
    0 & 0 & 0 & 0 
\end{pmatrix}
\end{align*}

\noindent
The result indicates the analogous bit matrix for an \textsc{o}--\textsc{s} join on  \texttt{skos:broader}, with \texttt{:SW} (on row 3) connected to \texttt{:CS} (on column 1), which we would expect per Figure~\ref{fig:rdfg}.

Translating joins into linear algebra enables hardware acceleration, particularly involving GPUs and HPC architectures, which can process tensors with high levels of parallelism. Such an approach is followed by MAGiQ~\cite{JamourACK19}, which represents an RDF graph as a single $n \times n$ matrix $\mathfrak{M}$, where $n$ is the number of nodes ($n = |\so{G}|$) and $\mathfrak{M}_{i,j}$ encodes the id of the property connecting the $i$\textsuperscript{th} node to the $j$\textsuperscript{th} node (or 0 if no such property exists). One issue with this representation is that it does not support two nodes being connected by multiple edges with different labels, and thus a coordinate list representation can rather be used. Basic graph patterns with projection are translated into matrix multiplication, scalar multiplication, transposition, etc., which can be executed on a variety of hardware, including GPUs.

Other engines that translate SPARQL query features into linear algebra (or other operations within GPUs) include Wukong(+G)~\cite{10.5555/3026877.3026902,10.5555/3277355.3277418}, TripleID-Q~\cite{8314130}, and gSmart~\cite{chen2021gsmart}. Wukong+G~\cite{10.5555/3277355.3277418} proposes a number of caching, pipelining, swapping and prefetching techniques in order to reduce the GPU memory required when processing large graphs while maintaining efficiency, and also proposes a partitioning technique to distribute computation over multiple CPUs and GPUs. TripleID-Q~\cite{8314130} represents an RDF graph as a dictionary-encoded triple table that can be loaded into the GPU in order to search for solutions to individual triple patterns without indexing, but with high degrees of parallelism. On top of this GPU-based search, join and union operators are implemented using GPU libraries. gSmart~\cite{chen2021gsmart} proposes a variety of optimizations for evaluating basic graph patterns in such settings, including a multi-way join optimization for computing star-like joins more efficiently on GPUs, compact representations for sparse matrices, data partitioning to enable higher degrees of parallelism, and more besides.

\subsection{Join reordering}

The order of join processing can have a dramatic effect on computational costs. For Figure~\ref{fig:wco}, if we apply pairwise joins in the order $(t_1(G) \Join t_2(G)) \Join t_3(G)$, the first join $(t_1(G) \Join t_2(G))$ yields 25 intermediate results, with 5 final results produced with the second join. If we rather evaluate $(t_2(G) \Join t_3(G)) \Join t_1(G)$, the first join $(t_2(G) \Join t_3(G))$ produces only 5 intermediate results, before the second join produces the 5 final results. The second plan should thus be more efficient than the first; if considering a graph at larger scale, the differences may reach orders of magnitude. 

A good plan depends not only on the query, but also the graph. Selecting a good plan thus typically requires some assumptions or statistics over the graph. As in relational settings, the most important information relates to \textit{cardinalities}: how many (distinct) solutions a given pattern returns; and/or \textit{selectivity}: what percentage of solutions are kept when restricting variables with constants or filters. Stat\-istics can be used not only to select an ordering for joins, but also to decide which join algorithm to apply. For example, given an arbitrary (sub-)BGP $\{ t_1, t_2 \}$, if we estimate that $|t_2(G)| \ll |t_1(G)|$, we may prefer to evaluate $t_2(G) \Join t_1(G)$ as an index nested-loop join, rather than a hash or merge join, to avoid reading $t_1(G)$ in full. 

While cardinality and selectivity estimates can be managed in a similar way to relational database optimizers, a number of approaches have proposed custom statistics for RDF. Stocker et al.~\cite{StockerSBKR08} collect statistics relating to the number of triples, the number of unique subjects, and for each predicate, the number of triples and a histogram of associated objects. RDF-3X~\cite{Neumann2010} uses a set of \textit{aggregated indexes}, which store the cardinality of all triple patterns with one or two constants. RDF-3X~\cite{Neumann2010} further stores the exact cardinality of frequently encountered joins, while characteristic sets~\cite{NeumannM11} and extended characteristic sets~\cite{MeimarisPMA17} (discussed in Section~\ref{sec:propind}) capture the cardinality of star joins. 

Computing and maintaining such statistics incur costs in terms of space and updates. An alternative is to apply \textit{sampling} while evaluating the query. Vidal et al.~\cite{VidalRLMSP10} estimate the cardinality of star joins by evaluating all solutions for the first pattern of the join, thereafter computing the full solutions of the star pattern for a sample of the initial solutions; the full cardinality of the star pattern is then estimated from the samples. Another alternative is to use \textit{syntactic heuristics} for reordering. Stocker et al.~\cite{StockerSBKR08} propose heuristics such as assuming that triple patterns with fewer variables have lower cardinality, that subject constants are more selective than objects and predicates, etc. Tsialiamanis et al.~\cite{TsialiamanisSFCB12} further propose to prioritize rarer joins (such as \textsc{p}--\textsc{s} and \textsc{p}--\textsc{o} joins), and to consider literals as more selective than IRIs. 

Taking into account such heuristics and statistics, the simplest strategy to try to find a good join ordering is to apply a greedy metaheuristic~\cite{StockerSBKR08,MeimarisPMA17}, starting with the triple pattern $t_1$ estimated to have the lowest cardinality, and joining it with the triple pattern $t_2$ with the next lowest cardinality; typically a constraint is added such that $t_n$ ($n > 1$) should have a variable in common with some triple pattern in $\{ t_1, \ldots, t_{n-1} \}$ to avoid costly Cartesian products. Aside from considering the cardinality of triple patterns, Meimaris and Papastefanatos~\cite{MeimarisP17} propose a distance-based planning, where pairs of triple patterns with more overlapping nodes and more similar cardinality estimates have lesser distance between them; the query planner then tries to group and join triple patterns with the smallest distances first in a greedy manner. Greedy strategies will not, however, always provide the best ordering corresponding to an optimal plan. 

More generally, reordering joins is an optimization problem, where classical methods from the relational literature can be leveraged likewise for BGPs, including dynamic programming~\cite{SelingerACLP79} (used, e.g., by~\cite{10.5555/1785162.1785179,Neumann2010,Gubichev014}) and simulated annealing~\cite{Ioannidis87} (used, e.g., by~\cite{VidalRLMSP10}). Other metaheuristics that have been applied for join reordering in BGPs include genetic algorithms~\cite{HogenboomMFK09} and ant colony systems~\cite{HogenboomFK13,KalayciKB15}.

\subsection{Caching}

Another possible route for optimization -- based on the observation that queries in practice may feature overlapping or similar patterns -- is to reuse work done previously for other queries. Specifically, we can consider \textit{caching} the results of queries. In order to increase cache hit rates, we can further try to reuse the results of subqueries, possibly generalizing them to increase usability. Ideally the cache should store solutions for subqueries that (a) have a high potential to reduce the cost of future queries; (b) can reduce costs for many future queries; (c) do not have a high space overhead; and (d) will remain valid for a long time. Some of these aims can be antagonistic; for example, caching solutions for triple patterns satisfies (b) and (c) but not (a), while caching solutions for complex BGPs satisfies (a) but not (b), (c) or (d). 

Lampo et al.~\cite{LampoVDR11} propose caching of solutions for star joins, which may strike a good balance in terms of reducing costs, being reusable, and not having a high space overhead (as they share a common variable). Other caching techniques try to increase cache hit rates by detecting similar (sub)queries. Stuckenschmidt~\cite{Stuckenschmidt04} uses a similarity measure for caching -- based on the edit distance between BGPs -- that estimates the amount of computational effort needed to compute the solutions for one query given the solutions to the other. Lorey and Naumann~\cite{LoreyN13a} propose a technique for grouping similar queries, which enables a pre-fetching strategy based on predicting what a user might be interested in based on their initial queries. Another direction is to normalize (sub)queries to increase cache hit rates. Wu et al.~\cite{WuY12} propose various algebraic normalizations in order to identify common subqueries~\cite{LoreyN13a}, while Papailiou et al.~\cite{PapailiouTKK15} generalize subqueries by replacing selective constants with variables and thereafter canonically labeling variables (modulo isomorphism) to increase cache hit rates. Addressing dynamic data, Martin et al.~\cite{MartinUA10} propose a cache where results for queries are stored in a relational database but are invalidated when a triple matching a query pattern changes. Williams and Weaver~\cite{WilliamsW11} add last-updated times to their RDF index to help invalidate cached data.

Given that an arbitrary BGP can produce an exponential number of results, Zhang et al.~\cite{ZhangSTQ15} propose to cache frequently accessed ``hot triples'' from the RDF graph in memory, rather than caching (sub-)query results. This approach limits the space overhead at the cost of recomputing joins.

\subsection{Discussion} 

Techniques for processing BGPs are often based on techniques for processing relational joins. Beyond standard pairwise joins, multiway joins can help to emulate some of the benefits of property table storage by evaluating star joins more efficiently. Another recent and promising approach is to apply wco join algorithms whose runtime is bounded theoretically by the number of results that the BGP could generate. More and more attention has also been dedicated to computing joins in GPUs by translating relational algebra (e.g., joins) into linear algebra (e.g., matrix multiplication). Aside from specific algorithms, the order in which joins are processed can have a dramatic effect on runtimes. Statistics about the RDF graph help to find a good ordering at the cost of computing and maintaining those statistics; more lightweight alternatives include runtime sampling, or syntactic heuristics that consider only the query. To decide the ordering, options range from simple greedy strategies to complex metaheuristics; while simpler strategies have lower planning times, more complex strategies may find more efficient plans. Another optimization is to cache results across BGPs, for which a time--space trade-off must be considered.

\section{Query Processing}
\label{sec:query}

While we have defined RDF stores as engines capable of storing, indexing and processing joins over RDF graphs, SP\-ARQL engines support various features beyond joins. We describe  techniques for efficiently evaluating such features, including the relational algebra (beyond joins) and property paths. We further include some general extensions proposed for SPARQL to support recursion and analytics.


\subsection{Relational algebra (beyond joins)}

Complex (navigational) graph patterns CGPs introduce additional relational operators beyond joins.

Like in relational databases, \textit{algebraic rewriting rules} can be applied over CGPs in SPARQL to derive equivalent but more efficient plans.
Schmidt et al.~\cite{schmidt2010foundations} present a set of such rules for SPARQL under set semantics, such as:
\begin{align*}
\sigma_{R_1 \vee R_2}(M) &~\equiv \sigma_{R_1}(M) \cup \sigma_{R_2}(M) \\
\sigma_{R_1 \wedge R_2}(M) &~\equiv \sigma_{R_1}(\sigma_{R_2}(M)) \\
\sigma_{R_1}(\sigma_{R_2}(M)) &~\equiv \sigma_{R_2}(\sigma_{R_1}(M)) \\
\sigma_{R}(M_1 \cup M_2) &~\equiv \sigma_{R}(M_1) \cup \sigma_{R}(M_2) \\
\sigma_{R}(M_1^* \bowtie M_2) &~\equiv \sigma_{R}(M_1^*) \bowtie M_2 \\
\sigma_{R}(M_1^* \loj M_2) &~\equiv \sigma_{R}(M_1^*) \loj M_2 \\
\sigma_{R}(M_1^* \vartriangleright M_2) &~\equiv \sigma_{R}(M_1^*) \vartriangleright M_2
\end{align*}
\noindent where for each $\mu \in M_1^*$, it holds that $\vars{R} \subseteq \dom{\mu}$. The first two rules split filters, meaning that they can be pushed further down in a query in order to reduce intermediary results. The third rule allows the order in which filters are applied to be swapped. Finally the latter four rules describe how filters can be pushed ``down'' inside various operators.

Another feature of importance for querying RDF graphs are optionals ($\loj$), as they facilitate returning partial solutions over incomplete data. Given that an optional can be used to emulate a form of negation (in Table~\ref{tab:ra} it is defined using an anti-join), it can lead to jumps in computational complexity~\cite{10.1145/1567274.1567278}. Works have thus studied a fragment called \textit{well-designed patterns}, which forbid using a variable on the right of an optional that does not appear on the left but does appear elsewhere in the query; taking an example, the CGP 
$(\{ (\vbx,\cnp,\vby) \} \texttt{ OPTIONAL } \{ (\vbx,\cnq,\vbz) \}) \texttt{ . }\{ (\vbx,\cnr,\vbz) \}$ is not well designed as the variable $\vbz$ appears on the right of an \texttt{OPTIONAL} and not on the left, but does appear elsewhere in the query. Such variables may or may not be left unbound after the left outer join is evaluated, which leads to complications if they are used outside the optional clause. Most SPARQL queries using optionals in practice are indeed well-designed, where rewriting rules have been proposed specifically to optimize such queries~\cite{10.1145/1567274.1567278,Letelier0PS13}.

\subsection{Property paths}

Navigational graph patterns (NGPs) extend BGPs with property paths, which are extensions of (2)RPQs that allow for matching paths of arbitrary length in the graph.

Some approaches evaluate property paths using graph search algorithms. Though not part of SPARQL, Gubichev and Neumann~\cite{GubichevN11} implement single-source shortest paths 
by applying Dijsktra's search algorithm over B-Trees. 
Baier et al.~\cite{BaierDRV17} propose to use the A* search algorithm, where search is guided by a heuristic that measures the minimum distance from the current node to completing a path. 

Extending RDF-3X, Gubichev et al.~\cite{GubichevBS13} build a FERRARI index~\cite{SeufertABW13} (see Section~\ref{sec:pathind}) for each property \iri{p} in the graph that forms a directed path of length at least 2. The indexes are used to evaluate paths \iri{p*} or \iri{p+}. 
Paths of the form $(\iri{p}\texttt{/}\iri{q})*$, $(\iri{p}\texttt{|}\iri{q})*$, etc., are not directly supported.

Koschmieder and Leser~\cite{KoschmiederL12}, and Nguyen and Kim~\cite{NguyenK17} optimize property paths by splitting them according to ``rare labels'': given a property path $\iri{p}*\texttt{/}\iri{q}/\iri{r}*$, if \iri{q} has few triples in the graph, the path can be split into $\iri{p}*\texttt{/}\iri{q}$ (evaluated right-to-left) and $\iri{q}/\iri{r}*$ (evaluated left-to-right), subsequently joining the results. Splitting paths can enable parallelism: Miura et al.~\cite{MiuraAK19} evaluate such splits on field programmable gate arrays (FPGAs), enabling hardware acceleration. Wadhwa et al.~\cite{WadhwaPRBB19} rather use bidirectional random walks from candidate endpoints on both sides of the path, returning solutions when walks from each side coincide.

Another way to support property paths is to use recursive queries. Stuckenschmidt et al.~\cite{StuckenschmidtVBH05} evaluate property paths such as $\texttt{:p}+$ using recursive nested-loop and hash joins. Dey et al.~\cite{DeyC0GWL13}, Yakovets et al.~\cite{YakovetsGG13} and Jachiet et al.~\cite{JachietGGL20} propose translations of more general property paths (or RPQs) to extensions of the relational algebra with recursive or transitive operators. Paths can be evaluated by SQL engines using \texttt{WITH RECURSIVE}; however Yakovets et al.~\cite{YakovetsGG13} note that highly nested SQL queries may result, and that popular relational database engines cannot (efficiently) detect cycles. Dey et al~\cite{DeyC0GWL13} alternatively explore the evaluation of RPQs via translations to recursive Datalog.

\begin{figure*}
\centering
\begin{tabular}{|c@{~}|c@{~}|c@{~~}|c@{~}|c@{~}c@{~~~}|}
\hline
(\var{x},\texttt{s:n/(s:r|s:n)*},\var{z})
& 
\setlength{\vgap}{0.3cm}
\setlength{\hgap}{0.2cm}
\begin{tikzpicture}
\node (r) {\texttt{s:r}};

\node[right=\hgap of r] (n2) {\texttt{s:n}};

\node[right=0.5\hgap of r] (m1) {};

\node[above=\vgap of m1] (d) {\texttt{|}}
   edge (r) edge (n2);
   
\node[above=\vgap of d] (s) {\texttt{*}}
   edge (d);

\node[left=\hgap of s] (n1) {\texttt{s:n}};

\node[right=0.5\hgap of n1,anchor=mid] (m2) {};

\node[above=\vgap of m2] (c) {\texttt{/}}
   edge (s) edge (n1);
\end{tikzpicture}
&
\setlength{\vgap}{0.5cm}
\setlength{\hgap}{0.5cm}
\begin{tikzpicture}[baseline=-2.3cm]
\node (i) {};

\node[circle,inner sep=2pt,draw,below=1.2\vgap of i,anchor=mid] (q0) {$q_0$}
  edge [arrin] (i);
  
\node[circle,inner sep=2pt,double,draw,below=1.8\vgap of q0,anchor=mid] (q1) {$q_1$}
  edge [arrin] node[lab] {s:n} (q0)
  edge [in=150,out=210,looseness=7,arrout] node[lab] {s:r} (q1)
  edge [in=30,out=330,looseness=7,arrout] node[lab] {s:n} (q1);
\end{tikzpicture}
&
\setlength{\vgap}{0.3cm}
\setlength{\hgap}{0.2cm}
\begin{tikzpicture}
\node (r) {(\var{y},\texttt{s:r},\var{z})};

\node[right=\hgap of r] (n2) {(\var{y},\texttt{s:n},\var{z})};

\node[right=0.5\hgap of r] (m1) {};

\node[above=\vgap of m1] (d) {$\cup$}
   edge (r) edge (n2);
   
\node[above=\vgap of d] (s) {\texttt{$\alpha$}}
   edge (d);

\node[left=\hgap of s] (n1) {(\var{x},\texttt{s:n},\var{y})};

\node[right=0.5\hgap of n1,anchor=mid] (m2) {};

\node[above=\vgap of m2,xshift=-0.2cm] (c) {$\bowtie$}
   edge (s) edge (n1);
\end{tikzpicture}
&
\setlength{\vgap}{0.3cm}
\setlength{\hgap}{0.2cm}

\setlength{\vgap}{0.5cm}
\setlength{\hgap}{0.5cm}
\begin{tikzpicture}[baseline=-2.3cm]
\node (i) {};

\node[circle,inner sep=2pt,draw,below=1.2\vgap of i,anchor=mid] (q0) {$q_0$}
  edge [arrin] (i);
  
\node[circle,inner sep=2pt,double,draw,below=1.8\vgap of q0,anchor=mid] (q1) {$q_1$}
  edge [arrin,bend left=40] node[lab] {s:n} (q0)
  edge [arrin,bend right=40] node[lab] {s:r} (q0);
\end{tikzpicture}
&
\setlength{\vgap}{0.5cm}
\setlength{\hgap}{0.5cm}
\begin{tikzpicture}[baseline=-2.3cm]
\node (i) {};

\node[circle,inner sep=2pt,draw,below=1.2\vgap of i,anchor=mid] (q0) {$q_0$}
  edge [arrin] (i);
  
\node[circle,inner sep=2pt,double,draw,below=1.8\vgap of q0,anchor=mid] (q1) {$q_1$}
  edge [arrin] node[lab] {s:n} (q0)
  edge [in=150,out=210,looseness=7,arrout] node[lab] {$\mathrm{WP}'$} (q1);
\end{tikzpicture}
\\[2pt]

Property Path & PT & ~~FA & $\alpha$-RA & ~~~~WP$'$ & ~~~WP~ \\[2pt]\hline
\end{tabular}

\caption{An example property path with its parse tree (PT) and three plans based on finite automata (FA), relational algebra with transitive closure ($\alpha$-RA), and a waveplan (WP) that uses a memoized waveplan (WP$'$) \label{fig:waveguide}}
\end{figure*}
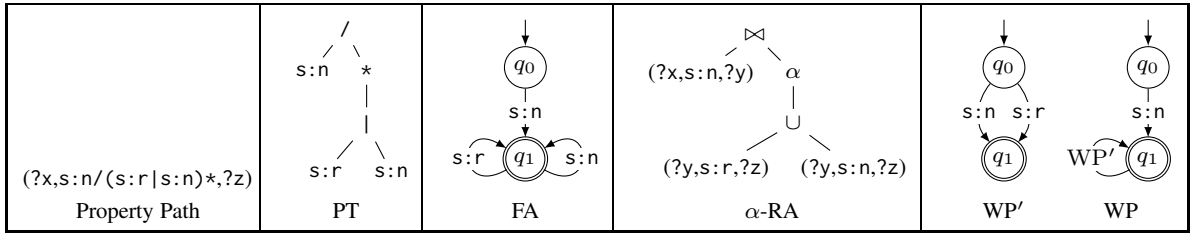

In later work, Yakovets et al.~\cite{YakovetsGG16} propose Waveguide, which first converts the property path into a parse tree, from which plans can be built based on finite automata (FA), or relational algebra with transitive closure ($\alpha$-RA, where $\alpha$ denotes transitive closure). Figure~\ref{fig:waveguide} gives an example of a parse tree and both types of plans. Although there is overlap, FA can express physical plans that $\alpha$-RA cannot, and vice versa. For example, in FA we can express non-deterministic transitions (see $q_0$ in Figure~\ref{fig:waveguide}), while in $\alpha$-RA we can materialize (cache) a particular relation in order to apply transitive closure over it. Waveguide then uses hybrid \textit{waveplans}, where breadth-first search is guided in a similar manner to FA, but where the results of an FA can be \textit{memoized} (cached) and reused multiple times like in $\alpha$-RA.

Evaluating complex property paths can be costly, but property paths in practice are often quite simple. Martens and Trautner~\cite{MartensT18} propose a class of RPQs called \textit{simple transitive expressions} (STEs) that are found to cover 99.99\% of the queries found in Wikidata SPARQL logs, and have desirable theoretical properties. Specifically, they define \textit{atomic expressions} of the form $p_1 \texttt{|} \ldots \texttt{|} p_n$, where $p_1,\ldots,p_n$ are IRIs and $n \geq 0$; and also \textit{bounded expressions} of the form $a_1 \texttt{/} \ldots \texttt{/} a_k$ or $a_1\texttt{?} \texttt{/} \ldots \texttt{/} a_k\texttt{?}$ where $a_1, \ldots, a_k$ are atomic expressions and $k \geq 0$. Then an expression of the form $b_1\texttt{/}a^*\texttt{/}b_2$, is a \textit{simple transitive expression} (\textit{STE}), where $b_1$ and $b_2$ are bounded expressions, and $a$ is an atomic expression. 
They then show that simple paths for STEs can be enumerated more efficiently than arbitrary RPQs. 

\subsection{Recursion} \label{sec:recursion}

Property paths offer a limited form of recursion. While extended forms of property paths have been proposed to include (for example) path intersection and difference~\cite{FiondaPC19}, more general extensions of SPARQL have also been proposed to support graph-based and relation-based recursion. 

Reutter et al.~\cite{ReutterSV15} propose to extend SPARQL with graph-based recursion, where a temporary RDF graph is built by recursively adding triples produced through \texttt{CONSTRUCT} queries over the base graph and the temporary graph up to a fixpoint; a \texttt{SELECT} query can then be evaluated over both graphs. The authors discuss how key features (including property paths) can then be supported through \textit{linear recursion}, meaning that each new triple only needs to be joined with the base graph, not the temporary graph, to produce further triples, leading to better performance. Corby et al.~\cite{CorbyFG17} propose LD-Script: a SPARQL-based scripting language supporting various features, including for-loops that can iterate over the triples returned by a \texttt{CONSTRUCT} query.

Hogan et al. \cite{10.1007/978-3-030-62419-4_29} propose SPARQAL: a lightweight language that supports relation-based (i.e., \texttt{SELECT}-based) recursion over SPARQL. The results of a \texttt{SELECT} query can be stored as a variable, and injected into a future query. Do--until loops can be called until a particular condition is met, thus enabling recursion over \texttt{SELECT} queries. 

\subsection{Analytics} 

SPARQL engines often focus on transactional (OLTP) workloads involving selective queries that are efficiently solved through lookups on indexes. Recently, however, a number of approaches have looked at addressing analytical (OLAP) workloads for computing slices, aggregations, etc.~\cite{call2021motivations}.

One approach is to rewrite SPARQL queries to languages executable in processing environments suitable for analytical workloads, including PigLatin (e.g., Pig\-SPARQL~\cite{Schtzle2011PigSPARQLMS}, RAPID+~\cite{10.1007/978-3-642-21064-8_4}), Hadoop (e.g., Sempala~\cite{Schtzle2014SempalaIS}), Spark (e.g., S2RDF~\cite{Schtzle2015S2RDFRQ}), etc. Such frameworks are better able to handle analytical (OLAP) workloads, but not all SPARQL features are easily supported on existing distributed frameworks.

Conversely, one can also translate from analytical languages to SPARQL queries, allowing for \textit{in-database analytics}, where  analytical workloads are translated into queries run by the SPARQL engine/database. Papadaki et al.~\cite{a14020034} propose the high-level functional query language HIFUN for applying analytics over RDF data. Rules for translating analytical HIFUN queries to SPARQL are then presented.

There has also been growing interest in combining \textit{graph analytics} -- such as centrality measures, shortest paths, graph clustering, etc. -- with SPARQL. In this way, SPARQL can be used as a declarative language to construct sub-graphs over which analytics are applied, and can further express queries involving the results of analytics. Unlike OLAP-style analytics, graph analytics often require recursion. 
One approach is to extend SPARQL to include imperative functions for invoking common graph algorithms. Abdelaziz et al. \cite{7959641} propose Spartex: an extension of SPARQL that allows for invoking common graph algorithms -- such as PageRank, shortest paths, etc. -- as well as user-defined procedures (UDPs) written in a custom procedural language. An alternative approach is to support graph analytics through a more general recursive language based on SPARQL (as discussed in Section~\ref{sec:recursion}). Hogan et al.~\cite{10.1007/978-3-030-62419-4_29} show how the recursive language SPARQAL allows for expressing and evaluating in-database graph analytics, including breadth-first search, PageRank, local clustering coefficient, etc.

\subsection{Graph query rewriting} 

We have seen approaches that rewrite SPARQL queries into languages such as SQL~\cite{Erling2010,YakovetsGG13}, PigLatin~\cite{Schtzle2011PigSPARQLMS,10.1007/978-3-642-21064-8_4}, etc. Other works rewrite SPARQL into the query languages of (other) graph databases. SPARQL--Gremlin~\cite{ThakkarARM020} rewrites SPARQL to Gremlin, allowing SPARQL queries to be evaluated on graph database engines that support Gremlin, while Semantic Property Graph~\cite{DBLP:journals/corr/abs-2009-07410} describes how reified RDF graphs can be projected into the property graph model supported by many graph database engines.

\subsection{Multi-query optimization}

While the techniques discussed thus far optimize queries individually, multi-query optimization evaluates batches of queries efficiently by exploiting their commonalities. Le et al.~\cite{10.1109/ICDE.2012.37} propose to first cluster a set of queries into groups with maximal common edge subgraphs; for example, the BGP $\{(\vbw_1,\cnp,\vbx_1),(\vbw_1,\cnq,\vby_1),(\vbw_1,\cnr,\vbz_1),(\vby_1,\cns,\vbz_1)\}$ and the BGP $\{(\vbw_2,\cnp,\vbx_2),(\vbw_2,\cnq,\vby_2),(\vbz_2,\cnr,\vbw_2)\}$ may form a cluster. A query is then constructed for each cluster by extending its maximal common sub-BGP with optional patterns needed by a proper subset of the queries; for example, $(\{(\vbw,\cnp,\vbx),(\vbw,\cnq,\vby)\} \loj \{ (\vbw,\cnr,\vbz),(\vby,\cns,\vbz) \}) \loj \{ (\vbz,\cnr,\vbw) \}$ would be used for the previous cluster. Individual query results are then computed from the cluster-level results. Optimizing for multiple property paths, Abul-Basher~\cite{Abul-Basher17} proposes to find a maximum common sub-automaton that can be evaluated and reused across multiple queries. More recent works further address multi-query optimization in specific settings, including federated systems~\cite{PengGZOXZ21}, and continuous querying over streaming RDF data~\cite{ZervakisSTH20}.

\subsection{Discussion}

SPARQL supports various features beyond joins that ideally should be implemented in an efficient manner. One option is to rewrite SPARQL queries into a target language and evaluate them using an existing engine for that language. However, it is unlikely that an existing language/engine will support all features of SPARQL in an efficient manner. Better performance for a wider range of features can be achieved with custom implementations and optimizations, where property paths have been the focus of many works. Other features that have been targeted for optimization are filters and optionals, noting that optionals are quite frequently used in the context of querying incomplete RDF data. Multi-query optimization can further help to evaluate multiple queries at once. More recent works have addressed recursion and analytics for SPARQL in order to support additional RDF data management scenarios and knowledge graph use-cases.

\section{Partitioning}
\label{sec:partition}

In distributed RDF stores and SPARQL engines, the data are partitioned over a cluster of machines in order to enable horizontal scale, where additional machines can be allocated to the cluster to handle larger volumes of data. However, horizontal scaling comes at the cost of network communication costs. Thus a key optimization is to choose a partitioning scheme that reduces communication costs by enforcing various forms of locality, principally allowing certain types of (intermediate) joins to be processed on each individual machine~\cite{akhter2018empirical}. Formally, given an RDF graph $G$ and $n$ machines, an $n$-partition of $G$ is a tuple of subgraphs $(G_1, \ldots, G_n)$ such that $G = \bigcup\limits_{i=1}^n G_i$, with the idea that each subgraph $G_i$ will be stored on machine $i$.\footnote{We relax the typical requirement for a set partition that $G_i \cap G_j = \emptyset$ for all $1 \leq i < j \leq n$ to allow for the possibility of replication or other forms of redundancy.} We now discuss different high-level alternatives for partitioning.

\subsection{Triple/Quad-based Partitioning}

A first option is to partition based on individual triples or quads without considering the rest of the graph. For simplicity we will speak about triples as the discussion generalizes straightforwardly to quads. The simplest option is to use \textit{round robin} or \textit{random partitioning}, which effectively places triples on an arbitrary machine. This ensures even load balancing, but does not support any locality of processing, and does not allow for finding the particular machine storing triples that match a given pattern.

An alternative is to partition according to a deterministic function over a given key; for example, a partition key of \textsc{s} considers only the subject, while a partition key of \textsc{po} considers both the predicate and object. Later given a triple pattern that covers the partition key (e.g., with a constant subject if the key is \textsc{s}), we can find the machine(s) storing all triples that match that pattern. We show some examples using different functions and partition keys in Figure~\ref{fig:partTQ} considering four machines. \textit{Range-based partitioning} assigns a range over the partition key to each function, where the example of Figure~\ref{fig:partTQ} splits \textsc{s} into [\texttt{a:1},\texttt{a:3}], [\texttt{a:4},\texttt{a:6}], [\texttt{b:1},\texttt{b:3}], [\texttt{c:1},\texttt{d:1}]. This approach allows for range-based queries to be pushed to one machine, but requires maintaining a mapping of ranges to machines, and can be complicated to keep balanced. An alternative is \textit{hash-based partitioning} where we compute the hash of the partition key modulo the number of machines, where the second example of Figure~\ref{fig:partTQ} splits \textsc{p} by hash. This does not require storing any mapping, and techniques such as consistent hashing can be used to rebalance load when a machine enters or leaves; however, if partition keys are skewed (e.g., one predicate is very common), it may lead to an unbalanced partition. A third option is to apply a \textit{hierarchical-based partition} based on prefixes, where the third example of Figure~\ref{fig:partTQ} partitions \textsc{o} by their namespace. This may lead to increased locality of data with the same prefix~\cite{10.1145/3066911.3066915}, where different levels of prefix can be chosen to enable balancing, but choosing prefixes that offer balanced partitions is non-trivial.

Any such partitioning function will send any triple with the same partition key to the same machine, which ensures that (equi-)joins on partition keys can be pushed to individual machines. Hash-based partitioning is perhaps the most popular among distributed RDF stores (e.g., YARS2~\cite{10.5555/1785162.1785179}, SHARD~\cite{10.1145/1940747.1940751}, etc.). Often triples will be hashed according to multiple partition keys in order to support different index permutations, triple patterns, and joins (e.g, with \textsc{s} and \textsc{o} as two partition keys, we can push \textsc{s}--\textsc{s}, \textsc{o}--\textsc{o} and \textsc{s}--\textsc{o} joins to each machine). 
Care must be taken to avoid imbalances caused by frequent terms, such as the \texttt{rdf:type} predicate, or frequent objects such as classes, countries, etc. Omitting partitioning on highly-skewed partition keys may be advantageous for balancing purposes~\cite{10.5555/1785162.1785179}.

\newcommand{\pa}[2]{\begin{tabular}{@{}c@{}}#1\\[-1ex]\tiny(#2)\end{tabular}}

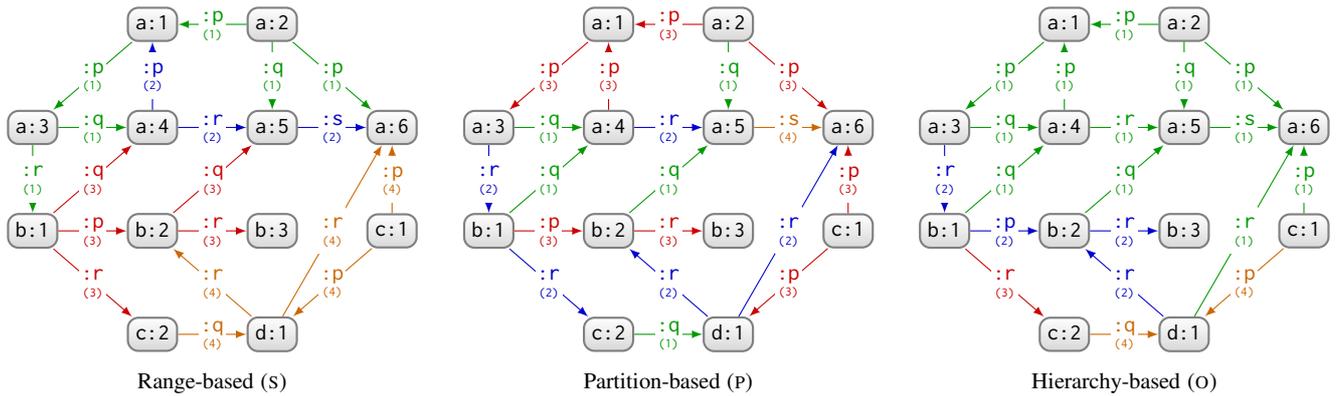
\begin{figure*}[t]
\setlength{\hgap}{0.9cm}
\setlength{\vgap}{0.9cm}

\begin{tikzpicture}[baseline]
\node[iri,anchor=mid] (a1) {a:1};

\node[iri,anchor=mid,right=\hgap of a1] (a2) {a:2}
  edge[arrout,m1] node[lab] {\pa{:p}{1}} (a1);
  
\node[iri,anchor=mid,below=\vgap of a1] (a4) {a:4}
  edge[arrout,m2] node[lab] {\pa{:p}{2}} (a1);
  
\node[iri,anchor=mid,left=\hgap of a4] (a3) {a:3}
  edge[arrin,m1] node[lab] {\pa{:p}{1}} (a1)
  edge[arrout,m1] node[lab] {\pa{:q}{1}} (a4);
  
\node[iri,anchor=mid,right=\hgap of a4] (a5) {a:5}
  edge[arrin,m1] node[lab] {\pa{:q}{1}} (a2)
  edge[arrin,m2] node[lab] {\pa{:r}{2}} (a4);
  
\node[iri,anchor=mid,right=\hgap of a5] (a6) {a:6}
  edge[arrin,m1] node[lab] {\pa{:p}{1}} (a2)
  edge[arrin,m2] node[lab] {\pa{:s}{2}} (a5);
  
\node[iri,anchor=mid,below=\vgap of a3] (b1) {b:1}
  edge[arrin,m1] node[lab] {\pa{:r}{1}} (a3)
  edge[arrout,m3] node[lab] {\pa{:q}{3}} (a4);

\node[iri,anchor=mid,right=\hgap of b1] (b2) {b:2}
  edge[arrin,m3] node[lab] {\pa{:p}{3}} (b1)
  edge[arrout,m3] node[lab] {\pa{:q}{3}} (a5);
  
\node[iri,anchor=mid,right=\hgap of b2] (b3) {b:3}
  edge[arrin,m3] node[lab] {\pa{:r}{3}} (b2);

\node[iri,anchor=mid,right=\hgap of b3] (c1) {c:1}
  edge[arrout,m4] node[lab] {\pa{:p}{4}} (a6);
  
\node[iri,anchor=mid,below=\vgap of b2] (c2) {c:2}
  edge[arrin,m3] node[lab] {\pa{:r}{3}} (b1);

\node[iri,anchor=mid,right=\hgap of c2] (d1) {d:1}
  edge[arrout,m4] node[lab] {\pa{:r}{4}} (a6)
  edge[arrout,m4] node[lab] {\pa{:r}{4}} (b2)
  edge[arrin,m4] node[lab] (bm) {\pa{:q}{4}} (c2)
  edge[arrin,m4] node[lab] {\pa{:p}{4}} (c1);

\node[below=0.1\vgap of bm] {Range-based (\textsc{s})};
\end{tikzpicture}
\hfill
\begin{tikzpicture}[baseline]
\node[iri,anchor=mid] (a1) {a:1};

\node[iri,anchor=mid,right=\hgap of a1] (a2) {a:2}
  edge[arrout,m3] node[lab] {\pa{:p}{3}} (a1);
  
\node[iri,anchor=mid,below=\vgap of a1] (a4) {a:4}
  edge[arrout,m3] node[lab] {\pa{:p}{3}} (a1);
  
\node[iri,anchor=mid,left=\hgap of a4] (a3) {a:3}
  edge[arrin,m3] node[lab] {\pa{:p}{3}} (a1)
  edge[arrout,m1] node[lab] {\pa{:q}{1}} (a4);
  
\node[iri,anchor=mid,right=\hgap of a4] (a5) {a:5}
  edge[arrin,m1] node[lab] {\pa{:q}{1}} (a2)
  edge[arrin,m2] node[lab] {\pa{:r}{2}} (a4);
  
\node[iri,anchor=mid,right=\hgap of a5] (a6) {a:6}
  edge[arrin,m3] node[lab] {\pa{:p}{3}} (a2)
  edge[arrin,m4] node[lab] {\pa{:s}{4}} (a5);
  
\node[iri,anchor=mid,below=\vgap of a3] (b1) {b:1}
  edge[arrin,m2] node[lab] {\pa{:r}{2}} (a3)
  edge[arrout,m1] node[lab] {\pa{:q}{1}} (a4);

\node[iri,anchor=mid,right=\hgap of b1] (b2) {b:2}
  edge[arrin,m3] node[lab] {\pa{:p}{3}} (b1)
  edge[arrout,m1] node[lab] {\pa{:q}{1}} (a5);
  
\node[iri,anchor=mid,right=\hgap of b2] (b3) {b:3}
  edge[arrin,m3] node[lab] {\pa{:r}{3}} (b2);

\node[iri,anchor=mid,right=\hgap of b3] (c1) {c:1}
  edge[arrout,m3] node[lab] {\pa{:p}{3}} (a6);
  
\node[iri,anchor=mid,below=\vgap of b2] (c2) {c:2}
  edge[arrin,m2] node[lab] {\pa{:r}{2}} (b1);

\node[iri,anchor=mid,right=\hgap of c2] (d1) {d:1}
  edge[arrout,m2] node[lab] {\pa{:r}{2}} (a6)
  edge[arrout,m2] node[lab] {\pa{:r}{2}} (b2)
  edge[arrin,m1] node[lab] (bm) {\pa{:q}{1}} (c2)
  edge[arrin,m3] node[lab] {\pa{:p}{3}} (c1);

\node[below=0.1\vgap of bm] {Partition-based (\textsc{p})};
\end{tikzpicture}
\hfill
\begin{tikzpicture}[baseline]
\node[iri,anchor=mid] (a1) {a:1};

\node[iri,anchor=mid,right=\hgap of a1] (a2) {a:2}
  edge[arrout,m1] node[lab] {\pa{:p}{1}} (a1);
  
\node[iri,anchor=mid,below=\vgap of a1] (a4) {a:4}
  edge[arrout,m1] node[lab] {\pa{:p}{1}} (a1);
  
\node[iri,anchor=mid,left=\hgap of a4] (a3) {a:3}
  edge[arrin,m1] node[lab] {\pa{:p}{1}} (a1)
  edge[arrout,m1] node[lab] {\pa{:q}{1}} (a4);
  
\node[iri,anchor=mid,right=\hgap of a4] (a5) {a:5}
  edge[arrin,m1] node[lab] {\pa{:q}{1}} (a2)
  edge[arrin,m1] node[lab] {\pa{:r}{1}} (a4);
  
\node[iri,anchor=mid,right=\hgap of a5] (a6) {a:6}
  edge[arrin,m1] node[lab] {\pa{:p}{1}} (a2)
  edge[arrin,m1] node[lab] {\pa{:s}{1}} (a5);
  
\node[iri,anchor=mid,below=\vgap of a3] (b1) {b:1}
  edge[arrin,m2] node[lab] {\pa{:r}{2}} (a3)
  edge[arrout,m1] node[lab] {\pa{:q}{1}} (a4);

\node[iri,anchor=mid,right=\hgap of b1] (b2) {b:2}
  edge[arrin,m2] node[lab] {\pa{:p}{2}} (b1)
  edge[arrout,m1] node[lab] {\pa{:q}{1}} (a5);
  
\node[iri,anchor=mid,right=\hgap of b2] (b3) {b:3}
  edge[arrin,m2] node[lab] {\pa{:r}{2}} (b2);

\node[iri,anchor=mid,right=\hgap of b3] (c1) {c:1}
  edge[arrout,m1] node[lab] {\pa{:p}{1}} (a6);
  
\node[iri,anchor=mid,below=\vgap of b2] (c2) {c:2}
  edge[arrin,m3] node[lab] {\pa{:r}{3}} (b1);

\node[iri,anchor=mid,right=\hgap of c2] (d1) {d:1}
  edge[arrout,m1] node[lab] {\pa{:r}{1}} (a6)
  edge[arrout,m2] node[lab] {\pa{:r}{2}} (b2)
  edge[arrin,m4] node[lab] (bm) {\pa{:q}{4}} (c2)
  edge[arrin,m4] node[lab] {\pa{:p}{4}} (c1);

\node[below=0.1\vgap of bm] {Hierarchy-based (\textsc{o})};
\end{tikzpicture}

\caption{Examples of triple-based partitioning schemes \label{fig:partTQ}}
\end{figure*}

\subsection{Graph-based Partitioning}

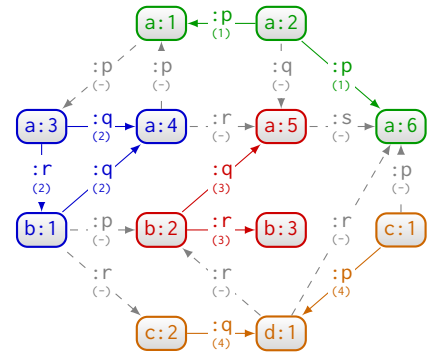
\begin{figure}

\setlength{\hgap}{0.9cm}
\setlength{\vgap}{0.9cm}

\centering
\begin{tikzpicture}[baseline]
\node[iri,m1,anchor=mid] (a1) {a:1};

\node[iri,m1,anchor=mid,right=\hgap of a1] (a2) {a:2}
  edge[arrout,m1] node[lab] {\pa{:p}{1}} (a1);
  
\node[iri,m2,anchor=mid,below=\vgap of a1] (a4) {a:4}
  edge[arrout,nm] node[lab] {\pa{:p}{-}} (a1);
  
\node[iri,m2,anchor=mid,left=\hgap of a4] (a3) {a:3}
  edge[arrin,nm] node[lab] {\pa{:p}{-}} (a1)
  edge[arrout,m2] node[lab] {\pa{:q}{2}} (a4);
  
\node[iri,m3,anchor=mid,right=\hgap of a4] (a5) {a:5}
  edge[arrin,nm] node[lab] {\pa{:q}{-}} (a2)
  edge[arrin,nm] node[lab] {\pa{:r}{-}} (a4);
  
\node[iri,m1,anchor=mid,right=\hgap of a5] (a6) {a:6}
  edge[arrin,m1] node[lab] {\pa{:p}{1}} (a2)
  edge[arrin,nm] node[lab] {\pa{:s}{-}} (a5);
  
\node[iri,m2,anchor=mid,below=\vgap of a3] (b1) {b:1}
  edge[arrin,m2] node[lab] {\pa{:r}{2}} (a3)
  edge[arrout,m2] node[lab] {\pa{:q}{2}} (a4);

\node[iri,m3,anchor=mid,right=\hgap of b1] (b2) {b:2}
  edge[arrin,nm] node[lab] {\pa{:p}{-}} (b1)
  edge[arrout,m3] node[lab] {\pa{:q}{3}} (a5);
  
\node[iri,m3,anchor=mid,right=\hgap of b2] (b3) {b:3}
  edge[arrin,m3] node[lab] {\pa{:r}{3}} (b2);

\node[iri,m4,anchor=mid,right=\hgap of b3] (c1) {c:1}
  edge[arrout,nm] node[lab] {\pa{:p}{-}} (a6);
  
\node[iri,m4,anchor=mid,below=\vgap of b2] (c2) {c:2}
  edge[arrin,nm] node[lab] {\pa{:r}{-}} (b1);

\node[iri,m4,anchor=mid,right=\hgap of c2] (d1) {d:1}
  edge[arrout,nm] node[lab] {\pa{:r}{-}} (a6)
  edge[arrout,nm] node[lab] {\pa{:r}{-}} (b2)
  edge[arrin,m4] node[lab] (bm) {\pa{:q}{4}} (c2)
  edge[arrin,m4] node[lab] {\pa{:p}{4}} (c1);
\end{tikzpicture}

\caption{Example of optimal $k$-way partitioning ($k = 4$) \label{fig:partKW}}
\end{figure}

Graph-based partitioning takes into consideration the entire graph when computing a partition. A common strategy is to apply a \textit{$k$-way partition} of the RDF graph $G$~\cite{10.5555/305219.305248}. Formally, letting $V = \so{G}$ denote the nodes of $G$, the goal is to compute a node partition $V_1, \ldots, V_n$ such that $V = \bigcup\limits_{i=1}^k V_n$, $V_i \cap V_j = \emptyset$ for all $1 \leq i < j \leq k$, $\lfloor \frac{|V|}{k} \rfloor \leq |V_i| \leq \lceil\frac{|V|}{k} \rceil$ for all $1 \leq i \leq k$, and the number of triples $(s,p,o) \in G$ such that $s$ and $o$ are in different node partitions is minimized. In Figure~\ref{fig:partKW}, we show the optimal 4-way partitioning of the graph seen previously, where each partition has 3 nodes, there are 10 edges between partitions (shown dashed), and no other such partition leads to fewer edges ($<$10) between partitions. Edges between partitions may be replicated in the partitions they connect. Another alternative is to $k$-way partition the \textit{line graph} of the RDF graph: an undirected graph where each triple is a node, and triples sharing a subject or object have an edge between them.

Finding an optimal $k$-way partition is intractable\footnote{Given a graph, deciding if there is a $k$-way partition with fewer than $n$ edges between partitions is NP-complete.}, where approximations are thus necessary for large-scale graphs, including \textit{spectral methods}, which use the eigenvectors of the graph's Laplacian matrix to partition it; \textit{recursive bisection}, which recursively partitions the graph into two; \textit{multilevel partitioning}, which ``coarsens'' the graph by computes a hierarchical graph summary (similar to a multilevel quotient graph, per Figure~\ref{fig:quotient}), then partitions the smaller graph summary (using, e.g., spectral methods), and finally ``uncoarsens'' by expanding back out to the original graph maintaining the partitions; etc. We refer for more details to Bulu\c{c} et al.~\cite{BulucMSS016}, who argue that multilevel partitioning is ``\textit{clearly the most successful heuristic for partitioning large graphs}''. Such techniques have been used by H-RDF-3x~\cite{huang2011scalable}, EAGRE~\cite{6544856}, Koral~\cite{Janke2017KoralAG}, and more besides.

\subsection{Query-based Partitioning}

While the previous partitioning schemes only consider the data, other partitioning methods are (also) based on queries. Workload-based partitioning schemes identify common joins in query logs that can be used to partition or replicate parts of the graph in order to ensure that high-demand joins can be pushed to individual machines. Partitioning can then be \textit{a priori}, for example, based on a query log; or \textit{dynamic} (aka.\ adaptive), where the partitions change as queries are received.  Such strategies are used by systems that include WARP~\cite{6547414}, Partout~\cite{Galarraga:2014:PDE:2567948.2577302}, WORQ~\cite{10.1007/978-3-030-00671-6_34}, and AdPart~\cite{Harbi:2016:ASQ:2944220.2944335}.

\subsection{Replication}

Rather than partitioning data, data can also be replicated across partitions. This may vary from replicating the full graph on each machine, such that queries can be answered in full by any machine to increase query throughput (used, e.g., by DREAM~\cite{Hammoud:2015:DDR:2735703.2735705}), to replicating partitions that are in high-demand (e.g., containing schema data, central nodes, etc.) so that more queries can be evaluated on individual machines and/or machines have equal workloads that avoid hot-spots (used, e.g., by Blazegraph~\cite{ThompsonPC14} and Virtuoso~\cite{Erling2010}).

\subsection{Discussion}

Triple/quad-based partitioning is the simplest to compute and maintain, being dependent only on the data present in an individual tuple, allowing joins on the same partition key to be pushed to individual machines. Graph-based partitions allow for evaluating more complex graph patterns on individual machines, but are more costly to compute and maintain (considering, e.g., dynamic data). Information about queries, where available, can be used for the purposes of workload-based partitioning, which partitions or replicates data in order to enable locality for common sub-patterns. Replication can further improve load balancing, locality and fault-tolerance at the cost of redundant storage.

\section{Systems and Benchmarks}
\label{sec:systems}

\ja{In an extended version online~\cite{online}}{In \autoref{sec:engines}} we present a comprehensive survey of 135 individual RDF stores and SPARQL query engines -- both distributed and local -- in terms of the techniques discussed herein that they use. \ja{We}{In \autoref{sec:benchmarking}, we} further present the synthetic and real-world benchmarks available for evaluating these systems under a variety of criteria. 

\section{Summary}
\label{sec:conclusion}
In order to conclude this survey paper, we first summarize some of the current high-level trends that we have observed while preparing this survey, and then summarize the open research challenges that are left to address.

\subsection{Current trends}
\label{sec:trends}

While RDF stores and SPARQL engines have traditionally relied on relational databases and relational-style optimizations to ensure scalability and efficiency, we see a growing trend towards (1) native graph-based storage, indexing and query processing techniques, along with (2) exploiting modern hardware and data management/processing.

Native storage techniques for graphs move away from relational-style schemata for RDF, and rather focus on optimizing for the compression and navigation of RDF as a graph, with techniques such as index-free adjacency, tensor-based storage, and other graph-based representations. Indexing likewise has evolved to consider entity-based (i.e., node-based) schemes, path indexes, and structural indexes based on summarizing the graph structure of RDF data. While join processing over RDF is still largely inspired by techniques for relational databases, algorithms based on sideways information passing, multi-way joins, worst-case optimal joins, etc., have been shown to work particularly well on RDF graphs (e.g., given their fixed arity). In terms of query processing, features such as property paths and graph-based recursion go beyond what is considered in typical relational database management, with increased attention being paid to supporting graph analytics in the RDF/SPARQL setting.

Regarding modern hardware, following broader trends, many works now leverage NoSQL systems and distributed processing frameworks in order to scale RDF stores across multiple machines and handle new types of workloads. A similar trend is to better exploit modern hardware, where a variety of compact data structures have been proposed for storing RDF graphs in main memory, possibly across multiple machines, following a general trend of exploiting the growing RAM capacity of modern hardware. Recent techniques for processing graphs -- represented as matrices/tensors -- further enable hardware acceleration by leveraging GPUs and HPC architectures, per machine learning.

Such trends seem set to continue, where we expect to see further proposals of ``native'' techniques for RDF/SPARQL, further works that bridge from the RDF/SPARQL setting to related data management and processing settings in order to better support other types of workloads, as well as techniques that better leverage modern hardware, including increased RAM capacity, solid-state disks, GPUs, clusters of machines, and HPC architectures.

\subsection{Research Challenges and Future Directions}
\label{sec:problems}

Though major advances have been made in terms of the \textit{scale} and \textit{efficiency} of RDF stores in recent years, these will remain central challenges as the scale of RDF graphs and demand for querying them in more complex ways increases. Other challenges have only been occasionally or partially addressed by the literature, where we highlight:

\textit{Dynamics}: Many of the surveyed works assume static data, and do not handle updates gracefully. Thus, more work is needed on efficiently querying dynamic RDF graphs with SPARQL, including storage that efficiently supports reads and writes, incremental indexing, caching, etc.

\textit{Query optimizations (beyond joins):} Most works focus on optimizing joins and basic graph patterns. We found relatively few works optimizing features of SPARQL~1.1, such as property paths, negation, etc., where more work is needed. The expressivity of the SPARQL language is sure to grow (e.g., in the context of SPARQL~1.2), where these new features will likewise call for new techniques.

\textit{Query volume:} Leading SPARQL endpoints process millions of queries per day. This challenge motivates further research on workload-aware or caching strategies that leverage frequent sub-queries. Another research challenge is on how to ensure effective policies for serving many clients while avoiding server overload, where methods such as \textit{preemption}~\cite{MinierSM19}, which allows for pausing and resuming costly query requests, are promising ideas for further development.

\textit{Evaluation:} Various benchmarks are now available for comparing different RDF stores, but they tend to focus on system-level comparisons, thus conflating techniques. More fine-grained evaluation at the level of individual techniques in the RDF/SPARQL setting would be very useful to understand the different trade-offs that exist. Also many benchmarks were proposed for SPARQL~1.0, where there is a lack of benchmarks including features such as property paths.

\textit{Integration}: RDF and SPARQL are widely adopted on the Web, and for managing and querying knowledge graphs. However, in such settings, additional types of tasks are often considered, including \textit{federated querying}, \textit{reasoning}, \textit{enrichment}, \textit{refinement}, \textit{learning}, \textit{analytics}, etc. More work is needed on supporting or integrating features for these tasks in SPARQL. Interesting questions relate to efficiently supporting RDFS/OWL/Datalog reasoning, graph algorithms, knowledge graph embeddings, graph neural networks, etc., for RDF graphs within SPARQL engines.

\paragraph{Acknowledgments} This work was partially funded by a grant from the European Union’s Horizon 2020 research and innovation programme under the Marie Skłodowska-Curie grant agreement No 860801.  Hogan was supported by Fondecyt Grant
No. 1181896 and ANID – Millennium Science Initiative Program – Code ICN17\_002. Bin Yao was supported by the NSFC (61922054, 61872235, 61832017,
61729202, 61832013), the National Key Research and Development Program of China (2020YFB1710202, 2018YF\-C1504504), the Science and Technology Commission of S\-hanghai Municipality (STCSM) AI under Project 1951112\-0300. This work was also supported by the German Federal Ministry of Education and Research (BMBF) within the EuroStars project E1114681 3DFed under the grant No 01QE2114 and project KnowGraphs (No 860801).

\paragraph{Competing interests: The authors declare no competing interests.} 

\balance
\bibliographystyle{abbrv} 
\bibliography{refs}  

\ja{}{\appendix
\newpage
\normalsize

\section{Survey of RDF Stores}\label{sec:engines}

We now present a survey of local and distributed RDF stores, and how they use the aforementioned techniques. At the end of this section, we will discuss some general trends for RDF stores. We include here systems for which we could find technical details regarding (at least) the storage, indexing and processing of joins over RDF graphs.\footnote{We thus exclude systems -- such as TPF-based systems, and stores such as Fabric, Fluree, and TriplePlace~\cite{arndt2011tripleplace} -- that do not (yet) describe direct support for joins or basic graph patterns. We also exclude systems -- such as Attean, Kineo, KiWi, librdf.sqlite, NitroBase, Oxigraph, Pointrel, Profium Sense, RedStore, RDF::Trine, TerminusDB~\cite{van2020succinct} and TriplyDB -- for which we could not find key technical details (e.g., indexes or join algorithms supported) at the time of writing.} In the case of distributed RDF stores, we expect similar technical details, along with the type of partitioning and/or replication used. We include systems with associated publications, as well as systems that are unpublished but widely known in practice. Both local and distributed systems are presented in approximate chronological order, based on the year of publication, or an approximate year in which the system was released. For unpublished local stores, we include the year where RDF was first supported. For unpublished distributed stores, we include the approximate year when distributed features were added. Some stores that are often deployed in local environments also support distribution; they are included in both sections. Some systems are unnamed; if it is a distributed store that extends an existing local store, we append the suffix ``-D'' or ``-D2'' to the local store's name; otherwise we use an abbreviation based on authors and year. Where systems change name, we prefer the more modern name. The papers sometimes use different terminology to refer to similar concepts; we often map the original terminology to that used in the body of the survey in order to increase coherency and improve readability.

\subsection{Local RDF Stores}

The local RDF stores we include, and the techniques they use, are summarized in Table~\ref{tab:crcfe}. 

\makeatletter
\newcommand{\wline}{%
    \noalign {\ifnum 0=`}\fi\color{white} \hrule height 1.1pt
    \futurelet \reserved@a \@xhline
}
\newcolumntype{"}{@{\color{white}\hskip\tabcolsep\vrule width 1.1pt\hskip\tabcolsep}}

\begin{table*}[p]
\centering
\captionof{table}{Categorization of local RDF Engines. \\[1ex]
\scriptsize
\textbf{Storage}: \texttt{T} = Triple Table, \texttt{Q} = Quad Table,  \texttt{V} = Vertical Partitioning, \texttt{P} = Property table, \texttt{G} = Graph-based, \texttt{E} = Matrix/Tensor-based, \texttt{M} = Miscellaneous  \\
\textbf{Indexing}: \texttt{T} = Triple, \texttt{Q} = Quad, \texttt{E} = Entity, \texttt{P} = Property, \texttt{N} = Path/Navigational, \texttt{J} = Join, \texttt{S} = Structural, \texttt{M} = Miscellaneous \\
\textbf{Join P.}: \texttt{P} = Pairwise, \texttt{M} = Multiway, \texttt{W} = Worst case optimal, \texttt{L} = Linear algebra \\
\textbf{Query P.}: \texttt{R} = Relational,  \texttt{N} = Paths/Navigational, \texttt{Q} = Query rewriting
}

\scriptsize
\setlength{\tabcolsep}{5pt}
\renewcommand{\arraystretch}{0.93}

\begin{tabular}{ny"a"a"a"a"a"a"a"b"b"b"b"b"b"b"b"d"d"d"d"e"e"e} 

\toprule
\cellcolor{white} & \cellcolor{white} & \multicolumn{7}{c}{\textbf{Storage}} 
& \multicolumn{8}{c}{\textbf{Indexing}}
& \multicolumn{4}{c}{\textbf{Join P.}}
& \multicolumn{3}{c}{\textbf{Query P.}} \\

\multirow{-2}{*}{\cellcolor{white} \textbf{Engine}} & \multirow{-2}{*}{\cellcolor{white} \textbf{Year}} & \cellcolor{white} \texttt{T} & \cellcolor{white}  \texttt{Q} & \cellcolor{white} \texttt{V} & 	\cellcolor{white}  \texttt{P} & \cellcolor{white} \texttt{G} & \cellcolor{white} \texttt{E} & \cellcolor{white} \texttt{M} & 	\cellcolor{white} \texttt{T} & \cellcolor{white}  \texttt{Q} & \cellcolor{white} \texttt{E} & 	\cellcolor{white}  \texttt{P} & 	\cellcolor{white} \texttt{N} & \cellcolor{white} \texttt{J} & \cellcolor{white}  \texttt{S} & 	\cellcolor{white} \texttt{M} & \cellcolor{white}  \texttt{P} & 	\cellcolor{white} \texttt{M}  & \cellcolor{white}  \texttt{W} & \cellcolor{white}  \texttt{L} & 	\cellcolor{white} \texttt{R}  & \cellcolor{white}  \texttt{N} &  \cellcolor{white} \texttt{Q} \\
 \midrule
 
Redland \cite{redland} & 2001 & \cmark &  &  &  &  &  &  & \cmark &  &  &  &  &  &  &  &  &  &  &  &  &  & \\\wline
Jena \cite{1067737} & 2002 & \cmark &  &  &  &  &  &  &  &  &  &  &  &  &  & \cmark & \cmark &  &  &  &  &  & \cmark\\\wline
RDF4J \cite{sesame2002} & 2002 & \cmark &  & \cmark &  &  &  &  &  &  &  &  &  &  &  & \cmark & \cmark &  &  &  &  &  & \cmark\\\wline
RSSDB \cite{KarvounarakisACPS02,KarvounarakisMACPST03} & 2002 & \cmark &  &  &  &  &  &  &  &  &  &  &  &  &  & \cmark & \cmark &  &  &  &  &  & \cmark\\\wline
3store \cite{DBLP:conf/psss/HarrisG03} & 2003 & \cmark &  &  &  &  &  &  &  &  &  &  &  &  &  & \cmark & \cmark &  &  &  &  &  & \cmark\\\wline
AllegroGraph & 2003 &  & \cmark &  &  &  &  &  &  & \cmark &  &  &  &  &  &  & \cmark &  &  &  & \cmark & \cmark & \\\wline
Jena2 \cite{jena2} & 2003 & \cmark &  &  & \cmark &  &  &  &  &  &  &  &  &  &  & \cmark & \cmark & \cmark &  &  &  &  & \cmark\\\wline
CORESE \cite{CorbyDF04,CorbyF07} & 2004 & \cmark &  &  &  &  &  &  & \cmark &  &  &  &  &  &  &  & \cmark &  &  &  & \cmark &  & \\\wline
Jena TDB & 2004 & \cmark & \cmark &  &  &  &  &  & \cmark & \cmark &  &  &  &  &  &  & \cmark &  &  &  & \cmark & \cmark & \\\wline
RStar  \cite{ma2004rstar} & 2004 & \cmark &  &  &  &  &  &  &  &  &  &  &  &  &  & \cmark & \cmark &  &  &  &  &  & \cmark\\\wline
BRAHMS \cite{janik2005brahms} & 2005 & \cmark &  &  &  &  &  &  & \cmark &  &  &  &  &  &  &  &  &  &  &  &  & \cmark & \\\wline
GraphDB \cite{KiryakovOM05,BishopKOPTV11} & 2005 & \cmark & \cmark &  &  &  &  &  & \cmark & \cmark &  &  &  &  &  &  & \cmark &  &  &  & \cmark & \cmark & \\\wline
Mulgara \cite{kowari} & 2005 &  & \cmark &  &  &  &  &  &  & \cmark &  &  &  &  &  &  & \cmark &  &  &  & \cmark & \cmark & \\\wline
RAP \cite{oldakowski2005rap} & 2005 & \cmark & & & & & & & \cmark & & & & & & & \cmark & \cmark & & & & \cmark &  & \cmark \\\wline
RDF\_MATCH \cite{10.5555/1083592.1083734} & 2005 &  & \cmark &  & \cmark &  &  &   &  & \cmark & \cmark &  &  & \cmark &  & \cmark & \cmark & \cmark &  &  & \cmark &  & \\\wline
YARS \cite{Harth:2005:OIS:1114687.1114857} & 2005 &  & \cmark &  &  &  &  &  &  & \cmark &  &  &  &  &  &  & \cmark &  &  &  & \cmark &  & \\\wline
ARC \cite{Nowack06} & 2006 & \cmark & \cmark &  &  &  &  &  & \cmark & \cmark &  &  &  &  &  &  & \cmark &  &  &  & \cmark &  & \cmark \\\wline
RDFBroker \cite{SintekK06} & 2006 &  &  &  & \cmark &  &  &  &  &  &  &  &  &  &  & \cmark & \cmark & \cmark &  &  & \cmark &  & \\\wline
Virtuoso \cite{Erling2010} & 2006 & \cmark & \cmark &  &  &  &  &  &  &  &  &  &  &  &  & \cmark & \cmark &  &  &  & \cmark & \cmark & \\\wline
GRIN \cite{UdreaPS07} & 2007 &  &  &  &  & \cmark &  &  &  &  &  &  &  &  & \cmark &  & \cmark &  &  &  &  &  & \\\wline
SW-Store \cite{swstore} & 2007 &  &  & \cmark &  &  &  &  &  &  &  &  &  &  &  &  &  &  &  &  &  &  & \\\wline
Blazegraph \cite{ThompsonPC14} & 2008 & \cmark & \cmark &  &  &  &  &  & \cmark & \cmark &  &  &  &  &  &  & \cmark & \cmark &  &  & \cmark & \cmark & \\\wline
Hexastore \cite{Weiss2008HexastoreSI} & 2008 &  &  &  &  & \cmark &  &  & \cmark &  &  &  &  &  &  &  & \cmark &  &  &  & \cmark &  & \\\wline
RDF-3X \cite{Neumann2010} & 2008 & \cmark &  &  &  &  &  &  & \cmark &  &  &  &  &  &  &  & \cmark & \cmark &  &  & \cmark &  & \\\wline
BitMat \cite{atrebitmat} & 2009 &  &  &  &  &  & \cmark &  & \cmark &  & \cmark &  &  &  &  &  &  &  &  & \cmark &  &  & \\\wline
DOGMA \cite{dogma2009} & 2009 &  &  &  &  & \cmark &  &  &  &  & \cmark &  &  &  & \cmark &  &  & \cmark &  &  &  &  & \\\wline
LuposDate \cite{GroppeGSL09} & 2009 & \cmark &  &  &  &  &  &  & \cmark &  &  &  &  &  &  &  & \cmark &  &  &  &  &  & \\\wline
Parliament \cite{parliament} & 2009 & \cmark & \cmark &  &  &  &  &  & \cmark & \cmark &  &  &  &  &  &  &  &  &  &  &  &  & \\\wline
RDFJoin & 2009 & \cmark &  &  &  &  &  &  & \cmark &  &  &  &  & \cmark &  &  & \cmark & \cmark &  &  & \cmark &  & \cmark\\\wline
System $\Pi$ \cite{WuLHW09} & 2009 &  &  &  &  & \cmark &  &  & \cmark &  &  &  & \cmark &  &  &  & \cmark &  &  &  & \cmark &  & \\\wline
HPRD \cite{LiuH10} & 2010 & \cmark & \cmark &  &  &  &  &  & \cmark & \cmark &  & \cmark &  &  &  &  & \cmark & \cmark &  &  &  &  & \\\wline
Stardog & 2010 &  & \cmark &  &  &  &  &  &  & \cmark &  &  &  &  &  &  & \cmark &  &  &  & \cmark & \cmark & \\\wline
StrixDB & 2010 & \cmark & & & & & & & \cmark & & & & & & & & & & & & & & \cmark \\\wline
dipLODocus \cite{10.5555/2063016.2063066} & 2011 &  &  &  &  & \cmark &  &  & \cmark &  & \cmark &  &  &  &  &  & \cmark &  &  &  & \cmark &  & \\\wline
gStore \cite{Zou2011} & 2011 &  &  &  &  & \cmark &  &  & \cmark &  & \cmark &  &  &  &  &  &  & \cmark &  &  &  &  & \\\wline
SpiderStore \cite{10.1007/978-3-319-25010-6_1} & 2011 &  &  &  &  & \cmark &  &  &  &  & \cmark &  &  &  &  &  &  &  &  &  &  &  & \\\wline
SAINT-DB \cite{PicalausaLFHV12} & 2012 &  &  &  &  & \cmark &  &  &  &  &  &  &  &  & \cmark &  & \cmark &  &  &  &  &  & \\\wline
Strabon \cite{10.1007/978-3-642-35176-1_19} & 2012 &  &  & \cmark &  &  &  & \cmark & \cmark &  &  &  &  &  &  & \cmark & \cmark &  &  &  & \cmark &  & \\\wline
BrightstarDB & 2013 & \cmark & \cmark & & & & & & \cmark & \cmark & & & & & & & \cmark & & & & \cmark & \cmark & \\\wline
DB2RDF \cite{10.1145/2463676.2463718} & 2013 &  &  &  & \cmark &  &  &  &  &  & \cmark &  &  &  &  & \cmark & \cmark &  &  &  & \cmark &  & \cmark\\\wline
OntoQuad \cite{PotockiPDHKU13} & 2013 &  & \cmark &  &  &  &  &  &  & \cmark &  &  &  &  &  &  & \cmark & \cmark &  &  & \cmark &  & \\\wline
OSQP \cite{TranLR13} & 2013 &  &  &  &  & \cmark &  &  &  &  &  &  &  &  & \cmark &  & \cmark &  &  &  & \cmark &  & \\\wline
Triplebit \cite{10.14778/2536349.2536352} & 2013 &  &  &  &  &  & \cmark &  & \cmark &  &  &  &  &  &  &  & \cmark & \cmark &  &  &  &  & \\\wline
R3F \cite{KimMK14,KimMK15} & 2014 & \cmark &  &  &  & \cmark &  &  & \cmark &  &  & \cmark & \cmark &  &  &  & \cmark &  &  &  &  &  & \\\wline
RQ-RDF-3X \cite{LeekaB14} & 2014 &  & \cmark &  &  &  &  &  &  & \cmark &  &  &  &  &  &  & \cmark &  &  &  & \cmark &  & \\\wline
SQBC \cite{ZhengZLZ0Z14} & 2014 &  &  &  &  & \cmark &  &  &  &  & \cmark &  &  &  &  &  &  & \cmark &  &  &  &  & \\\wline
WaterFowl \cite{CureBRF14} & 2014 & \cmark &  &  &  &  &  &  & \cmark &  &  &  &  &  &  &  & \cmark &  &  &  & \cmark &  & \\\wline
GraSS \cite{LyuWLFW15} & 2015 &  &  &  &  & \cmark &  &  & \cmark &  & \cmark &  &  &  &  &  &  & \cmark &  &  &  &  & \\\wline
$k^2$-triples \cite{Alvarez-GarciaB15} & 2015 &  &  & \cmark &  &  &  &  & \cmark &  &  &  &  &  &  &  & \cmark &  &  &  &  &  & \\\wline
RDFCSA \cite{BrisaboaCFN15,BrisaboaCBFN20} & 2015 & \cmark &  &  &  &  &  &  & \cmark &  &  &  &  &  &  &  & \cmark &  &  &  &  &  & \\\wline
RDFox \cite{10.1007/978-3-319-25010-6_1} & 2015 & \cmark &  &  &  &  &  &  & \cmark &  &  &  &  &  &  &  & \cmark &  &  &  & \cmark & \cmark & \\\wline
Turbo$_{\mathsf{HOM++}}$ \cite{turbo2015} & 2015 &  &  &  &  & \cmark &  &  &  &  & \cmark &  &  &  &  &  &  & \cmark &  &  &  &  & \\\wline
ClioPatria \cite{WielemakerBHO16} & 2016 & & \cmark & & & & & & & \cmark & & & & & & & & & & & \cmark & \cmark & \cmark \\\wline
LevelGraph \cite{MaccioniC16} & 2016 & \cmark & & & & & & & \cmark & & & & & & & & \cmark & \cmark & & & & & \\\wline
RIQ \cite{katib2016riq} & 2016 &  & \cmark &  &  & \cmark &  &  & \cmark &  &  &  &  &  & \cmark &  &  & \cmark &  &  &  &  & \\\wline
axonDB \cite{MeimarisP17} & 2017 & \cmark &  &  &  &  &  &  & \cmark &  &  & \cmark &  &  &  &  & \cmark & \cmark &  &  &  &  & \\\wline
HTStore \cite{LiZRCF17} & 2017 &  &  &  &  & \cmark &  &  &  &  & \cmark &  &  &  &  &  &  &  &  &  &  &  & \\\wline
Ontop \cite{CalvaneseCKKLRR17,XiaoLKKKDCCCB20} & 2017 &  &  &  &  &  &  & \cmark &  &  &  &  &  &  &  & \cmark &  &  &  &  & \cmark &  & \cmark \\\wline
Quadstore & 2017 &  & \cmark &  &  &  &  & &  & \cmark &  &  &  &  &  & \cmark &  \cmark & &  &  & \cmark & \cmark & \\\wline
AMBER \cite{doi:10.1002/9781119528227.ch5} & 2018 &  &  &  &  & \cmark &  &  &  &  &  &  &  &  &  &  &  & \cmark &  &  &  &  & \\\wline
TripleID-Q \cite{8314130} & 2018 & \cmark &  &  &  &  & \cmark &  &  &  &  &  &  &  &  &  &  &  &  & \cmark & \cmark &  & \\\wline
Jena-LTJ \cite{HRRSiswc19} & 2019 & \cmark &  &  &  &  &  &  & \cmark &  &  &  &  &  &  &  &  &  & \cmark &  &  &  & \\\wline
MAGiQ \cite{JamourACK19} & 2019 &  &  &  &  &  & \cmark &  &  &  &  &  &  &  &  &  &  &  &  & \cmark &  &  & \\\wline
BMatrix \cite{BrisaboaCBF20} & 2020 &  &  &  &  &  & \cmark &  & \cmark &  &  &  &  &  &  &  & \cmark &  &  &  &  &  & \\\wline
Tentris \cite{tentris2020} & 2020 &  &  &  &  &  & \cmark &  & \cmark &  &  &  &  &  &  &  &  &  & \cmark & \cmark & \cmark &  & \\\wline
Ring \cite{ArroyueloHNRRS21} & 2021 & \cmark &  &  &  &  &  &  & \cmark &  &  &  &  &  &  &  &  &  & \cmark &  &  &   & \\
\bottomrule
\end{tabular}
\label{tab:crcfe}
\end{table*}

\newcommand{\sys}[3]{\paragraph{#1} \cite{#2} ({#3})}
\newcommand{\sysf}[4]{\paragraph{#1} \footnote{\label{#3}\url{#2}} ({#4})}
\newcommand{\sysn}[2]{\paragraph{#1} {(#2)}}

\sys{Redland}{redland}{2001} is a set of RDF libraries for native RDF storage that has seen various developments over the years. The original paper describes triple-table like storage based on creating three hash maps -- \textsc{sp}$\rightarrow$\textsc{o}, \textsc{po}$\rightarrow$\textsc{s}, \textsc{so}$\rightarrow$\textsc{p} -- which, given two elements of an RDF triple, allow for finding the third element; for example, using \textsc{po}$\rightarrow$\textsc{s}, we can find the subjects of triples with a given predicate and object. The hash maps can be stored either in-memory or on persistent storage. Support for the RDQL and SPARQL query languages were later added with the Rasqal query library.

\sys{Jena}{1067737}{2002} uses relational databases to store RDF graphs as triple tables, with entries for subject, predicate, object IRIs, and object literals. IRIs and literals are encoded as IDs and two separate dictionaries are created for both. Indexing is delegated to an underlying relational DBMS (e.g., Postgresql, MySQL, Oracle, etc.). RDQL is used as a query language and is translated into SQL and run against the underlying relational DBMS. The Jena store would later be extended in various directions, with SDB referring to the use of relational-style storage (per the original system), and TDB referring to the use of native storage.

\sys{RDF4J}{sesame2002}{2002}, known originally as Sesame, provides persistent storage and querying of RDF data. RDF4J provides storage-independent solutions and can be deployed on top of a variety of storage engines such as RDBMSs and object-oriented databases. Graph queries can be expressed with the RQL language. The storage, indexing, and query processing techniques depend on the underlying storage engine used by RDF4J. Recent versions of RDF4J features improved functionalities such as both in-memory and persistent data storage, SeRQL and SPARQL support, etc. 

\sys{RSSDB}{KarvounarakisACPS02,KarvounarakisMACPST03}{2002} stores an RDF graph using a vertical partitioning approach with Postgres as the underlying database. Two variants are considered for class instances: creating a unary table per class (named after the class, with rows indicating instances), or creating one binary table called instances (with rows containing both the instance and the class) in order to reduce the number of tables. Four tables are also added to model RDFS definitions (classes, properties with their domain and range, sub-classes and sub-properties). The system supports queries in RQL (proposed in the same paper~\cite{KarvounarakisMACPST03}), which are translated to SQL by an RQL interpreter and evaluated over Postgres.

\sys{3store}{DBLP:conf/psss/HarrisG03}{2003} uses MySQL as a back-end, sorting RDF graphs in four tables, namely a triple table, a models table, a resource table, and a literal table. The triple table stores RDF triples (one per row) with additional information: (1) the model the triple belongs to, (2) a boolean value to indicate if the object is a literal, and (3) a boolean value to indicate if this triple is inferred. The models, resource, and literal tables are two-column tables that dictionary encode models, resources, and literals, respectively. Queries expressed in RDQL are rewritten to SQL for execution over MySQL.

\sysf{AllegroGraph}{https://franz.com/agraph/allegrograph/}{fn:allegrograph}{2003} is a general purpose store for semi-structured data that can be used to query documents (e.g., JSON) and graph data (e.g., RDF). RDF data is stored and indexed in six permutations as quads, which are additionally associated with a triple identifier. SPARQL queries are supported, where the most recent version provides an option for two query engines: SBQE, which is optimized for SPARQL~1.0-style queries, and MJQE, which features merge joins and caching techniques optimized for property paths.

\sys{Jena2}{jena2}{2003} is a revised version of the original Jena database schema, with support for both triple and property tables. Unlike the original version, IRIs and literals are stored directly in the tables, unless they exceed a certain length, in which case they are dictionary encoded by two separate tables; this allows filter operations to be directly performed on the triple and property tables, thus reducing dictionary lookups, but increasing storage sizes as string values are stored multiple times. Indexing is handled by an underlying relational database, and graph queries in RDQL are rewritten to SQL queries evaluated over the database.

\sys{CORESE}{CorbyDF04,CorbyF07}{2004} began as a search engine with path-finding functionality and inference~\cite{CorbyDF04}, but was extended to support SPARQL query features~\cite{CorbyF07}. CORESE models RDF graphs as conceptual graphs; for simplicity we discuss their methods in terms of the RDF model. RDF graphs are indexed according to the terms, enabling the efficient evaluation of triple patterns. Given a basic graph pattern, the triple patterns are reordered based on heuristics -- such as the number of constants or filters associated with the triple pattern, or the number of variables bound by previous triple patterns in the order -- as well as cardinality estimates. A nested-loop style algorithm is then applied to perform joins. Filters are evaluated as soon as possible to reduce intermediate results.

\sysf{Jena TDB}{https://jena.apache.org/documentation/tdb/}{fn:jenatdb}{2004} is a native RDF store that has seen continuous development in the past decades. A TDB instance consists of three tables: a node table (a dictionary, allowing to encode/decode RDF terms to/from 8-byte identifiers), a triple/quad table (with dictionary-encoded terms), and a prefixes table (used to store common prefixes used for abbreviations). Storage is based on custom B+trees used to build indexes for various triple/quad permutations. Join processing uses pairwise (nested-loop) joins, with a variety of statistic- and heuristic-based methods available for join reordering. SPARQL 1.1 query processing is implemented in the custom Jena~ARQ query processor. Jena TDB has become the recommended RDF store for Jena, with older relational-based storage (later named Jena~SDB) having been deprecated.

\sys{RStar}{ma2004rstar}{2004} stores (RDFS-style) ontology information and instance data using multiple relations in the IBM DB2 RDBMS. Five two-column tables are used to store ontological data (property dictionary, sub-property relations, class dictionary, sub-class relations, and domain and range relations). Another five two-column tables are used to store instance-related data (literal dictionary, IRI dictionary, triples, class instances, namespace dictionary). RStar pushes indexing and other tasks to the underlying database. The RStar Query Language (RSQL) is used and translated into SQL.

\sys{BRAHMS}{janik2005brahms}{2005} is an in-memory RDF store. The RDF graph is indexed in three hash tables -- \textsc{s}$\rightarrow$\textsc{po}, \textsc{o}$\rightarrow$\textsc{sp}, \textsc{p}$\rightarrow$\textsc{so} -- which allow for finding triples that use a particular constant. The motivating use-case of BRAHMS is to find semantic associations -- i.e., paths between two subject/object nodes -- in large RDF graphs. This path-search functionality was implemented in BRAHMS using depth-first search and breadth-first search algorithms. 

\sys{GraphDB}{KiryakovOM05,BishopKOPTV11}{2005} (formerly known as OWLIM) stores RDF graphs using a mix of triple and quad tables. In the most recent version, indexes are built for two triple permutations (\textsc{pos} and \textsc{pso}) as well as a quad permutation (\textsc{gpso}). Predicate lists (\textsc{sp} and \textsc{op}) can also be indexed in order to quickly find the predicates associated with a given subject or object. Terms are dictionary encoded. Joins are reordered according to cardinality estimations. SPARQL~1.1 is supported, along with a wide range of other features, including spatial features, full-text indexing, inference, semantic similarity, integration with MongoDB, and more besides.

\sys{Mulgara}{kowari,muys2006building}{2005}, a fork of an earlier RDF store known as Kowari, implements native RDF storage in the form of quads tables using AVL trees. Dictionary encoding based on 64-bit longs is used. Support for transactions is provided using immutable arrays that store quads on disk in compressed form, with skiplists enabling fast search; insertions and deletions lead to a new immutable array being generated on disk. Indexing is based on six permutations of quads (which is sufficient to efficiently evaluate all sixteen possible quad patterns). Joins are evaluated pairwise and reordered (possibly on-the-fly) based on cardinality estimations. Queries are expressed in the iTQL language, where SPARQL support was added later.

\sys{RAP}{oldakowski2005rap}{2005} is a general-purpose PHP-based API for RDF that includes an RDF store. Two forms of storage are provided. An in-memory store collects triples in an array, with three indexes provided on \textsc{s}, \textsc{p} and \textsc{o} to find triples by a given term and position. Alternatively, persistent storage is supported through database backends, where triples are stored in a triple table in raw form (i.e., as RDF terms). Early versions of RAP supported RDQL queries. To evaluate such queries over the in-memory store, individual triple patterns are first evaluated and their results joined. If persistent storage is rather used, the query is rewritten in full to SQL and evaluated by the underlying database. Later versions would include support for SPARQL.

\sys{RDF\_MATCH}{10.5555/1083592.1083734}{2005} is an RDF store that uses the Oracle RDBMS as an underlying database. It stores RDF data in two different tables: a dictionary table, and a quads table. Indexes are defined based on B-trees. Queries are evaluated as self-joins over the quads table, which is further joined with the dictionary. Materialized views can further be used to index \textsc{s--s}, \textsc{s--p}, \textsc{s--o}, \textsc{p--p}, \textsc{p--o} and \textsc{o--o} joins on-demand, as well as selected property tables. Support for Datalog-style rules is also provided. Queries are expressed in an SQL-style syntax, with functions used to express graph patterns, which in turn are interpreted and evaluated by Oracle.

\sys{YARS}{Harth:2005:OIS:1114687.1114857}{2005} is a native RDF store that indexes quad tables in B+trees using dictionary encoded terms. It uses four types of indexes: dictionary indexes, keyword indexes, quad indexes (with six permutations), and cardinality indexes that count occurrences of quad patterns. YARS implements pairwise (nested-loop) joins that are reordered by cardinality. Basic graph patterns in Notation3 (N3) syntax are supported.

\sys{ARC}{Nowack06}{2006} was first proposed as a general-purpose PHP library for processing RDF, but would be extended in subsequent years with a variety of additional features, including RDF storage and querying. These features are implemented on top of an underlying relational store, with support for persistent storage through MariaDB and MySQL, and in-memory storage with SQLite. Data are stored as triple or quad tables. A fragment of the SPARQL standard is supported, including some selected features from SPARQL~1.1.

\sys{RDFBroker}{SintekK06}{2006} is an RDF store that follows a property table approach. For each subject in the graph, its signature (equivalent to the notion of characteristic sets that would come later) is extracted, with a property table defined for each signature, including a column for the subject, and a column for each property in the signature. Support for RDFS reasoning is also described. An index over signatures is proposed based on a lattice that models set containment between signatures. Given a signature extracted from the query, the lattice can be used to find tables corresponding to signatures that subsume that of the query. A prototype based on in-memory storage is described, implementing typical relational query optimizations such as join reordering.

\sys{Virtuoso}{Erling2010}{2006} stores RDF data as a quad table, where in the most recent version, by default, the quad table includes five indexes: \textsc{psog} (for \textsc{p}, \textsc{ps}, \textsc{pso} and \textsc{psog} prefixes), \textsc{pogs} (for \textsc{po} and \textsc{pog}), \textsc{sp} (for \textsc{s}), \textsc{op} (for \textsc{o}), and \textsc{gs} (for \textsc{g}). The table is sorted by the primary key \textsc{psog}. If only the subject or object are known, the \textsc{sp}/\textsc{op} index can be used to identify predicates, allowing the \textsc{psog}/\textsc{pogs} index to be subsequently used. In the case that only the graph is known, the \textsc{gs} index is joined with \textsc{sp} and then with \textsc{psog}. Subjects, predicates and graph names are dictionary encoded; objects are stored in raw form (for fast filtering). Configurations for row-wise and column-wise storage are provided. Query execution is based on translating SPARQL queries into SQL to be executed on a custom underlying database.

\sys{GRIN}{UdreaPS07}{2007} is an RDF store based on a structural index (see Section~\ref{sec:structure}). This index is a binary tree, where the root refers to all the nodes of the graph, and both children divide the nodes of its parents based on a given distance from a given node in the graph. The leaves can then be seen as forming a partition of the triples in the graph induced by the nodes in its division. The structural index is used to find small subgraphs that may generate results for a query, over which an existing subgraph matching algorithm is applied.

\sys{SW-Store}{AbadiMMH07,swstore}{2007} is an RDF store based on vertical partitioning. SW-Store relies on a column-oriented DBMS called C-store \cite{10.5555/1083592.1083658}, which is shown to pair well with vertical partitioning in terms of performance (e.g., the object column of a \texttt{foaf:age} table will have integers in an interval $[0,150]$, which are highly compressible). Each table is indexed by both subject and object. An ``overflow'' triple table is used for inserts alongside the compressed, vertically partitioned tables. Jena ARQ is used to translate SPARQL queries into SQL for evaluation over C-Store. Pairwise joins are used, preferring merge joins when data are sorted appropriately, otherwise using index nested-loop joins. Materialization of \textsc{s--o} joins is also discussed. 

\sys{Blazegraph}{ThompsonPC14}{2008}, formerly known as \emph{BigData}, is a native RDF store supporting SPARQL~1.1. Blazegraph allows for either indexing triples to store RDF graphs, or quads to store SPARQL datasets. Three index permutations are generated for triples, and six permutations are generated for quads; indexes are based on B+trees. Both row and column data storage models are supported, which can be saved both in-memory or on-disk. Dictionary encoding with 64-bit integers is used for compressed representation of RDF triples. Two query optimization strategies are available: the default approach uses static analysis and cardinality estimation; the second approach uses runtime sampling of join graphs. Supported joins include hash joins, index nested-loop joins, merge joins, and multiway star joins.

\sys{Hexastore}{Weiss2008HexastoreSI}{2008} is an in-memory RDF store based on adjacency lists similar to \autoref{fig:adj}. Six indexes are built for all $3! = 6$ permutations of the elements of a triple. For example, in the \textsc{spo} index, each subject $s$ is associated with an ordered vector of predicates, wherein each $p$ in turn points to an ordered vector of objects. In the \textsc{pso} index, each $p$ points to a vector of subjects, wherein each $s$ points to the same vector of objects as used for $sp$ in the \textsc{spo} index. Terms are dictionary encoded. Having all six index orders allows for (pairwise) merge joins to be used extensively.

\sys{RDF-3X}{NeumannW08,NeumannW09,Neumann2010}{2008} stores RDF graphs as triple tables in compressed indexes, which are based on clustered B+trees whose values are delta-encoded to reduce space. Triples are indexed in all six possible ways. RDF-3X also indexes counts for all triple patterns with one or two (distinct) constants; thus, it can find the exact number of triples matching, for example, ($\vbs$,$\cnp$,$\cno$), ($\cns$,$\vbp$,$\vbo$), etc. Counts are also maintained for frequent path and star joins. Join reordering is based on a dynamic programming procedure, which uses the aforementioned counts as cardinality statistics, and leverages the complete indexing of all six permutations to enable merge joins, further evaluating multiway joins in the case of star patterns. A later extension adds sideways information passing strategies~\cite{NeumannW09} in order to filter intermediate results based on global heuristics.

\sys{BitMat}{atrebitmat,AtreCZH10}{2009} considers a one-hot encoding of an RDF graph, i.e., a 3-dimensional bit array (or matrix/tensor) of dimension $|S| \times |P| \times |O|$, where $S$, $P$, $O$, indicate the set of subjects, predicates and objects of the graph; and where index $s,p,o$ contains a $1$ if the corresponding (dictionary encoded) triple $(s,p,o)$ exists, or $0$ otherwise. The system stores slices of this array (called BitMats), where for each predicate, $\textsc{so}$ and $\textsc{os}$ BitMats are stored; for each subject, a $\textsc{po}$ BitMat is stored; and for each object, a $\textsc{ps}$ BitMat is stored. Each BitMat is a 2-dimensional bit array; e.g., the $\textsc{os}$ matrix of a given predicate $p$ enables finding all objects matching $\vbo$ in $(\vbs,\cnp,\vbo)$ or all subjects matching  $\vbs$ in $(\vbs,\cnp,\cno)$. Though $\textsc{op}$ and $\textsc{sp}$ BitMats could be indexed for subjects and objects, resp., the authors argue they would be rarely used. BitMats also store the count of 1's (triples) they contain, a row vector indicating which columns contain a 1, and a column vector indicating which rows contain a 1. In total, $2|P||S||O|$ BitMats are generated, gap compressed, stored on disk, and loaded in memory as needed. Bitwise AND/OR/NOT operators are used for multiway joins.

\sys{DOGMA}{dogma2009}{2009} is a graph-based RDF store, where an RDF graph is first decomposed into subgraphs using a graph partitioning algorithm. These subgraphs are indexed as the leaves of a balanced binary tree stored on disk. Each non-leaf node in this tree encodes the $k$-merge of its two children, which is a graph with $k$ nodes that is isomorphic to a quotient graph (see Section~\ref{sec:structure}) of both children. DOGMA proposes a variety of algorithms for evaluating basic graph patterns with constant predicates. The basic algorithm generates a set of candidate results for each individual variable node based on its incoming and outgoing edges; starting with the node with the fewest candidates, the algorithm then proceeds to check the edges between them in a depth-first manner (similar to wco joins). Further algorithms prune the sets of candidates based on their distance from the candidates of other nodes in the query based on the distance between nodes in the subgraphs in the leaves of the binary tree.  

\sys{LuposDate}{GroppeGSL09}{2009} stores RDF in a triple table. Seven hash indexes are added for \textsc{s}, \textsc{p}, \textsc{o}, \textsc{sp}, \textsc{so}, \textsc{po}, \textsc{spo}, enabling efficient evaluation of all eight triple patterns. Triples are also annotated with their rank (position in the order) with respect to the six permutations of the triple; for example, the \textsc{ops} rank  indicates the position of the triple when the graph is sorted in \textsc{ops} order. These ranks are used for fast sorting of intermediate results when applying sort-merge joins.

\sys{Parliament}{parliament}{2009} stores RDF graphs in three tables: a resource table encoding details of individual terms, a statement table encoding triples, and a dictionary table. These tables are stored as linked lists. For each RDF term, the resource table stores the first triple containing the term in the statement table; the number of triples that use it in the subject, predicate and object position; a pointer to its entry in the dictionary; and a bitvector encoding metadata about the term. The statement table contains eight components: a statement identifier; three identifiers for the subject, predicate, and object of the triple; three statement identifiers pointing to the next triple with the same subject, predicate, and object; and a bitvector for encoding metadata of the statement. This storage scheme avoids the need for multiple orders and enables fast lookups when triple patterns have one constant; for triple patterns with multiple constants, however, the most selective constant is looked up, with filters run to check the other constants. SPARQL query processing is enabled through existing libraries, such as Jena ARQ.

\sys{RDFJoin}{rdfjoin}{2009} stores RDF graphs using three types of tables. Two dictionary tables are used to encode and decode subjects/objects and predicates. Three triple tables are used, where each has two positions of the triple as the primary key, and the third position is encoded as a bit vector; for example, in the \textsc{po} table, predicate and object are used as a primary key, and for each predicate--object pair, a bit vector of length $|\nodes{G}|$ is given, with a 1 at index $k$ encoding a triple with the subject identifier $k$ for the given predicate and object. Join tables store the results of \textsc{s--s}, \textsc{o--o}, and \textsc{s--o} joins, encoded with the two predicates of both triples as primary key (joins using the same predicate twice are excluded), and a bit vector to encode the join terms for that predicate pair (the subject/object that matches the join variable). MonetDB and LucidDB are used as underlying databases. SPARQL is supported, where joins are evaluated using the join indexes and pairwise algorithms. Inference would later be added in the extended RDFKB system~\cite{rdfkb}. 

\sys{System $\Pi$}{WuLHW09}{2009} is a graph-based RDF store. Nodes of the graph are indexed with an identifier, value and type, while edges are indexed as triples, with their subject, predicate and object identifiers. Nodes are then linked to their inward and outward edges, which enables lookups for triples with a given subject or object. A more compressed version where outward edges only store the predicate and object, or inward edges only store the subject and predicate, is also proposed. Edges are then grouped by their vertex (outward edges by subject, inward edges by object). The compressed variant is thus similar to an adjacency list. Indexes are built for three triple permutations and for reachability queries on a single predicate (PLSD; see Section~\ref{sec:pathind}). SPARQL~(1.0) queries are supported over the proposed indexes using pairwise joins, with the PLSD index used to support entailment over transitive properties.

\sys{HPRD}{LiuH10}{2010} is an RDF store based on three types of index over the dictionary-encoded graph. Triple indexes are based on B+trees and cover three triple permutations: \textsc{spo}, \textsc{po}, \textsc{os}. A path index is built using suffix arrays, and is used only to cache paths that are commonly accessed in queries; the paths are indexed by their predicates. Context indexes are based on B+Trees and are used to support temporal data, versioning, or named graphs; six permutations are covered, namely \textsc{cspo}, \textsc{spo}, \textsc{poc}, \textsc{ocs}, \textsc{cp} and \textsc{so}. Cardinality statistics are further stored for triple patterns, and used for greedy join reordering. RDQL queries supported.

\sysf{Stardog}{https://docs.stardog.com/}{fn:stardog}{2010} is a commercial RDF store built upon the RocksDb key-value store. Stardog indexes quads in various permutations using RocksDB.  Different types of pairwise joins -- such as hash join, bind join, merge join etc. -- are used. Stardog supports SPARQL 1.1, full-text search through Lucene, ACID transactions, versioning, and a variety of other features, including support for property graphs.

\sysf{StrixDB}{http://opoirel.free.fr/strixDB/}{fn:strixdb}{2010} is a native RDF store with transactional support. Storage is built upon the kernel of Gigabase, which is an object-relational embedded database. One write transaction is supported in combination with multiple read transactions. Triples are indexed in B-trees under three permutations: \textsc{spo}, \textsc{pos} and \textsc{osp}. SPARQL queries are translated into executable byte-code. Rules in Datalog or Turtle-like syntax are supported though a variant of SLG resolution.

\sys{dipLODocus}{10.5555/2063016.2063066}{2011} is an RDF store based on the notion of a ``molecule'', which is a subgraph surrounding a particular ``root'' node. The root nodes are defined based on matching triple patterns provided by the administrator. The molecule of a root node is then the subgraph formed by expanding outward in the graph until another root node is encountered. Dictionary encoding is used. Indexes are further built that map nodes and the values of properties indicated by the administrator to individual molecules. SPARQL is supported through the Rasqal query library, with joins pushed within individual molecules where possible; otherwise hash joins are used. Aggregate queries are pushed to the indexes on values of individual properties (which offers benefits similar to column-wise storage).

\sys{gStore}{Zou2011}{2011} is a graph-based RDF store. The RDF graph is stored using adjacency lists (see Section~\ref{sec:gstore}) where each node is associated with a bit vector -- which serves as a vertex signature (see Section~\ref{sec:estore}) -- that encodes the triples where the given node is the subject. gStore then indexes these signatures in a vertex signature tree (VS-tree) that enables multi-way joins. The leaves of the VS-tree encode signatures of nodes, and non-leaf nodes encode the bitwise OR of their children; the leaves are further connected with labeled edges corresponding to edges between their corresponding nodes in the graph. Basic graph patterns can then be encoded in a similar manner to the graph, where gStore then evaluates the pattern by matching its signature with that of the indexed graph. 

\sys{SpiderStore}{10.1007/978-3-319-25010-6_1}{2011} is an in-memory graph store based on adjacency lists. Specifically, for each node in the RDF graph, an adjacency list for incoming and outgoing edges is stored. Likewise, for each predicate, a list of subject nodes is stored. Rather than storing the constants directly in these lists, pointers are stored to the location of the term (with the adjacency lists for the node or the subjects of the predicate). Alongside these pointers, cardinality metadata are stored. (Though SPARQL queries with basic graph patterns and filters are evaluated in the experiments, the types of join algorithms used are not described.)

\sys{SAINT-DB}{PicalausaLFHV12}{2012} is an RDF store with a structural index that organizes triples in the graph according to the type of join that exists between them (\textsc{s}--\textsc{s}, \textsc{p}--\textsc{o}, etc.). The index itself is then a directed edge-labeled graph whose nodes represent a set of triples from the graph, edges indicate that some pair of triples in both nodes are joinable, and edge labels indicate the type of join that exists (which makes the graph directed as \textsc{s}--\textsc{o} differs from \textsc{o}--\textsc{s}). The nodes of the index then form a partition of the graph: no triple appears in more than one node, and their union yields the graph. This index can range from a single node with all triples in the graph (with loops for each type of join present), to singleton nodes each with one triple of the graph. A condition based on semi-joins is used to strike a balance, minimizing the intermediate results generated for individual triple patterns. Given a basic graph pattern, each triple pattern is then mapped to nodes in the structural index, where the triple patterns it joins with must match some triple in a neighbor on an edge whose label corresponds to the type of join.

\sys{Strabon}{10.1007/978-3-642-35176-1_19}{2012} is an RDF store that supports custom features for indexing and querying geospatial data (specifically in the form of stRDF~\cite{10.1007/978-3-642-13486-9_29} data). Strabon is built upon Sesame/RDF4J, which is chosen as an open-source solution that can easily integrate with PostGIS: a DBMS with spatial features. Strabon then stores RDF using a vertical partitioning scheme with dictionary encoding; an identifier for each triple is also included. B+tree indexes are built for the three columns of each table (subject, predicate, identifier). Strabon supports an extension of SPARQL, called stSPARQL \cite{10.1007/978-3-642-13486-9_29}, for querying stRDF based datasets, with spatial features supported through PostGIS. 

\sysf{BrightstarDB}{https://brightstardb.readthedocs.io/_/downloads/en/latest/pdf/}{fn:brightstardb}{2013} is a persistent RDF store that indexes dictionary-encoded RDF datasets using B-trees and/or B+trees. Two types of persistence are supported: in append-only mode, writes are made to pages at the end of the index files, while in rewritable mode, writes are made to copies of index pages that are made active upon a commit. The system further supports querying over multiple named graphs. SPARQL~1.1 queries are processed over BrightstarDB's storage using dotNetRDF's Leviathan library, which supports hash joins and uses a heuristic-based join reordering based on which elements of the triple patterns are constant.

\sys{DB2RDF}{10.1145/2463676.2463718}{2013} 
uses a relational schema similar to property tables to store RDF data. However, rather than having a column for each property/predicate associated with a given subject, DB2RDF uses a ``primary hash'' table with columns \textsc{s}, \textsc{p}$_1$, \textsc{o}$_1$, \ldots, \textsc{p}$_k$, \textsc{o}$_k$, where each \textsc{p}$_i$, \textsc{o}$_i$ pair of columns indicates the $i$\textsuperscript{th} predicate--object pair associated with the subject listed in the \textsc{s} column. A binary ``spill'' column is added, with a $1$ indicating that a subject has more than $k$ triples, in which case it will occupy more than one row of the table. Rows for subjects with fewer than $k$ triples are completed with \textsc{null}s. A second table is used to deal with multi-valued properties, where if a subject $s$ has multiple values for the property $p$ -- say $o_1, \ldots, o_n$ -- then a single fresh value $v$ is used in the primary hash table, which is mapped to $o_1, \ldots, o_n$ (as a binary relation) in the second table. Two orders are indexed: in the second order, subjects and objects are reversed. IBM DB2 is used as the underlying database for experiments, with SPARQL queries being optimized and translated to SQL.
\medskip 

\sys{OntoQuad}{PotockiPDHKU13}{2013} is an RDF store that extends the triple-based representation of Hexastore to additionally support quads. A structure similar to a trie is used, where the top layer is a vector of values for \textsc{s}, \textsc{p}, \textsc{o}, \textsc{g}; the second level encodes $\textsc{sp}, \ldots, \textsc{go}$, etc., with three children for each parent in the top layer (e.g., \textsc{sp}, \textsc{so}, \textsc{sg} for \textsc{s}); the third layer has two children for each parent in the second layer encoding $\textsc{spo}, \ldots, \textsc{gop}$; the fourth layer has one child for each parent in the third layer, completing the quad permutation. B-trees are then used for indexing. Both pairwise and multiway joins are supported using zig-zag joins that seek forward to the maximum compatible join value across the triple patterns. Cardinality estimates and query rewriting rules are used to optimize SPARQL query plans.

\sys{OSQP}{TranLR13}{2013} is an RDF store based on a structural index using various notions of bisimulation, where two nodes in the graph are bisimilar if they cannot be distinguished from their paths. The nodes of the graph are then partitioned into sets that are pairwise bisimilar. The index is then based on a quotient graph, where supernodes correspond to a set of bisimilar nodes in the input graph. In order to reduce index sizes, a parameter corresponding to path lengths is added, such that bisimulation only considers paths within a bounded region of the graph rather than the entire graph.  A basic graph pattern is then matched over the quotient graph (kept in-memory), where the triples corresponding to each matched node are retrieved (from the disk) and used to compute the final results. Custom optimizations are considered for triples with unprojected variables, whose triple patterns can be definitively ``satisfied'' and thus pruned based on the index; and selective triple patterns, which are evaluated directly over the RDF graph.

\sys{Triplebit}{10.14778/2536349.2536352}{2013} represents a dictionary-encoded RDF graph as a compressed 2-dimensional bit matrix. Each column of the matrix represents a triple, and each row represents a subject/object node. The subject and object rows are assigned 1 for the corresponding column of the triple. Columns are sorted by predicate, where a range of columns corresponds to the triples for that predicate. The columns for triples are sparse (at most two 1's) and thus the two identifiers for subjects and objects are used, rather than storing 1's; two orders are maintained for \textsc{so} and \textsc{os} (thus effectively covering \textsc{pso} and \textsc{pos} orders). Two auxiliary indexes are used in TripleBit. Given a subject or object node and a predicate node, the first index (called \textit{ID-Chunk}) supports lookups for finding the range for the unspecified object or subject. Given a subject or object node alone, the second index (called \textit{ID-predicate}) finds predicates associated with that subject or object. Basic graph patterns are evaluated using multiway merge-joins for star joins, with semi-joins used to reduce the number of intermediate results across star joins. Join ordering uses a greedy strategy on selectivity.

\sys{R3F}{KimMK14,KimMK15}{2014} is an extension of RDF-3X with path-based indexes and novel join processing techniques. The first addition is the ``RP-index'', which indexes all nodes with a given incoming path expression up to a certain length; for example, the incoming path expression $pqr$ (of length 3) indexes all nodes $z$ such that there exists $w,x,y$, such that $(w,p,x), (x,q,y), (y,r,z)$ are all triples of the graph. The RP-index is structured as a trie indexing the prefixes of the incoming path expressions, whose leaves are the list of nodes (which are dictionary encoded, sorted and delta encoded). Virtual inverse predicates are added to the RDF graph to support paths in both directions. The second extension is a modification to the sideways information passing strategy of RDF-3X to incorporate information about paths for filtering additional intermediate results.

\sys{RQ-RDF-3X}{LeekaB14}{2014} is an extension of RDF-3X towards support for quads. The extension follows the same principles and techniques for RDF-3X, but the extension to quads requires covering additional permutations. Indexes are built for all $4! = 24$ quad permutations, similar to how RDF-3X indexes all $3! = 6$ triple permutations; having all permutations enables reading the results for any variable of any triple pattern in sorted order, which in turn enables merge joins. The delta encoding used by RDF-3X is extended to the fourth element. Like in RDF-3X, counts are indexed for all quad patterns with 1, 2, or 3 constants, requiring 4, 12 and 24 indexes, respectively (40 in total). Join and query processing use RDF-3X's techniques. RQ-RDF-3X then offers optimized support for reification using named graphs/triple identifiers.

\sys{SQBC}{ZhengZLZ0Z14}{2014} is a graph store -- with support for RDF graphs -- inspired by existing subgraph matching techniques for efficiently finding subgraph isomorphisms.\footnote{The evaluation of basic graph patterns in SPARQL is defined in terms of homomorphisms, rather than subgraph isomorphisms as supported by SQBC, with the difference being that two or more variables in a basic graph pattern can match one node in the RDF graph.} In order to index the graph, codes are extracted for each node that capture structural information about it, including its label, the largest clique containing it, the degrees of its neighbours, etc. Given a basic graph pattern, candidates are identified and filtered for variable nodes. If the basic graph pattern has no cliques, degree information is used; otherwise clique sizes can be used to filter candidate matches.

\sys{WaterFowl}{CureBRF14}{2014} is a compact RDF store based on succinct data structures. The RDF graph is dictionary encoded and sorted in \textsc{spo} order, and represented as a trie: the first layer denotes subjects, connected to their predicates in a second layer, connected to their objects in the third layer. This trie structure is encoded in a compact representation using a combination of bit strings that indicate the number of children for a parent (e.g., for predicates, $\texttt{100101}\ldots$ tells us that the first subject has three children (unique predicates) and the second has two); and wavelet trees that encode the sequence of terms themselves (e.g., the sequence of predicates). Pairwise joins are evaluated in terms of left-deep plans, with further support for SPARQL~(1.0) features. RDFS inference is also supported.

\sys{GraSS}{LyuWLFW15}{2015} is an RDF store that is based on decomposing basic graph patterns into subgraph patterns forming star joins (considering \textsc{s--s}, \textsc{s--o}, or \textsc{o--o} joins). An ``FFD-index'' for star joins is proposed, where for each node, a bit-string signature is computed that encodes its incoming and outgoing edges, i.e., the triples in which it appears as subject or object. A neighbourhood table is constructed: each row denotes a node, which is associated with its signature and edges. Five triple permutations are further indexed (covering \textsc{sp*}, \textsc{op*}, \textsc{s*}, \textsc{p*}, \textsc{o*}), where in the \textsc{sp*} permutation, for example, $(s,p)$ pairs are mapped to a list of objects and their degrees. A basic graph pattern is then decomposed into sub-patterns forming star joins, which are evaluated using the available indexes.

\sys{$k^2$-triples}{Alvarez-GarciaB15}{2015} is a compact in-memory RDF store based on $k^2$ trees. The RDF graph is first dictionary encoded.  For each predicate, a $k^2$ tree is used to index its subjects and objects. In order to support variable predicates in triple patterns, \textsc{sp} and \textsc{op} indexes are used to map subjects and objects, respectively, to their associated predicates; these indexes are encoded using compressed predicate lists. For processing basic graph patterns, \textsc{s}--\textsc{s}, \textsc{s}--\textsc{o}, \textsc{o}--\textsc{s} and \textsc{o}--\textsc{o} index nested-loop joins and merge joins are supported. A sideways information passing optimization is supported where two $k^2$ trees involved in a join can be descended in a coordinated fashion to filter intermediate results.

\sys{RDFCSA}{BrisaboaCFN15,BrisaboaCBFN20}{2015} is a compact in-memory RDF store based on text indexes. Specifically, triples of the RDF graph are dictionary encoded and considered to be strings of length $3$. The graph is thus sorted and encoded as a string of length $3n$, where $n$ is the number of triples. This string is indexed in a compressed suffix array (CSA): a compact data structure commonly used for indexing text. The CSA is modified by shifting elements so that instead of indexing a string of $3n$ elements, triples cycle back on themselves, giving $n$ circular strings of length 3. Thus in an \textsc{spo} permutation, after reading the object of a triple, the next integer will refer to the subject of the same triple rather than the next one in the order. With cyclical strings, one triple permutation is sufficient to support all triple patterns; \textsc{spo} is in fact equivalent to \textsc{pos} and \textsc{osp}. Merge joins, sort--merge joins and a variant of index nested-loop joins (called ``chain joins'') are supported. 

\sys{RDFox}{10.1007/978-3-319-25010-6_1}{2015} is an in-memory RDF engine that supports Datalog reasoning. The RDF graph is stored as a triple table implemented as a linked list, which stores identifiers for subject, predicate and object, as well as three pointers in the list to the next triple with the same subject, predicate and object (similar to Parliament~\cite{parliament}). Four indexes are built: a hash table for three constants, and three for individual constants; the indexes for individual constants offer pointers to the first triple in the list with that constant, where patterns with two constants can be implemented by filtering over this list, or (optionally) by using various orderings of the triple list to avoid filtering (e.g., a triple list ordered by \textsc{spo} can be used to evaluate patterns with constant subject and predicate without filtering). These in-memory indexes support efficient parallel updates, which are key for fast materialization. According to the implementation, (index) nested-loop joins are supported; optionally join plans can be generated based on tree decompositions. SPARQL~1.1 is further supported over the engine.

\sys{Turbo$_{\mathsf{HOM++}}$}{turbo2015}{2015} is an in-memory, graph-based RDF store. The RDF graph is stored as the combination of adjacency lists for incoming and outgoing triples (see Section~\ref{sec:gstore}), and an index that allows for finding nodes of a particular type (based on \texttt{rdf:type}). Evaluation of basic graph patterns is then conducted by generating candidates for an initial node of the query graph based on local information (intersecting adjacency lists and type information in order to match all triple patterns that the node appears in), where the neighbors of the candidates are explored recursively in the graph guided by the graph pattern, generating candidates for further query nodes (in a manner akin to DOGMA~\cite{dogma2009}). A number of optimizations are included, including a multiway join that can check if a new candidate is connected to the candidates of multiple query nodes in one operation in a way that satisfies the query.

\sys{ClioPatria}{WielemakerBHO16}{2016} is an RDF store based on SWI-Prolog. RDF quads are stored in SWI-Prolog's main memory store. Nine custom hash-based indexes are defined using \textsc{s}, \textsc{p}, \textsc{o}, \textsc{sp}, \textsc{po},  \textsc{spo},  \textsc{g},  \textsc{sg} and \textsc{pg} as keys.  Persistence is enabled though on-disk journal files. SPARQL queries are supported through rewriting to Prolog, and evaluated using SLD resolution. Inference can also be supported through translation of rules to Prolog.

\sys{LevelGraph}{MaccioniC16}{2016} is an RDF store that can be used client-side (in-the-browser) with, for example, Node.js applications. LevelDB is used for underlying storage. Triples are stored and indexed in all six permutations -- namely \textsc{spo}, \textsc{sop}, \textsc{pso}, \textsc{pos}, \textsc{osp} and \textsc{ops} -- which are stored in a sorted key--value store. LevelGraph also provides adapters to work with external NoSQL stores, such as DynamoDB, Redis, MongoDB, and MySQL. A query optimizer supports merge joins (on star-shared patterns) and nested loop joins, which are used to evaluate basic graph graphs.

\sys{RIQ}{katib2016riq}{2016} provides a layer on top of an existing RDF store that indexes similar named graphs in a SPARQL dataset. A bit vector -- called a ``pattern vector'' -- is computed for each named graph in the dataset. The pattern vector consists of seven vectors for \textsc{s}, \textsc{p}, \textsc{o}, \textsc{sp}, \textsc{so}, \textsc{po} and \textsc{spo}, where, e.g., the \textsc{sp} vector hashes all subject--predicate pairs in the named graph. An index over the pattern vectors (PV-index) is constructed by connecting similar pattern vectors (based on locality-sensitive hashing) into a graph; each connected component of the graph forms a group of similar graphs. The union of the graphs in each group is further encoded into Bloom filters. In order to evaluate a basic graph pattern, a pattern vector is computed combining the triple patterns (e.g., a triple pattern $(\cns,\cnp,\vbo)$ will generate a single \textsc{sp} sub-vector). The PV-index is then used to optimize an input query by narrowing down the candidate (named) graphs that match particular basic graph patterns before evaluating the optimized query over the underlying SPARQL store.

\sys{axonDB}{MeimarisP17}{2017} uses two dictionary-encoded triple tables to store RDF graphs. In the first table, each triple is additionally associated with the characteristic set (CS) of its subject (see Section~\ref{sec:propind}). The CS is assigned a unique identifier and one-hot encoded, i.e., represented by a bit vector with an index for each property that carries a 1 if the property is part of the CS, or a 0 otherwise. Triples are then sorted by their CS, grouping subjects with the same CS together. A second triple table stores each triple, along with the corresponding extended characteristic set (ECS; again see Section~\ref{sec:propind}). The ECS is encoded with a unique identifier, and the identifiers for the subject and object CSs. The triple table is sorted by ECS. When evaluating a basic graph pattern, its analogous CSs and ECSs are extracted, along with the paths that connect them. The CSs and ECSs are matched with those of the graph, enabling multiway joins; binary hash joins are used to join the results of multiple CSs/ECSs.

\sys{HTStore}{LiZRCF17}{2017} uses hash-based indexes to build an RDF store. The RDF graph is indexed in a hash tree whose top layer forms a hash table over the nodes of the graph. The hash tree is based on a sequence of prime numbers. When hashing a node, the first prime number is used, and if no collision is detected, the node is inserted in the first layer. Otherwise the second prime number is used, and if no collision is detected, it is inserted in that layer as a child of the bucket of the first layer that caused the collision. Otherwise the third prime number is used, and so forth. Nodes in the hash tree then point to their adjacency lists in the graph. To evaluate queries, constant nodes in the query are hashed in the same manner in order to retrieve the data for the node. SPARQL queries are supported, though details about join and query processing are omitted.

\sys{Ontop}{CalvaneseCKKLRR17}{2017} is a open source Ontology-Based Data Access (OBDA) system based on relational (and potentially decentralised) storage. The underlying data are mapped to RDF graphs and/or ontologies using languages such as the R2RML standard. SPARQL queries are rewritten to SQL queries following such mappings, which are evaluated over the underlying database; a more recent version rather translates SPARQL into an intermediate algebraic query that is subsequently optimised and translated into SQL~\cite{XiaoLKKKDCCCB20}. Entailment for RDFS and OWL~2~QL are additionally supported through query rewriting techniques that expand the given query to capture solutions over entailments. 

\sysf{Quadstore}{https://github.com/beautifulinteractions/node-quadstore}{fn:quadstore}{2017} is a client-side RDF store that can be used with Node.js for in-browser management of RDF quads. The system also supports a variety of underlying storage options through the Level-down interface, such as LevelDB and RocksDB for persistent storage, and MemDown for in-memory storage. By default, indexes are generated for six quad permutations, namely \textsc{spog}, \textsc{ogsp}, \textsc{gspo}, \textsc{ospg}, \textsc{pogs} and \textsc{gpos}, though these indexes are configurable by the user. SPARQL~1.1 queries and updates are supported.

\sys{AMBER}{doi:10.1002/9781119528227.ch5}{2018} stores RDF graphs in a ``multigraph'' representation, where IRIs form nodes, whereas predicate--literal pairs form ``attributes'' on nodes. All nodes, predicates and attributes are dictionary encoded. AMBER then generates three indexes: the first stores the set of nodes for each attribute, the second stores vertex signatures that encode meta-data about the triples where a given node is subject or object, and the third stores adjacency lists. Basic graph patterns are evaluated by classifying query nodes with degree greater than one as core nodes, and other nodes as satellite nodes. Core nodes are processed first, where candidates are produced for each query node based on the available indexes, recursively producing candidates for neighbors; the algorithm starts with the core query node with the most satellite nodes attached, or the highest degree. For each solution over the code nodes, each satellite node is then evaluated separately as they become disconnected once the core nodes are bound to constants.

\sys{TripleID-Q}{8314130}{2018} is an RDF store that uses a compact representation called TripleID for RDF graphs such that query processing can be conducted on GPUs. The TripleID representation is based on a dictionary-encoded triple table. Rather than indexing the triple table, chunks of the table can be loaded into GPUs, which, given a particular triple pattern, will scan the triple table in parallel looking for matching triples in the RDF graph. Other operators such as union, join, filter, distinct, etc., are then implemented on top of this GPU search; specifically, these operators are translated into functions that are executed in the GPU over the results of the search. RDFS entailment is further supported.

\sys{Jena-LTJ}{HRRSiswc19}{2019} extends the Jena TDB RDF store with the ability to perform worst-case optimal (wco) joins (see Section~\ref{sec:wcojoins}). Specifically, Jena TDB is extended with an algorithm similar to Leapfrog TrieJoin (LTJ), which is adapted from a relational setting for the RDF/SPARQL settings. The algorithm evaluates basic graph patterns variable-by-variable in a manner that ensures that the overall cost of enumerating all of the results is proportional to the number of results that it can return in the worst case. In order to reach wco guarantees, the three-order index of Jena TDB -- based on B+trees -- is extended to include all six orders. This ensures that for any triple pattern, the results for any individual variable can be read in sorted order directly from the index, which in turn enables efficient intersection of the results for individual variables across triple patterns. Thus Jena-LTJ uses twice the space of Jena TDB, but offers better query performance, particularly for basic graph patterns with cycles.

\sys{MAGiQ}{JamourACK19}{2019} is an RDF store that can use a variety of compressed sparse matrix/tensor representations for RDF graphs in order to translate basic graph patterns into linear algebra operations. These representations include compressed sparse column, doubly compressed sparse column, and coordinate list encodings of the graph as a matrix/tensor. Basic graph patterns are then translated into operations such as matrix multiplication, scalar multiplication, transposition, etc., over the associated matrices/tensor, which can be expressed in the languages provided by libraries such as GraphBLAS, Matlab, CombBLAS, and ultimately evaluated on CPUs and GPUs for hardware acceleration.

\sys{BMatrix}{BrisaboaCBF20}{2020} is a compact in-memory RDF store, where the RDF graph is first dictionary encoded and sorted by predicate. Two binary matrices are created: an $s \times n$ matrix called \textsc{st} and an $o \times n$ matrix called \textsc{ot}, where $s$, $o$ and $n$ are the number of unique subjects, objects and triples respectively. The \textsc{st}/\textsc{ot} matrix contains a $1$ in index $i,j$ if the subject/object of the $j$\textsuperscript{th} triple corresponds to the $i$\textsuperscript{th} term (or a $0$ otherwise). Both matrices are indexed with $k^2$-trees, while a bit string of length $n$ encodes predicate boundaries with a $1$, i.e., in which columns of the matrix (denoting triples sorted or grouped by predicate) the predicate changes. These indexes are sufficient to cover all eight possible triple patterns. Further compression can be applied to the leaf matrices of the $k^2$.tree in order to trade space for time. The authors mention that joins can be supported in a similar fashion as used for RDFCSA and $k^2$-triples.

\sys{Tentris}{tentris2020}{2020} is an in-memory RDF store wherein an RDF graph is viewed as a one-hot encoded 3-order tensor (equivalent to the 3-dimensional array used in BitMat~\cite{AtreCZH10}), which in turn is viewed as a trie of three levels for \textsc{s}, \textsc{p} and \textsc{o}. However, rather than storing tries for all permutations, a \textit{hypertrie} is used with three levels. The leaves in the third level correspond to all possible combinations of two constants in a triple: for each triple $(s,p,o)$, there exists a leaf that maps $(s,p,?)$ to the set of all objects that replace $?$ in the graph (including $o$), with analogous leaves for $(?,p,o)$ and $(s,?,o)$. The second level refers to single constants, where three nodes are present for $(s,?,?)$, $(?,p,?)$, $(?,?,o)$ such that $(s,?,?)$ will map to the leaves for $(s,p,?)$ and $(s,?,o)$, and likewise for $(?,p,?)$ and $(?,?,o)$. Finally, the top level -- the root, representing zero constants -- maps to all the second level elements. Basic graph patterns (with projection) are translated into tensor operations that can be evaluated on the hypertrie using a worst-case optimal join algorithm.

\sys{Ring}{ArroyueloHNRRS21}{2021} is an in-memory RDF store that uses FM-indexes (a text-indexing technique) in order to represent and index RDF graphs in a structure called a ``ring''. Specifically, a dictionary-encoded RDF graph is sorted lexicographically by subject-predicate-object; then the triples are concatenated to form a string $s_1p_1o_1\ldots s_np_no_n$, where $(s_i,p_i,o_1)$ indicates the $i$\textsuperscript{th} (dictionary-encoded) triple in the order and $n = |G|$. A variant of a Burrows--Wheeler Transform is applied over this string, which allows for finding triples given any constant and position (or sequence of constants and positions), and for traversing to other elements of a triple in any direction. The result is a bidirectional circular index that covers all triple permutations with one index that encodes the graph and requires sub-linear space additional to the graph. For basic graph pattern queries, a variant of Leapfrog-Trie Join is implemented, offering worst-case optimal joins.


\subsection{Distributed RDF Engines}

\begin{table*}[p]
\centering
\captionof{table}{Categorization of distributed RDF Engines. \\[1ex]
\scriptsize
\textbf{Storage}: \texttt{T} = Triple Table, \texttt{Q} = Quad Table,  \texttt{V} = Vertical Partitioning, \texttt{P} = Property table, \texttt{G} = Graph-based, \texttt{E} = Matrix/Tensor-based, \texttt{M} = Miscellaneous  \\
\textbf{Indexing}: \texttt{T} = Triple, \texttt{Q} = Quad, \texttt{E} = Entity, \texttt{P} = Property, \texttt{N} = Path/Navigational, \texttt{J} = Join, \texttt{S} = Structural, \texttt{M} = Miscellaneous \\
\textbf{Join P.}: \texttt{P} = Pairwise, \texttt{M} = Multiway, \texttt{W} = Worst case optimal, \texttt{L} = Linear algebra \\
\textbf{Query P.}:  \texttt{R} = Relational, \texttt{N} = Paths/Navigational, \texttt{Q} = Query rewriting\\
\textbf{Partitioning}: \texttt{S} = Statement (Triple/Quad)-based, \texttt{G} = Graph-based, \texttt{Q} = Query-based, \texttt{R} = Replication
}

\scriptsize
\setlength{\tabcolsep}{3.2pt}
\renewcommand{\arraystretch}{0.9}

\begin{tabular}{ny"a"a"a"a"a"a"a"b"b"b"b"b"b"b"b"d"d"d"d"e"e"e"g"g"g"g"f} \toprule
\cellcolor{white} & \cellcolor{white} & \multicolumn{7}{c}{\textbf{Storage}} 
& \multicolumn{8}{c}{\textbf{Indexing}}
& \multicolumn{4}{c}{\textbf{Join P.}}
& \multicolumn{3}{c}{\textbf{Query P.}}
& \multicolumn{5}{c}{\textbf{Partitioning}}\\

\multirow{-2}{*}{\cellcolor{white}\textbf{Engine}} & \multirow{-2}{*}{\cellcolor{white}\textbf{Year}} & \cellcolor{white} \texttt{T} & \cellcolor{white}  \texttt{Q} & \cellcolor{white} \texttt{V} & \cellcolor{white}  \texttt{P} & \cellcolor{white} \texttt{G} & \cellcolor{white} \texttt{E} & \cellcolor{white} \texttt{M} & 	\cellcolor{white} \texttt{T} & \cellcolor{white}  \texttt{Q} & \cellcolor{white} \texttt{E} & 	\cellcolor{white}  \texttt{P} & 	\cellcolor{white} \texttt{N} & \cellcolor{white} \texttt{J} & \cellcolor{white}  \texttt{S} & \cellcolor{white} \texttt{M} & \cellcolor{white}  \texttt{P} & 	\cellcolor{white} \texttt{M}  & \cellcolor{white}  \texttt{W} & \cellcolor{white}  \texttt{L} & 	\cellcolor{white} \texttt{R}  & \cellcolor{white}  \texttt{N} &  \cellcolor{white} \texttt{Q} & \cellcolor{white} \texttt{S} & \cellcolor{white} \texttt{G} & \cellcolor{white} \texttt{Q} & \cellcolor{white} \texttt{R} &  \multicolumn{1}{c}{\cellcolor{white} Store~~} \\
\midrule
YARS2 \cite{10.5555/1785162.1785179}  & 2007 &   & \cmark  &   &   &   &  &   &   & \cmark  &   &   &   &   &   &   & \cmark  &   &  &   & \cmark  &   &   & \cmark  &   &   &   & Custom\\\wline
Clustered TDB \cite{OwensSG}  & 2008 & \cmark  &   &   &   &   &  &   & \cmark  &   &   &   &   &   &   &   & \cmark  &   &  &   &   &   &   & \cmark  &   &   &   & \textit{Jena TDB}\\\wline
Virtuoso EE \cite{Erling2010}  & 2008 & \cmark  & \cmark  &   &   &   &  &   &   & \cmark  &   &   &   &   &   & \cmark  & \cmark  &   &  &   & \cmark  & \cmark  & \cmark  & \cmark  &   &   & \cmark  & Custom\\\wline
4store \cite{harris4store}  & 2009 &   & \cmark  &   &   &   &  &   &   & \cmark  &   &   &   &   &   &   & \cmark  &   &  &   & \cmark  &   &   & \cmark  &   &   &   & Custom\\\wline
Blazegraph  & 2009 & \cmark  & \cmark  &   &   &   &  &   & \cmark  & \cmark  &   &   &   &   &   &   & \cmark  & \cmark  &  &   & \cmark  & \cmark  &   & \cmark  &   &   & \cmark  & Custom\\\wline
SHARD \cite{10.1145/1940747.1940751}  & 2009 &   &   &   &   & \cmark  &  &   &   &   &   &   &   &   &   &   & \cmark  & \cmark  &  &   & \cmark  &   &   & \cmark  &   &   &   & HDFS\\\wline
Allegrograph  & 2010 &   & \cmark  &   &   &   &  &   &   & \cmark  &   &   &   &   &   &   & \cmark  &   &  &   & \cmark  & \cmark  &   & \cmark  &   &   & \cmark  & Custom\\\wline
GraphDB \cite{KiryakovOM05,BishopKOPTV11}  & 2010 & \cmark  & \cmark  &   &   &   &  &   & \cmark  & \cmark  &   &   &   &   &   &   & \cmark  &   &  &   & \cmark  & \cmark  &   &   &   &   & \cmark  & Custom\\\wline
AnzoGraph  & 2011 &   & \cmark  &   &   &   &  &   &   &   &   &   &   &   &   &   & \cmark  &   &  &   & \cmark  & \cmark  &   & \cmark  &   &   &   & Custom\\\wline
CumulusRDF \cite{Ladwig_cumulusrdf:linked}  & 2011 & \cmark  & \cmark  &   &   &   &  &   & \cmark  & \cmark  &   &   &   &   &   & \cmark  & \cmark  &   &  &   & \cmark  &   &   & \cmark  &   &   &   & Cassandra\\\wline
H-RDF-3X \cite{huang2011scalable}  & 2011 & \cmark  &   &   &   &   &  &   & \cmark  &   &   &   &   &   &   &   & \cmark  & \cmark  &  &   & \cmark  &   &   & \cmark  &   &   & \cmark  & \textit{RDF-3X}\\\wline
PigSPARQL \cite{Schtzle2011PigSPARQLMS}  & 2011 & \cmark  &   &   &   &   &  &   &   &   &   &   &   &   &   &   & \cmark  & \cmark  &  &   & \cmark  &   & \cmark  & \cmark  &   &   &   & HDFS\\\wline
Rapid+ \cite{10.1007/978-3-642-21064-8_4}  & 2011 &   &   & \cmark  &   &   &  &   &   &   &   &   &   &   &   &   & \cmark  & \cmark  &  &   & \cmark  &   & \cmark  & \cmark  &   &   &   & HDFS\\\wline
AMADA \cite{10.1145/2396761.2398749}  & 2012 & \cmark  &   &   &   &   &  &   &   &   &   &   &   &   &   & \cmark  & \cmark  &   &  &   & \cmark  &   &   & \cmark  &   &   &   & SimpleDB\\\wline
H2RDF(+) \cite{10.1145/2187980.2188058}  & 2012 & \cmark  &   &   &   &   &  &   & \cmark  &   &   &   &   &   &   & \cmark  & \cmark  & \cmark  &  &   &   &   &   & \cmark  &   &   &   & HBase\\\wline
Jena-HBase \cite{10.5555/2887379.2887401}  & 2012 & \cmark  &   & \cmark  &   &   &  &   & \cmark  &   &   &   &   &   &   & \cmark  & \cmark  &   &  &   & \cmark  &   &   & \cmark  &   &   &   & HBase\\\wline
Rya \cite{10.1145/2347673.2347677}  & 2012 & \cmark  &   &   &   &   &  &   & \cmark  &   &   &   &   &   &   & \cmark  & \cmark  &   &  &   & \cmark  &   &   & \cmark  &   &   &   & Accumulo\\\wline
Sedge \cite{YangYZK12}  & 2012 &   &   &   &   & \cmark  &  &   &   &   &   &   &   &   &   &   & \cmark  &   &  &   &   &   &   &   & \cmark  & \cmark  & \cmark  & Pregel\\\wline
chameleon-db \cite{Alu2013chameleondbAW}  & 2013 &   &   &   &   & \cmark  &  &   & \cmark  &   &   &   &   & \cmark  &   &   & \cmark  &   &  &   & \cmark  &   &   &   & \cmark  & \cmark  &   & Custom\\\wline
D-SPARQ \cite{10.5555/2874399.2874465}  & 2013 & \cmark  &   &   &   &   &  &   & \cmark  &   &   &   &   &   &   & \cmark  & \cmark  & \cmark  &  &   &   &   &   & \cmark  &   &   & \cmark  & MongoDB\\\wline
EAGRE \cite{6544856}  & 2013 & \cmark  &   &   &   & \cmark  &  &   & \cmark  &   & \cmark  &   &   &   &   &   & \cmark  & \cmark  &  &   & \cmark  &   &   &   & \cmark  &   &   & HDFS\\\wline
MR-RDF \cite{DuBCD13}  & 2013 & \cmark  &   &   &   &   &  &   & \cmark  &   &   &   &   &   &   &   & \cmark  & \cmark  &  &   & \cmark  &   &   &   & \cmark  &   &   & \textit{RDF-3X}\\\wline
SHAPE \cite{Lee:2013:SQO:2556549.2556571}  & 2013 & \cmark  &   &   &   &   &  &   & \cmark  &   &   &   &   &   &   &   & \cmark  & \cmark  &  &   &   &   &   & \cmark  &   &   &   & \textit{RDF-3X}\\\wline
Trinity.RDF \cite{Zeng:2013:DGE:2488329.2488333}  & 2013 & \cmark  &   &   &   &   &  &   & \cmark  &   & \cmark  &   &   &   &   &   & \cmark  & \cmark  &  &   & \cmark  &   &   &   & \cmark  &   &   & Trinity\\\wline
TripleRush \cite{10.5555/2874551.2874555}  & 2013 & \cmark  &   &   &   &   &  &   & \cmark  &   &   &   &   &   &   &   & \cmark  &   &  &   &   &   &   & \cmark  &   &   &   & Signal/Collect\\\wline
WARP \cite{6547414}  & 2013 & \cmark  &   &   &   &   &  &   & \cmark  &   &   &   &   &   &   &   & \cmark  & \cmark  &  &   &   &   &   &   &   & \cmark  & \cmark  & \textit{RDF-3X}\\\wline
AKZ14 \cite{AbiriKZ14}  & 2014 & \cmark  &   &   & \cmark  &   &  &   & \cmark  &   &   &   &   &   &   & \cmark  & \cmark  & \cmark  &  &   & \cmark  &   & \cmark  & \cmark  & \cmark  &   &   & HBase\\\wline
Partout \cite{Galarraga:2014:PDE:2567948.2577302}  & 2014 & \cmark  &   &   &   &   &  &   & \cmark  &   &   &   &   &   &   &   & \cmark  &   &  &   & \cmark  &   &   &   &   & \cmark  &   & \textit{RDF-3X}\\\wline
P-LUPOSDATE \cite{GroppeKWHSG14}  & 2014 & \cmark  &   &   &   &   &  &   & \cmark  &   &   &   &   &   &   & \cmark  & \cmark  &   &  &   & \cmark  &   & \cmark  & \cmark  &   &   &   & HBase\\\wline
RDF-3X-MPI \cite{chirravuri2014rdf3x} & 2014 & \cmark & & & & & & & \cmark & & & & & & & & \cmark & & & & & & & \cmark & \cmark & & \cmark & \textit{RDF-3X}\\\wline
Sempala \cite{Schtzle2014SempalaIS}  & 2014 &   &   &   & \cmark  &   &  &   &   &   &   &   &   &   &   & \cmark  & \cmark  & \cmark  &  &   & \cmark  &   &   & \cmark  &   &   &   & HDFS\\\wline
SemStore \cite{10.1145/2661829.2661876}  & 2014 &   &   & \cmark  &   & \cmark  &  &   & \cmark  &   &   &   &   &   &   &   & \cmark  & \cmark  &  &   &   &   &   &   & \cmark  &   &   & \textit{TripleBit}\\\wline
SparkRDF \cite{10.5555/2878453.2878519}  & 2014 &   &   & \cmark  &   &   &  &   & \cmark  &   &   &   &   & \cmark  &   &   & \cmark  &   &  &   & \cmark  &   &   & \cmark  &   &   &   & HDFS/Spark\\\wline
TriAD \cite{Gurajada:2014:TDS:2588555.2610511}  & 2014 & \cmark  &   &   &   &   &  &   & \cmark  &   &   &   &   &   & \cmark  &   & \cmark  & \cmark  &  &   &   &   &   &   & \cmark  &   &   & Custom\\\wline
CK15 \cite{ChengK15}  & 2015 &   &   & \cmark  &   &   &  &   & \cmark  &   &   &   &   & \cmark  &   &   & \cmark  &   &  &   &   &   &   & \cmark  &   & \cmark  &   & Custom\\\wline
CliqueSquare \cite{7113332}  & 2015 & \cmark  &   &   &   &   &  &   &   &   &   &   &   &   &   &   & \cmark  & \cmark  &  &   &   &   &   & \cmark  &   &   & \cmark  & HDFS\\\wline
DREAM \cite{Hammoud:2015:DDR:2735703.2735705}  & 2015 & \cmark  &   &   &   &   &  &   & \cmark  &   &   &   &   &   &   &   & \cmark  &   &  &   & \cmark  &   &   &   &   &   & \cmark  & \textit{RDF-3X}\\\wline
AdPart \cite{Harbi:2016:ASQ:2944220.2944335}  & 2016 & \cmark  &   &   &   &   &  &   & \cmark  &   &   &   &   &   &   &   & \cmark  &   &  &   &   &   &   & \cmark  &   &   & \cmark  & Custom\\\wline
DiploCloud \cite{7323867}  & 2016 &   &   &   &   & \cmark  &  &   & \cmark  &   & \cmark  &   &   &   &   &   & \cmark  &   &  &   & \cmark  &   &   &   & \cmark  & \cmark  &   & Custom\\\wline
Dydra \cite{AndersonB16,Anderson19} & 2016 &  & \cmark   &   &   &  &  &   &  & \cmark &  &   &   &   &   &   & \cmark  & &  &   & \cmark  &   &   &   &  & & \cmark & Custom \\\wline
gStore-D \cite{10.1007/s00778-015-0415-0}  & 2016 &   &   &   &   & \cmark  &  &   & \cmark  &   & \cmark  &   &   &   &   &   &   & \cmark  &  &   & \cmark  &   &   &   & \cmark  &   &   & gStore\\\wline
Halyard \cite{SotonaN16} & 2016  & \cmark & \cmark & & & & & & \cmark & \cmark & & & & & & & \cmark & & & & \cmark & & & \cmark & & & \cmark & HBase \\\wline
JARS \cite{RajithNY16}  & 2016 & \cmark  &   &   &   &   &  &   & \cmark  &   &   &   &   &   &   &   & \cmark  & \cmark  &  &   &   &   &   & \cmark  &   &   &   & Custom\\\wline
S2RDF \cite{Schtzle2015S2RDFRQ}  & 2016 &   &   & \cmark  &   &   &  &   &   &   &   &   &   &   &   & \cmark  & \cmark  &   &  &   & \cmark  &   & \cmark  &   & \cmark  &   &   & HDFS\\\wline
S2X \cite{10.1007/978-3-319-41576-5_12}  & 2016 & \cmark  &   &   &   &   &  &   &   &   &   &   &   &   &   &   & \cmark  &   &  &   & \cmark  &   & \cmark  & \cmark  &   &   &   & Spark\\\wline
SPARQLGX \cite{10.1007/978-3-319-46547-0_9}  & 2016 &   &   & \cmark  &   &   &  &   &   &   &   &   &   &   &   &   & \cmark  &   &  &   & \cmark  &   &   & \cmark  &   &   &   & HDFS\\\wline
Wukong \cite{10.5555/3026877.3026902}  & 2016 & \cmark  &   &   &   &   &  &   & \cmark  &   &   &   &   &   &   & \cmark  & \cmark  &   &  &   &   &   &   &   &   &   &   & DrTM-KV\\\wline
CM-Well \cite{BennettEL17} & 2017 & \cmark & & & & \cmark & & & & & \cmark & & & & & \cmark & \cmark & & & & \cmark & \cmark & & \cmark & & & & Cassandra/ElasticS. \\ \wline
Koral \cite{Janke2017KoralAG}  & 2017 & \cmark  &   &   &   &   &  &   & \cmark  &   &   &   &   &   &   &   & \cmark  &   &  &   &   &   &   & \cmark  & \cmark  &   &   & Custom\\\wline
MarkLogic & 2017 & \cmark & & & & & & & \cmark & & & & & & & & \cmark & & & & \cmark & \cmark & & \cmark & & & & Custom \\\wline
SANSA \cite{10.1007/978-3-319-68204-4_15}  & 2017 & \cmark  &   & \cmark  &   &   &  &   &   &   &   &   &   &   &   &   & \cmark  &   &  &   &   &   & \cmark  & \cmark  &   &   &   & HDFS\\\wline
Spartex \cite{AbdelazizHSK17}  & 2017 &   &   &   &   & \cmark  &  &   & \cmark  &   &   &   &   &   &   &   &   & \cmark  &  &   &   &   &   & \cmark  &   &   &   & GSP/Custom\\\wline
Stylus \cite{10.14778/3149193.3149200}  & 2017 &   &   &   &   & \cmark  &  &   & \cmark  &   & \cmark  & \cmark  &   &   &   &   &   & \cmark  &  &   & \cmark  &   &   & \cmark  &   &   & \cmark  & Trinity\\\wline
Neptune \cite{BebeeCGGKKMMPRR18} & 2018 &   & \cmark  &   &   &   &  &   &   & \cmark  &   &   &   &   &   &   & \cmark  &   &  &   & \cmark  & \cmark  &   &   &   &   & \cmark  & Custom\\\wline
PRoST \cite{Cossu2018PRoSTDE}  & 2018 &   &   & \cmark  & \cmark  &   &  &   &   &   &   &   &   &   &   & \cmark  & \cmark  & \cmark  &  &   &   &   & \cmark  & \cmark  &   &   &   & HDFS\\\wline
RDFox-D \cite{PotterMNH18}  & 2018 & \cmark  &   &   &   &   &  &   & \cmark  &   &   &   &   &   &   &   & \cmark  &   &  &   &   &   &   &   & \cmark  &   &   & \textit{RDFox}\\\wline
WORQ \cite{10.1007/978-3-030-00671-6_34}  & 2018 &   &   & \cmark  &   &   &  &   &   &   &   &   &   & \cmark  &   &   & \cmark  & \cmark  &  &   &   &   &   &   &   &   &   & Spark\\\wline
Wukong+G \cite{10.5555/3277355.3277418} & 2018 & \cmark  &  &  &  & \cmark  &  &  & \cmark  &  &  &  &  &  &  &  & \cmark  &  &  &  &  &  &  &  & \cmark  &  &  & \textit{Wukong}\\\wline
Akutan & 2019 & \cmark & & & & & & & \cmark & & & & & & & & \cmark & & & & \cmark & \cmark & & \cmark & & & \cmark & RocksDB \\\wline
DiStRDF \cite{10.1145/3325135}  & 2019 & \cmark  &   &   & \cmark  &   &  &   &   &   &   &   &   &   &   & \cmark  & \cmark  &   &  &   & \cmark  &   &   & \cmark  &   &   &   & HDFS\\\wline
gStore-D2 \cite{8368253}  & 2019 &   &   &   &   & \cmark  &  &   &   &   &   &   &   & \cmark  &   &   & \cmark  & \cmark  &  &   &   &   &   &   & \cmark  & \cmark  &   & Custom\\\wline
Leon \cite{10.1007/978-3-030-18576-3_44}  & 2019 & \cmark  &   &   &   &   &  &   & \cmark  &   &   & \cmark  &   &   &   &   & \cmark  & \cmark  &  &   &   &   &   &   & \cmark  &   &   & Custom\\\wline
SPT+VP \cite{HassanB19}  & 2019 &   &   & \cmark  & \cmark  &   &  &   &   &   &   &   &   &   &   &   & \cmark  & \cmark  &  &   & \cmark  &   & \cmark  &   &   &   &   & Spark\\\wline
StarMR \cite{WangCXYLWC19}  & 2019 &   &   &   &   & \cmark  &  &   &   &   &   &   &   &   &   &   & \cmark  & \cmark  &  &   &   &   &   & \cmark  &   &   &   & HDFS\\\wline
DISE \cite{9031466}  & 2020 & \cmark  &   &   &   &   & \cmark  &   & \cmark  &   &   &   &   &   &   &   & \cmark  &   &  &   &   &   & \cmark  & \cmark  &   &   &   & Spark\\\wline
DP2RPQ \cite{WangWXYLW20}  & 2020 &   &   &   &   & \cmark  &  &   &   &   &   &   &   &   &   &   &   &   &  &   &   & \cmark  &   & \cmark  &   &   &   & Spark\\\wline
Triag \cite{NaackeC20}  & 2020 &   &   &   &   & \cmark  &  &   &   &   &   &   &   & \cmark  &   &   & \cmark  & \cmark  &  &   &   &   & \cmark  &   & \cmark  &   &   & Spark\\\wline
WISE \cite{GUO2020100161}  & 2020 & \cmark  &   &   &   &   &  &   & \cmark  &   &   & \cmark  &   &   &   &   & \cmark  & \cmark  &  &   &   &   & \cmark  &   & \cmark  & \cmark  &   & Leon\\\wline
gSmart \cite{chen2021gsmart} & 2021 &  &  &  &  &  & \cmark &  & \cmark &  &  &  &  &  &  &  &  & \cmark &  & \cmark &  &  &  & \cmark  &  & \cmark  &  & Custom\\
\bottomrule
\end{tabular}
\label{tab:crdfe}
\end{table*}

We now survey distributed RDF stores. Table~\ref{tab:crdfe} summarizes the surveyed systems and the techniques they use. We further indicate the type of underlying storage used, where italicized entries refer to local stores. Some systems that appear in the following may have appeared before in the local discussion if they are commonly deployed in both settings. 

\sys{YARS2}{10.5555/1785162.1785179}{2007} is an RDF store based on similar principles to YARS (see local stores) but for a distributed environment. The index manager in YARS2 uses three indexes namely a quad index, keyword index, and a join index for evaluating queries. The quad indexes cover six permutations of quads. The keyword index is used for keyword lookups. The join indexes help speed up query execution for common joins. The core index on quads is based on hashing the first element of the permutation, except in the case of predicates (e.g., for a \textsc{pogs} permutation), where hashing creates skew and leads to imbalance, and where random distribution is thus used. Indexed nested loop joins are used, with triple patterns being evaluated on one machine where possible (based on hashing), or otherwise on all machines in parallel (e.g., for constant predicates or keyword searches). Dynamic programming is used for join reordering in order to optimize the query.

\sys{Clustered TDB}{OwensSG}{2008} is a distributed RDF store based on Jena TDB storage (a local system). The system is based on a master--slave architecture where the master receives and processes queries, and slaves index parts of the graph and can perform joins. Hash-based partitioning is used to allocate dictionary-encoded tri\-ples to individual slaves based on each position of the triple; more specifically, distributed \textsc{spo}, \textsc{pos} and \textsc{osp} index permutations are partitioned based on \textsc{s}, \textsc{p} and \textsc{o}, respectively. An exception list is used for very frequent predicates, which are partitioned by \textsc{po} instead of \textsc{p}. Index-nested loop joins are supported and used to evaluate SPARQL basic graph patterns.

\sys{Virtuoso EE}{Erling2010}{2008} is a local RDF store whose enterprise edition also offers support for indexing over a cluster of machines. Recalling that Virtuoso stores RDF graphs as a quads table in a custom relational database, the most recent version of Virtuoso offers three options for each table: partitioned, replicated or local. Partitioning is based on partition columns specified by the administrator, which are used for hash-based partitioning; partitions can also be replicated, if specified. Replication copies the full table to each machine, which can be used for query-based partitioning, or to store a global schema that is frequently accessed by queries. Local tables are only accessible to the individual machine, and are typically used for local configuration.

\sys{4store}{harris4store}{2009} stores quads over a cluster of machines, where subject-based hash partitioning is used. Three types of indexes are used in 4Store namely R, M, and P indexes. The R index is a hash table that dictionary encodes and stores meta-data about individual RDF terms (called ``resources''). The M index is a hash table that maps graph names (called ``models'') to the corresponding triples in the named graph. The P Indexes consist of radix tries, with two for each predicate (similar to vertical partitioning): one for \textsc{sog} order and another for \textsc{osg} order. Joins are pushed, where possible, to individual machines. Join reordering uses cardinality estimations. SPARQL queries are supported.

\sys{Blazegraph}{ThompsonPC14}{2009}, discussed previously as a local store, also features partitioning in the form of key-range shards that allow for partitioning B+tree indexes, potentially across multiple machines. An alternative replication cluster is supported that indexes the full RDF graph or SPARQL dataset on each machine, allowing queries to be evaluated entirely on each machine without network communication.
 
\sys{SHARD}{10.1145/1940747.1940751}{2009} is a distributed, Hadoop-based RDF store. It stores an RDF graph in flat files on HDFS such that each line presents all the triples associated with a given subject resource of the RDF triple, which can be seen as an adjacency list. The graph is hash partitioned, so that every partition contains a distinct set of triples. As the focus is on batch processing of joins, rather than evaluating queries in real-time, there is no specific indexing employed in SHARD. Query execution is performed through MapReduce iterations: first, it collects the results for the subqueries, which are joined and finally filtered according to bound variables and to remove redundant (duplicate) results. 

\sysn{AllegroGraph}{2010}, discussed previously as a local store, offers a distributed version where data are horizontally partitioned into shards, which are indexed locally on each machine per the local version. Alongside these shards, ``knowledge bases'' can be stored, consisting of triples that are often accessed by all shards (e.g., schema or other high level data), such that queries can be evaluated (in a federated manner) over one shard, and potentially several knowledge bases. 

\sys{GraphDB}{KiryakovOM05,BishopKOPTV11}{2010}, also a local store, offers an enterprise edition that can store RDF graphs on a cluster of machines using a master--slave architecture, where each cluster has at least one master node that manages one or more worker nodes, each replicating the full database copy, thus allowing for queries to be evaluated in full on any machine. Updates are coordinated through the master.

\sysf{AnzoGraph}{https://docs.cambridgesemantics.com/anzograph/userdoc/features.htm}{fn:ag}{2011} is an in-memory, massively parallel processing (MPP) RDF store based on a master--slave architecture. The system indexes named graphs, where partitioning and replication are also organized by named graphs. By default, all triples involving a particular term are added into a named graph for that term. A dictionary is provided to map terms to named graphs. The query is issued at a master node, which features a query planner that decides the type of join (hash or merge joins are supported) or aggregation needed. Individual operations are then processed over the slaves in parallel, generating a stream of intermediate results that are combined on the master. 

\sys{CumulusRDF}{Ladwig_cumulusrdf:linked}{2011} works on top of Apache Cassandra: a distributed key-value store with support for tabular data. Three triple permutations -- \textsc{spo}, \textsc{pos}, \textsc{osp} -- and one quad permutation -- \textsc{gspo} -- are considered. A natural idea would be to index the first element as the row key (e.g., \textsc{s} for \textsc{spo}), the second (e.g., \textsc{p}) as the column key, and the third (e.g., \textsc{o}) as the cell value, but this would not work in multi-valued cases as columns are unique per row. Two other data storage layouts are thus proposed. Taking \textsc{spo}, the ``hierarchical layout'' stores \textsc{s} as the row key (hashed and used for partitioning), \textsc{p} as the supercolumn key (sorted), \textsc{o} as the column key (sorted), with the cell left blank. An alternative that outperforms the hierarchical layout is a ``flat layout'', where for \textsc{spo}, \textsc{s} remains the row key, but \textsc{po} is concatenated as the column key, and the cell is left blank. In the \textsc{pos} permutation, the \textsc{p} row key may create a massive row; hence \textsc{po} is rather used as the row key, with \textsc{p} being indexed separately. Join and query processing is enabled though Sesame.

\sys{H-RDF-3X}{huang2011scalable}{2011} is a Hadoop-based RDF store that uses RDF-3X on a cluster of machines. A graph-based partitioning (using the METIS software package) is used to distribute triples among multiple worker nodes. It also employs a $k$-hop guarantee, which involves replicating nodes and edges that are $k$ hops away from a given partition, thus increasing the locality of processing possible, and reducing communication costs. Local joins are optimized and evaluated on individual machines by RDF-3X, while joins across machines are evaluated using Hadoop. The use of Hadoop -- which involves expensive coordination across machines, and heavy use of the disk -- is minimized by leveraging the $k$-hop guarantee and other heuristics.

\sys{PigSPARQL}{Schtzle2011PigSPARQLMS}{2011} is a Hadoop-based RDF store that uses a vertical partitioning strategy. Data are stored on HDFS without indexes, and thus the focus is on batch processing. SPARQL queries are translated into PigLatin: an SQL-inspired scripting language that can be compiled into Hadoop tasks by the Pig framework. The Jena ARQ library is used to parse SPARQL queries into an algebra tree, where optimizations for filters and selectivity-based join reordering are applied. The tree is traversed in a bottom-up manner to generate PigLatin expressions for every SPARQL algebra operator. The resulting PigLatin script is then translated to -- and run as -- MapReduce jobs on Hadoop.

\sys{Rapid+}{10.1007/978-3-642-21064-8_4}{2011} is a Hadoop-based system that uses a vertical partitioning strategy for storing RDF data. Without indexing, the system targets batch processing. Specifically, Pig is used to generate and access tables under a vertical partitioning strategy. In order to translate SPARQL queries into PigLatin scripts, user-defined-functions are implemented that allow for optimizing common operations, such as loading and filtering in one step. Other optimizations include support for star joins using grouping, and a look-ahead heuristic that reduces and prepares intermediate results for operations that follow; both aim to reduce the number of Hadoop tasks needed to evaluate a query.

\sys{AMADA}{10.1145/2396761.2398749}{2012} is an RDF store based on the Amazon Web Services (AWS) cloud infrastructure. Indexes for the RDF graph are built using Amazon SimpleDB: a key-value storage solution that supports a subset of SQL. SimpleDB offers several indexing strategies, where ``attribute indexing'' can be used to create three indexes for the three elements of a triple. In AMADA, a query is submitted to a query processing module running on EC2, which in turn evaluates triple patterns using the SimpleDB-based indexes.

\sys{H2RDF(+)}{10.1145/2187980.2188058,Papailiou2013H2RDFHD}{2012} stores RDF graphs using the HBase distributed tabular NoSQL store. Three triple permutations (\textsc{spo}, \textsc{pos}, and \textsc{osp}) are created over HBase tables in the form of key-value pairs. A join executor module creates the query plan, which decides between the execution of joins in a centralized (local) and distributed (Hadoop-based) manner. It further reorders joins according to selectivity statistics. Multiway (sort-)merge joins are run in Hadoop.

\sys{Jena-HBase}{10.5555/2887379.2887401}{2012} (also known as \textit{HBase-RDF}\footnote{\url{https://github.com/castagna/hbase-rdf}}) is a distributed RDF store using HBase as its back-end.
Jena-HBase supports three basic storage layouts for RDF graphs in HBase namely ``simple'': three triple tables, the first indexed and partitioned by \textsc{s}, the second by \textsc{p}, the third by \textsc{o}; ``vertical partitioning'': two tables for each predicate, one indexed by \textsc{s}, the other by \textsc{o}; ``indexed'': six triple tables covering all permutations of a triple. Hybrid layouts are also proposed that combine the basic layouts, and are shown to offer better query times at the cost of additional space. Jena is used to process joins and queries.

\sys{Rya}{10.1145/2347673.2347677}{2012} is a distributed RDF store that employs Accumulo -- a key-value and tabular store -- as its back-end. However, it can also use other NoSQL stores as its storage component. Rya stores three index permutations namely \textsc{spo}, \textsc{pos}, and \textsc{osp}. Query processing is based on RDF4J, with index-nested loop joins being evaluated in a MapReduce fashion. The count of the distinct subjects, predicates, and objects is maintained and used during join reordering and query optimization. 

\sys{Sedge}{YangYZK12}{2012} is an RDF store based on Pregel: a distributed (vertex-centric) graph processing framework. Pregel typically assumes a strict partition of the nodes in a graph, where Sedge relaxes this assumption to permit nodes to coexist in multiple partitions. A complementary graph partitioning approach is proposed involving two graph partitionings, where the cross-partition edges of one are contained within a partition of the other, reducing cross-partition joins. Workload-aware query-based partitioning is also proposed, where commonly accessed partitions and frequently-queried cross-partition ``hotspots'' are replicated. The store is implemented over Pregel, where indexes are built to map partitions to their workloads and to their replicas, and to map nodes to their primary partitions.

\sys{chameleon-db}{Alu2013chameleondbAW}{2013} is a distributed RDF store using custom graph-based storage. Partitioning is graph-based and is informed by the queries processed, which may lead to dynamic repartitioning to optimize for the workload being observed. An incremental indexing technique -- using a decision tree -- is used to keep track of partitions relevant to queries. It also uses a hash-table to index the nodes in each partition, and a range-index to keep track of the minimum and maximum values for literals of each distinct predicate in each partition. The evaluation of basic graph patterns is delegated to a subgraph matching algorithm over individual partitions, whose results are then combined in a query processor per the standard relational algebra. Optimizations involve rewriting rules that preserve the equivalence of the query but reduce intermediate results. 

\sys{D-SPARQ}{10.5555/2874399.2874465}{2013} is a distributed RDF store built on top of MongoDB: a NoSQL store for JSON-like documents. D-SPARQ partitions the RDF graph by subject.  Partial data replication is used whereby selected triples are replicated across partitions to increase parallelism when executing (sub-)queries. Indexes are built for \textsc{sp} and \textsc{po} permutations. D-SPARQ optimizes multiway \textsc{s}--\textsc{s} (star) joins, taking advantage of the locality offered by the \textsc{s}-based partitioning; selectivity estimates are used to reorder joins. 

\sys{EAGRE}{6544856}{2013} stores RDF data on HDFS, where data are pre-processed using Hadoop to extract entities and their classes, thereafter applying graph-based data partitioning using METIS.  
For each entity class, EAGRE adopts a space-filling curve technique (see Section~\ref{sec:estore}): an in-memory index structure that is used to index high-dimensional data, and more specifically in this case, to decide where the data for a given entity should be stored.
Joins are pushed to individual nodes where possible, with multiway joins between nodes being evaluated using Hadoop. A strategy similar to a distributed form of sideways-information-passing is used to reduce network traffic, where nodes share information about the possible ranges of constants matching individual variables, filtering intermediate results outside those ranges before they are sent over the network.

\sys{MR-RDF}{DuBCD13}{2013} is a distributed RDF store that uses RDF-3X for local storage, and Hadoop for join processing. A partition is generated for each characteristic set, where a triple is added to the partition for the characteristic set of its subject. Given that this may give rise to a large number of partitions, similar characteristic sets are clustered together to form larger partitions corresponding to the number of machines available. The larger partition is then described by the union of the characteristic sets it contains, which can be used for matching star joins (with constant predicates) to partitions. Star joins are evaluated locally by RDF-3X, and their results are joined over Hadoop.

\sys{SHAPE}{Lee:2013:SQO:2556549.2556571}{2013} uses RDF-3X to store and index RDF triples on a distributed cluster of machines. Triples are partitioned using a semantic hash partitioning scheme that is based on the IRI prefix hierarchy: triples with the same subject or object prefixes are identified and are placed in the same partition. The intuition is that such triples are more likely to be queried together. A distributed query execution planner coordinates the intermediate results from different nodes, which are joined using Hadoop.

\sys{Trinity.RDF}{Zeng:2013:DGE:2488329.2488333}{2013} is an RDF store implemented on top of Trinity: a distributed memory-based key-value storage system. A graph-based storage scheme is used, where an inward and outward adjacency list is indexed for each node. Hash-based partitioning is then applied on each node such that the adjacency lists for a given node can be retrieved from a single machine; however, nodes with a number of triples/edges exceeding a threshold may have their adjacency lists further partitioned. Aside from sorting adjacency lists, a global predicate index is also generated, covering the \textsc{pos} and \textsc{pso} triple permutations.
Queries are processed through graph exploration, with dynamic programming over cardinality estimates used to choose a query plan. 

\sys{TripleRush}{10.5555/2874551.2874555}{2013} is based on the Signal/Collect distributed graph processing framework~\cite{10.1007/978-3-642-17746-0_48}. In this framework, TripleRush considers an in-memory graph with three types of nodes. Triple nodes embed an RDF triple with its subject, predicate, and object. Index nodes embed a triple pattern. Query nodes coordinate the query execution. The index graph is formed by index and triple nodes, which are linked based on matches. A query execution is initialized when a query node is added to a TripleRush graph. The query vertex emits a query particle (a message) which is routed by the Signal/Collect framework to index nodes for matching. Partitioning of triples and triple patterns is based on the order \textsc{s}, \textsc{o}, \textsc{p}, where the first constant in this order is used for hash-based partitioning. Later extensions explored workload-aware query-based partitioning methods~\cite{Tschanz14}.

\sys{WARP}{6547414}{2013} uses RDF-3X to store triples in partitions among a cluster of machines. Like H-RDF-3X, graph-based partitioning is applied along with a replication strategy for $k$-hop guarantees. Unlike H-RDF-3X, WARP proposes a query-based, workload-aware partitioning, whereby the value of $k$ is kept low, and selective replication is used to provide guarantees specifically with respect to the queries of the workload, reducing storage overheads. Sub-queries that can be evaluated on one node are identified and evaluated locally, with custom merge joins (rather than Hadoop, as in the case of H-RDF-3X) used across nodes. Joins are reordered to minimize the number of single-node subqueries. 

\sys{AKZ14}{AbiriKZ14}{2014} is a distributed RDF store based on the  HBase tabular store. A property table storage scheme is implemented over HBase, which is built based on clustering entities with similar properties. A secondary triple table is used for multi-valued properties and (infrequent) properties that do not appear in the clusters. Property tables are used to solve subject-based star joins, with other joins being evaluated over Hadoop by translating SPARQL queries to Hive (an SQL-like language for Hadoop). Metadata for the HBase tables are stored in a relational database (MySQL).

\sys{Partout}{Galarraga:2014:PDE:2567948.2577302}{2014} is a distributed store that uses RDF-3X for underlying storage on each machine. The RDF graph is partitioned using a workload-aware query-based partitioning technique, aiming to group together triples that are likely to be queried together. Each partition is indexed using standard RDF-3X indexing. The SPARQL query is issued to a query processing master, which uses  RDF-3X to generate a suitable query plan according to a global statistics file. The local execution plan of RDF-3X is transformed into a distributed plan, which is then refined by a distributed cost model that assigns subqueries to partitions. This query plan is executed by slave machines in parallel, whose results are combined in the master.

\sys{P-LUPOSDATE}{GroppeKWHSG14}{2014} is a distributed RDF store built on HBase. Triples are distributed according to six triple permutations -- partitioning on \textsc{s}, \textsc{p}, \textsc{o}, \textsc{sp}, \textsc{so}, \textsc{po} -- enabling lookups for any triple pattern. In order to reduce network communication, Bloom filters are pre-computed  for each individual variable of each triple pattern with at least one constant and one variable that produces some result; e.g., for \textsc{sp}, a Bloom filter is generated encoding the objects of each subject--predicate pair; for \textsc{s}, a Bloom filter is generated for each subject encoding its predicates, and optionally, another Bloom filter is generated for its objects. These Bloom filters are sent over the network in order to compute approximate semi-join reductions, i.e., to filter incompatible results before they are sent over the network. SPARQL~(1.0) queries are evaluated by translating them to PigLatin, which are compiled into Hadoop jobs by Pig.

\sys{RDF-3X-MPI}{chirravuri2014rdf3x}{2014} is a distributed RDF store build on top of RDF-3X and a Message Passing Interface (MPI). After dictionary-encoding the triples, they are initially partitioned based on hashes on graph nodes, where  the partitions are extended to ensure an $n$-hop guarantee: i.e., that any node reachable in $n$-hops from a node assigned to that partition will also be available on the same partition. The partitions are stored in RDF-3X on each machine, and basic graph patterns are evaluated independently on each partition (it is assumed that the value of $n$ is sufficient to enable this, with other queries left for future work).

\sys{Sempala}{Schtzle2014SempalaIS}{2014} stores RDF triples in a distributed setting, using the columnar Parquet format for HDFS that supports queries for specific columns of a given row (without having to read the full row). In this sense, Parquet is designed for supporting a single, wide (potentially sparse) table and thus Sempala uses a single ``unified property table'' for storing RDF triples with their original string values; multi-valued properties are stored using additional rows that correspond to a Cartesian product of all values for the properties of the entity. SPARQL queries are translated into SQL, which is executed over the unified property table using Apache Impala: a massively parallel processing (MPP) SQL engine that runs over data stored in HDFS. 

\sys{SemStore}{10.1145/2661829.2661876}{2014} is a distributed RDF store with a master--slave architecture. A custom form of graph partitioning is used to localize the evaluation of subqueries of particular patterns -- star, chain, tree, or cycle -- that form the most frequent elements of basic graph patterns. A $k$-means partitioning algorithm is used to assign related instances of patterns to a particular machine, further increasing locality. The master creates a global bitmap index over the partitions and collects global cardinality-based statistics. Slave nodes use the TripleBit local RDF engine for storage, indexing and query processing. The master node then generates the query plan using dynamic programming and global cardinality statistics, pushing joins (subqueries) to individual slave nodes where possible.
   
\sys{SparkRDF}{10.5555/2878453.2878519}{2014} is a Spark-based RDF engine that distributes the graph into subgraphs using vertical partitioning, adding tables for classes as well as properties. SparkRDF then creates indexes over the class and property tables, and further indexes class--property, property--class, and class--property-class joins. These indexes are loaded into an in-memory data structure in Spark (a specialized RDD) that implements query processing functionalities such as joins, filters, etc. Class information is used to filter possible results for individual variables, where a greedy selectivity-based strategy is used for reordering joins. Joins themselves are evaluated in a MapReduce fashion.

\sys{TrIAD}{Gurajada:2014:TDS:2588555.2610511}{2014} is a in-memory distributed RDF store based on a master--slave architecture. The master maintains a dictionary of terms, a graph summary that allows for pruning intermediate results, as well as global cardinality-based statistics that allow for query planning. The graph summary is a quotient graph using METIS' graph partitioning: each partition forms a supernode, and labeled edges between supernodes denote triples that connect nodes in different partitions; the graph summary is indexed in two permutations: \textsc{pso} and \textsc{pos}. The triples for each partition are stored on a slave; triples connecting two partitions are stored on both slaves. Each slave indexes their subgraph in all six triple permutations. Given a basic graph pattern, the graph summary is used to identify relevant partitions, which are shared with the slaves and used to prune results; dynamic programming uses the global statistics to optimize the query plan. Alongside distributed hash and merge joins, an asynchronous join algorithm using message passing is implemented.

\sys{CK15}{ChengK15}{2015} is a distributed in-memory RDF store that combines two types of partitioning: triple-based partitioning and query-based partitioning. The graph is initially divided over the machines into equal-size chunks and dictionary-encoded in a distributed manner (using hash-based partitioning of terms). The encoded triples on each machine are then stored using a vertical partitioning scheme, where each table is indexed by hashing on subject, and on object, providing $\textsc{p}\rightarrow\textsc{so}$, $\textsc{ps}\rightarrow\textsc{o}$ and $\textsc{po}\rightarrow\textsc{s}$ lookups. Parallel hash joins are proposed. Secondary indexes are then used to cache intermediate results received from other machines while processing queries, such that they can be reused for future queries. These secondary indexes can also be used for computing semi-join reductions on individual machines, thus reducing network traffic.
 
\sys{CliqueSquare}{7113332}{2015} is a Hadoop-based RDF engine used to store and process massive RDF graphs. It stores RDF data in a vertical partitioning scheme using semantic hash partitioning, with the objective of enabling co-located or partitioned joins that can be evaluated in the map phase of the MapReduce paradigm. CliqueSquare also maintains three replicas for fast query processing and increased data locality. In order to evaluate SPARQL queries, CliqueSquare uses a clique-based algorithm, which works in an iterative way to identify cliques in a query--variable graphs and to collapse them by evaluating joins on the common variables of each clique. The process will then terminate when the query--variable graph consists of only one node.
   
\sys{DREAM}{Hammoud:2015:DDR:2735703.2735705}{2015} is a distributed store using RDF-3X for its underlying storage and indexing. The entire RDF graph is replicated on every machine, with standard RDF-3X indexing and query processing being applied locally. To reduce communication, dictionary-encoded terms are shared within the cluster. In the query execution phase, the SPARQL query is initially represented as a directed graph, which is divided into multiple subqueries to be evaluated by different machines. The results of subqueries are combined using hash joins and eventually dictionary-decoded.
   
\sys{AdPart}{Harbi:2016:ASQ:2944220.2944335}{2016} is a distributed in-memory RDF store following a master--slave architecture. The master initially performs a hash-based partitioning based on the subjects of triples. The slave stores the corresponding triples using an in-memory data structure. Within each slave, AdPart indexes triples by predicate, predicate--subject, and predicate--object. Each slave machine also maintains a replica index that incrementally replicates data accessed by many queries; details of this replication are further indexed by the master machine. Query planning then tries to push joins locally to slaves (hash joins are used locally), falling back to distributed semi-joins when not possible. Join reordering then takes communication costs and cardinalities into account.

\sys{DiploCloud}{7323867}{2016} is a distributed version of the local RDF store dipLODocus. The store follows a master--slave architecture, where slaves store ``molecules'' (see the previous discussion on dipLODocus). The master provides indexes for a dictionary, for the class hierarchy (used for inference), as well as an index that maps the individual values of properties selected by the administrator to their molecule. Each slave stores the molecule subgraphs, along with an index mapping nodes to molecules, and classes to nodes. Query processing pushes joins where possible to individual slaves; if intermediate results are few, the master combines results, or otherwise a distributed hash join is employed. Molecules can be defined as a $k$-hop subgraph around the root node, based on input from an administrator, or based on a given workload of queries.

\sys{Dydra}{AndersonB16,Anderson19}{2016} is an RDF store that can leverage both local and remote storage, and provides support for versioned RDF graphs. In terms of local storage, RDF data are dictionary encoded and indexed in six permutations of quad tables -- namely \textsc{gspo}, \textsc{gpos}, \textsc{gosp}, \textsc{spog}, \textsc{posg}, \textsc{opsg} -- using on-disk B+trees. These B+trees offer support for both static and streaming data, and further capture information about revisions, enabling versioned queries and other RDF archival features. Support for replication through convergent replicated data types (CvRDTs) is also proposed~\cite{Anderson19}. A SPARQL query processor is layered on top of storage, providing support for SPARQL~1.1 queries and updates. Dydra further offers a multi-tenant cloud-based storage service.

\sys{gStore-D}{10.1007/s00778-015-0415-0}{2016} is a distributed RDF store that uses a variation of the local gStore RDF engine for local storage and indexing (it uses adjacency lists and vertex signatures, as discussed previously). A graph partitioning algorithm is applied over the RDF graph, with the subgraphs induced by each partition being assigned to individual machines. Edges that connect distinct partitions are indexed within the subgraphs of both partitions. Where possible, joins are then pushed to the individual subgraphs. Joins across subgraphs can be evaluated in a central (i.e., by a single master) or distributed (i.e., by several slaves) manner. Support for the relational features of SPARQL (1.0) is described.

\sys{Halyard}{SotonaN16}{2016} is a distributed RDF store that combines RDF4J with underlying HBase storage. MapReduce (Hadoop) is used to perform a bulk load of RDF quads into HBase tables. Sorted hash-based indexes are built in six permutations -- \textsc{spo}, \textsc{pos}, \textsc{osp}, \textsc{gspo}, \textsc{gpos}, \textsc{gosp} -- enabling efficient lookups for any quad pattern. HBase is connected with RDF4J as a Storage and Inferencing Layer (SAIL), enabling support for SPARQL~1.1. Nested-loop joins are used, and various optimizations are implemented for distributed evaluation, including push-based physical operators, priority queues for parallelizing operations, etc. Inferencing is supported by materializing entailments and bulk-loading them into HBase using the MapReduce framework.

\sys{JARS}{RajithNY16}{2016} is a distributed RDF store that combines triple-based and query-based partitioning. The graph is partitioned by hashing on subject, and hashing on object, constructing two distributed triple tables. The subject-hashed table is indexed on the \textsc{pos}, \textsc{pso}, \textsc{osp} and \textsc{spo} permutations, while the object-hashed table is indexed on \textsc{pos}, \textsc{pso}, \textsc{sop} and \textsc{ops}. Specifically, by hashing each triple on subject and object, the data for \textsc{s}--\textsc{s}, \textsc{o}--\textsc{o} and \textsc{s}--\textsc{o} are on one machine; the permutations then allow for such joins to be supported as merge joins on each machine. Basic graph patterns are then decomposed into subqueries answerable on a single machine, with a distributed hash join applied over the results. Jena ARQ is used to support SPARQL.
   
\sys{S2RDF}{Schtzle2015S2RDFRQ}{2016} is a distributed RDF store based on HDFS (with Parquet). The storage scheme is based on an extended version of vertical partitioning with semi-join reductions (see Section~\ref{sec:evp}). This scheme has a high space overhead, but ensures that only data useful for a particular (pairwise) join will be communicated over the network. In order to reduce the overhead, semi-join tables are not stored in cases where the selectivity of the join is high; in other words, semi-join tables are stored only when many triples are filtered by the semi-join (the authors propose a threshold of $0.25$, meaning that at least 75\% of the triples must be filtered by the semi-join for the table to be included). SPARQL queries are optimized with cardinality-based join reordering, and then translated into SQL and evaluated using Spark.
 
\sys{S2X}{10.1007/978-3-319-41576-5_12}{2016} runs SPARQL queries over RDF graphs using GraphX: a distributed graph processing framework built on the top of Spark. The triples are stored in-memory on different slave machines with Spark (RDDs), applying a hash-based partitioning on subject and objects (per GraphX's default partitioner). S2X does not maintain any custom indexing. For SPARQL query processing, graph pattern matching is combined with relational operators (implemented in the Spark API) to produce solution mappings.
   
\sys{SPARQLGX}{10.1007/978-3-319-46547-0_9}{2016} stores RDF data on HDFS per a vertical partitioning scheme. A separate file is created for each unique predicate in the RDF graph, with each file containing the subjects and objects of that triple. No indexes are provided, and thus the system is intended for running joins in batch-mode. SPARQL queries are first optimized by applying a greedy join reordering based on cardinality and selectivity statistics; the query plan is then translated into Scala code, which is then directly executed by Spark. 
  
\sys{Wukong}{10.5555/3026877.3026902}{2016} stores RDF graphs in DrTM-KV: a distributed key--value store using ``remote direct memory access'' (RDMA), which enables machines to access the main memory of another machine in the same cluster while by-passing the remote CPU and OS kernel. Within this store, Wukong maintains three kinds of indexes: a node index that maps subjects or (non-class) objects to their corresponding triples; a predicate index, which returns all subjects and objects of triples with a given predicate; and a type index, which returns the class(es) to which a node belongs. Hash-based partitioning is used for the node index, while predicate and type indexes are split and replicated to improve balancing. A graph-traversal mechanism is used to evaluate basic graph patterns, where solutions are incrementally extended or pruned. For queries involving fewer data, the data are fetched from each machine on the cluster and joined centrally; for queries involving more data, subqueries are pushed in parallel to individual machines. A work-stealing mechanism is employed to provide better load balancing while processing queries.

\sys{CM-Well}{BennettEL17}{2017} is a distributed RDF store developed by Thomson Reuters (Refinitiv) that combines Cassandra and ElasticSearch for underlying storage, further using Akka and Kafka for coordination and communication. RDF triples are grouped by node (subject), and stored in Cassandra, with inverted indexes for each node indexed by ElasticSearch. In order to process queries, two modes are considered. In sub-graph mode, candidate nodes are identified using ElasticSearch, and their associated triples are loaded in Jena in order to process. In full-graph mode, queries are translated directly to ElasticSearch operators.	

\sys{Koral}{Janke2017KoralAG}{2017} is a distributed RDF store based on a modular master--slave architecture that supports various options for each component of the system. Among these alternatives, various triple-based and graph-based partitioning schemes are supported. In order to evaluate basic graph patterns, joins are processed in an analogous way to TrIAD, using asynchronous execution, which makes the join processing strategy independent of the partitioning chosen. The overall focus of the system is to be able to quickly evaluate different alternatives for individual components -- particularly partitioning strategies -- in a distributed RDF store.

\sysf{MarkLogic}{https://docs.marklogic.com/guide/semantics}{fn:marklogic}{2017} is a multi-model distributed store with support for XML, JSON and RDF. Originally focusing on the storage and querying of XML documents, support for RDF and SPARQL was added in 2017 with the release of MarkLogic Server v.7.0, with SPARQL~1.1 support added in v.8.0. MarkLogic stores dictionary-encoded RDF triples in 4 kilobyte blocks, over which an LRU cache is implemented. Indexes are built for three permutations: \textsc{pso}, \textsc{sop} and \textsc{ops}. SPARQL~1.1 is supported, as well as rule-based inferencing implemented using backward chaining. Distribution is enabled through ``evaluator nodes'' that perform query processing, and ``data nodes'' that store and index data.

\sys{SANSA}{10.1007/978-3-319-68204-4_15}{2017} is a Spark-based distributed RDF store. RDF data are stored on HDFS, where triple-based partitioning -- such as predicate-based vertical partitioning -- is employed. Queries are transformed into Spark (or Flink) programs, using Sparklify~\cite{10.1007/978-3-030-30796-7_19}: a query engine for SPARQL-to-SQL translations, which can be run on Spark. SANSA is part of a larger stack that supports RDF-based inferencing and machine learning in a distributed environment.

\sys{Spartex}{AbdelazizHSK17}{2017} is a distributed RDF store with analytical capabilities. An extension of SPARQL queries is proposed with user-defined procedures for analytics (e.g., PageRank), among other features. The system is built on top of GPS: an open-source implementation of Pregel's distributed, vertex-centric graph processing framework. A master--slave architecture is employed. The master is responsible for query planning and manages global statistics. The RDF graph is partitioned among its slaves; namely each (subject/object) node and its incident edges (triples) is assigned to a slave. Each slave stores and indexes its subgraph in-memory using \textsc{ps} and \textsc{po} permutations. Basic graph patterns are then evaluated using graph traversals in GSP, with nodes (vertexes) sharing intermediate results as messages, which are joined with local data. Optimizations are based on minimizing duplicate traversals involving (non-Eulerian) cycles, as well as traditional cardinality estimates.

\sys{Stylus}{10.14778/3149193.3149200}{2017} is a distributed RDF store using Trinity: a graph engine based on an in-memory key--value store. Terms of the RDF graph are dictionary encoded. Each subject and object node is associated with a dictionary identifier and its characteristic set. A sorted adjacency list (for inward and outward edges) is then stored for each node that also encodes an identifier for the characteristic set of the node. Schema-level indexes for characteristic sets are replicated on each machine. Hash-based partitioning is employed on the data level. Indexes are used to efficiently find characteristic sets that contain a given set of properties, as well as to evaluate common triple patterns. Given a basic graph pattern, the characteristic sets are used to prune intermediate results on star joins, where candidates are kept for each variable node in the query. Cardinality-based join reordering is applied. Relational features of SPARQL~(1.0) are supported.

\sys{Neptune}{BebeeCGGKKMMPRR18}{2018}  is an RDF store that is hosted as a service on Amazon's S3 cloud storage infrastructure. Neptune stores SPARQL datasets in the form of quads with three index permutations: \textsc{spog}, \textsc{pogs} and \textsc{gspo}; this is sufficient to cover 9 out of 16 possible quad patterns. Neptune makes use of cardinality estimations and static analysis to rewrite queries. Partitioning is not supported, where Neptune rather offers up to 16 replicas of the full graph to increase query throughput; a primary replica is nominated to receive and coordinate updates. Graphs in Neptune can be queried (and processed) through the SPARQL 1.1, Apache TinkerPop and Gremlin languages. 

\sys{PRoST}{Cossu2018PRoSTDE}{2018} is a distributed RDF store using HDFS storage and Spark query processing. The storage scheme uses a combination of vertical partitioning and property tables that aims to leverage the strengths and minimize the weaknesses of both schemes. Like Sempala, the property table is stored in the column-wise Parquet format; multi-valued properties are supported by adding lists of values. Star joins on a common subject variable are evaluated on the property table, while other patterns and joins are addressed with the vertical partitioning tables. Selectivity-based heuristics are used to reorder joins. Queries are then rewritten into SQL for execution with Spark.

\sys{RDFox-D}{PotterMNH18}{2018} is a distributed in-memory RDF store based on RDFox that uses distributed index nested loop joins. A global index is built mapping the terms of the RDF graph to the partitions it appears in. The graph is partitioned by a weighted graph-based partitioning scheme, where nodes are weighted by the number of triples they appear in as subject. The partitioning minimizes cross-partition edges while balancing the sum of the node weights in each partition. Triples with a subject in the same partition are sent to the same machine; the weights used for partitioning then help to ensure more even balancing. Joins are evaluated in a pairwise manner, where each machine extends solutions asynchronously, without central coordination, based on its partition of the graph; it then sends the extended partial solution to the machines that can potentially extend it further (based on the global index). Termination occurs when all partial solutions have been forwarded. Various optimizations are discussed. Joins are reordered based on cardinalities.

\sys{WORQ}{10.1007/978-3-030-00671-6_34}{2018} is a distributed RDF store that uses a workload-aware approach to partition data. In order to reduce the number of intermediate results, Bloom filters are used to index the constants matching the variable of a given triple pattern, which are shared and used to filter results for that variable elsewhere. Bloom filters provide an approximate membership function (i.e., they may yield false positives), and thus a distributed join algorithm must be applied over the (reduced) intermediate results. Further Bloom filters can be computed for multiway joins, analogous to an approximate form of semi-join reduction (as used by S2RDF). These reductions can be cached for later re-use, where they are partitioned across machines based on the join element. WORQ is implemented over Spark.

\sys{Wukong+G}{10.5555/3277355.3277418}{2018} extends the distributed RDF store Wukong~\cite{10.5555/3026877.3026902} in order to additionally exploit GPUs (as well as CPUs) for processing queries in a distributed environment. One of the main design emphases of the system is to ensure that large RDF graphs can be processed efficiently on GPUs by ensuring effective use of the memory available, noting in particular that the local memory of GPUs has a much higher bandwidth for reading data into the GPU's cores, but a much lower capacity than typical for CPU RAM. Wukong+G thus employs a range of memory-oriented optimizations involving prefetching, pipelining, swapping, etc., to ensure efficient memory access when processing queries on the GPU. A graph partitioning algorithm is further employed to distribute storage, where lower-cost queries are processed on CPU (as per Wukong), but heavier loads are delegated to GPUs, where (like Wukong) efficient communication is implemented using RDMA primitives, allowing more direct access to remote CPU and GPU memory.

\sysf{Akutan}{https://tech.ebayinc.com/engineering/akutan-a-distributed-knowledge-graph-store/}{fn:akutan}{2019} (formerly known as \textit{Beam}) is a distributed RDF store developed by eBay. Triple storage is implemented on top of RocksDB, with indexes provided on \textsc{sp} $\rightarrow$ \textsc{o} and \textsc{op} $\rightarrow$ \textsc{s}. Triples are additionally associated with triple identifiers. Transactional logging is implemented using Apache Kafka, which coordinates read and write requests across machines. A SPARQL(-like) query processor is then layered on top of the underlying storage layer, which includes an optimizer that leverages statistics about the data to reorder joins. Hash joins and nested-loop joins are supported, and selected, as appropriate, by the query planner. Queries are then processed in streams and/or batches. A limited form of inference based on transitive closure is also supported.

\sys{DiStRDF}{10.1145/3325135}{2019} is a massively parallel processing (MPP) RDF store based on Spark with support for spatio-temporal queries. A special dictionary-encoding mechanism is used where the identifier concatenates a bit-string for spatial information, a bit-string for temporal information, and a final bit-string to ensure that the overall identifier is unique. Thus spatial and temporal processing can be applied directly over the identifiers. Storage based on both a triple table and property tables is supported, where range-based partitioning is applied to the triples (based on the spatio-temporal information). Data is stored on HDFS in CSV or Parquet formats. Query processing is implemented in Spark. Distributed hash joins and sort--merge joins are supported; selections and projections are also supported. Three types of query plans are proposed that apply RDF-based selections, spatio-temporal selections and joins in different orders.

\sys{gStore-D2}{8368253}{2019} is a distributed RDF store using workload-aware graph partitioning methods. Frequently accessed (subgraph) patterns are mined from the workload, where all subjects and objects are mapped to variables. Sub-graphs that instantiate these patterns are assigned DFS codes that are indexed as a tree, and associated with various meta-data, including identifiers for queries that use the pattern, cardinality estimations, partition identifiers, etc. Three partitioning methods are based on these patterns, with partitions stored locally in gStore. ``Vertical partitioning'' indexes all instances of a given pattern on the same machine. ``Horizontal partitioning'' distributes instances of the same pattern across various machines based on its constants. ``Mixed partitioning'' combines the two. Basic graph patterns are decomposed into frequent sub-patterns, where the join order and algorithms are selected to reduce communication costs. 

\sys{Leon}{10.1007/978-3-030-18576-3_44}{2019} is an in-memory distributed RDF store based on a master--slave architecture. Triples are partitioned based on the characteristic set of their subject; the characteristic sets are ordered in terms of the number of triples they induce, and assigned to machines with the goal of keeping a good balance. Indexes (similar to those of Stylus~\cite{10.14778/3149193.3149200}) are built, including a bidirectional index between subjects and their characteristic sets, an index to find characteristic sets that contain a given set of properties, and indexes to evaluate certain triple patterns. A multi-query optimization technique is implemented where, given a workload (a set) of queries, the method searches for an effective way to evaluate and share the results for common subqueries -- in this case, based on characteristic sets -- across queries.

\sys{SPT+VP}{HassanB19}{2019} is a distributed RDF store based on the principle of combining two partitioning techniques. First, a modified property table scheme is used for storage, where one table is maintained with a column for subject and all properties in the RDF graph; instead of storing multi-valued properties in multiple rows, as in Sempala's unified property table, such values are stored as lists nested in the given row. The property table is then split (vertically) into multiple tables, similar to a clustering-based definition of a property table, but where a subject may appear in multiple tables. This ``subset property table'' approach is combined, secondly, with vertical partitioning storage. Given a SPARQL query, joins are reordered based on global statistics, with the property tables used for \textsc{s}--\textsc{s} joins and vertical partitioning used for other joins. The query is then translated into Spark SQL for execution.

\sys{StarMR}{WangCXYLWC19}{2019} is a distributed RDF store that centers around optimizations for star joins. A graph-based storage scheme is employed, where for each node in the graph, its outward edges are represented in an adjacency list; this then supports efficient evaluation for \textsc{s}--\textsc{s} star joins. No indexing is provided, where the system targets batch-based (e.g., analytical) processing. A basic graph pattern is then decomposed into (star-shaped) sub-patterns, which are evaluated and joined. Hadoop is then used to join the results of these individual sub-patterns. Optimizations include the use of characteristic sets to help filter results, and the postponement of Cartesian products, which are used to produce the partial solutions for star joins including the non-join variables; these partial solutions are not needed if the corresponding join value is filtered elsewhere.

\sys{DISE}{9031466}{2020} is an in-memory, distributed RDF store that conceptualizes an RDF graph as a 3-dimensional binary tensor, similar to local approaches such as BitMat; however, physical representation and storage is based on dictionary encoded triples. Partitioning is based on slicing the tensor, which is equivalent to a triple-based partitioning. Joins are evaluated starting with the triple pattern with the fewest variables. SPARQL queries are supported through the Jena (ARQ) query library and evaluated using Spark.

\sys{DP2RPQ}{WangWXYLW20}{2020} is an RDF store built on a distributed graph processing framework with support for regular path queries (RPQs), which form the core of SPARQL's property paths. Unlike the standard RPQ semantics, the evaluation returns the ``provenance'' of the path, defined to be the subgraph induced by matching paths. Automata are used to represent the states and the potential transitions of paths while evaluating the RPQ, and are thus used to guide a navigation-based evaluation of the RPQ implemented by passing messages between nodes in the framework. Optimizations include methods to filter nodes and edges that cannot participate in the solutions to the RPQ, compression techniques on messages, as well as techniques to combine multiple messages into one. DP2RPQ is implemented on Spark's GraphX.

\sys{Triag}{NaackeC20}{2020} is a distributed RDF store that optimizes for triangle-based (sub)-patterns in queries. Two types of triangular RDF subgraphs are extracted using Spark: cyclic ones (e.g., $(a,p,b),(b,q,c),(c,r,a)$) and (directed) acyclic ones (e.g., $(a,p,b),(b,q,c),(a,r,c)$). The predicates of such subgraphs are extracted, ordered, hashed, and indexed in a distributed hash table using the predicate-based hash as key and the three nodes (e.g., $a,b,c$) as value. An encoding is used to ensure that the ordering of predicates is canonical for the pattern (assuming that nodes are variables) and that the subgraph can be reconstructed from the node ordering. Parallel versions of hash joins and nested loop joins are supported, where triangular subqueries can be pushed to the custom index. Queries are executed over Spark. Support for inferencing is also described.

\sys{WISE}{GUO2020100161}{2020} is a distributed RDF store using workload-aware query-based partitioning. The system follows a master--slave architecture. Queries processed by the master are also analyzed in terms of workload: common sub-patterns are extracted from a generalized version of the queries where constant subject and object nodes are first converted to variables. Query-based partitioning is applied so that common sub-patterns can be pushed to individual machines. Partitioning is dynamic, and may change as queries are received. A cost model is thus defined for the dynamic partitioning, taking into account the benefits of the change in partitioning, the cost of migrating data, and potential load imbalances caused by partition sizes; a greedy algorithm is then used to decide on which migrations to apply. The system uses Leon -- an in-memory distributed RDF store discussed previously -- for underlying storage and indexing.

\sys{gSmart}{chen2021gsmart}{2021} is a distributed RDF store that is capable of leveraging both GPUs and CPUs in a distributed setting. In order to take advantage of faster access for GPU memory despite its limited capacity, the LSpM storage system is used, which allows for loading compressed matrices for particular predicates and edge directions, as relevant for the query; matrices are encoded row-wise and column-wise, representing edge direction, in a compressed format, and can be partitioned for parallel computation. ``Heavy queries'' involving triple patterns with variable subjects and objects are then delegated to GPU computation, while ``light queries'' are run on CPU, where intermediate results are then combined to produce the final results on the CPU. Basic graph patterns are compiled into linear algebra operations that are efficiently computable on GPUs, with additional optimizations applied to process multi-way star joins.

\subsection{Trends}

We remark on some general trends based on the previous survey of local and distributed systems.

In terms of local systems, earlier approaches were based on underlying relational stores given that their implementations were already mature when interest began to coalesce around developing RDF stores. Thus, many of these earlier stores could be differentiated in terms of the relational schema (triple table, vertical partitioning, property tables, etc.) used to represent and encode RDF graphs. Systems that came later tended to rather build custom native storage solutions, optimizing for specific characteristics of RDF in terms of its graph structure, its fixed arity, etc.; relating to the fixed arity, for example, native stores began to develop complete indexes, by default, that would allow efficient lookups for any triple pattern possible. Also, many engines began to optimize for star-joins, which are often used to reconstruct $n$-ary relations from RDF graphs. Engines would soon start to explore graph-inspired storage and indexing techniques, including structural indexes, compressed adjacency lists, etc. A more recent trend -- likely following developments in terms of hardware -- has been an increased focus on in-memory stores using compact representations and compressed tensor-based representations of graphs that enable GPU-based hardware acceleration. Another recent development has been the application of worst-case optimal join algorithms for evaluating basic graph patterns, as well as techniques for translating queries into operations from linear algebra that can be efficiently evaluated on GPUs.

With respect to distributed RDF stores, in line with an increased demand for managing RDF graphs at very large scale, proposals began to emerge around 2007 regarding effective ways to store, index and query RDF over a cluster of machines.\footnote{We highlight that decentralized proposals for managing RDF graphs existed before this, including federated systems, P2P systems, etc., but are not considered in-scope here.} Initial proposals were based on existing native stores, which were extended with triple/quad-based partitioning and distributed join processing techniques to exploit a cluster of machines. A second trend began to leverage the maturation and popularity of ``Big Data'' platforms, including distributed processing frameworks like Hadoop and later Spark, and distributed NoSQL stores like Cassandra, HBase, MongoDB, etc., in order to build distributed RDF stores. During this time, graph-based and later query-based  partitioning methods began to emerge. Like in the local case, more and more in-memory distributed RDF stores began to emerge. Another trend was to explore the use of distributed graph processing frameworks -- that offer a vertex-based computation and messaging paradigm -- for evaluating queries over RDF. A very recent trend is towards using both CPUs and GPUs in a distributed environment in order to enable hardware acceleration on multiple machines.

While proposed solutions have clearly been maturing down through the years, and much attention has been given to evaluating basic graph patterns over RDF, some aspects of SPARQL query processing have not gained much attention. Most stores surveyed manage triples rather than quads, meaning that named graphs are often overlooked. A key feature of SPARQL -- and of graph query languages in general -- is the ability to query paths of arbitrary length, where optimizing property paths in SPARQL has not received much attention, particularly in the distributed setting. Many works also focus on a WORM (write once, read many) scenario, with relatively little attention paid (with some exceptions) to managing dynamic RDF graphs. 

A final aspect that is perhaps not well-understood is the trade-off that exists between different proposals, what precisely are their differences on a technical level (e.g., between relational- and graph-based conceptualizations), and which techniques perform better or worse in which types of settings. In this regard, a number of benchmarks have emerged to try to compare RDF stores in terms of performance; we will discuss these in the following section.

\section{SPARQL Benchmarks for RDF Stores}
\label{sec:benchmarking}

We now discuss a variety of SPARQL benchmarks for RDF stores. We speak specifically of SPARQL benchmarks since benchmarks for querying RDF either came after the standardization of SPARQL (and thus were formulated in terms of SPARQL), or they were later converted to SPARQL for modern use. The discussion herein follows that of Saleem et al.~\cite{triplestoreBench2019}, who analyze different benchmarks from different perspectives. We first discuss the general design principles for benchmarks, and then survey specific benchmarks.

\subsection{SPARQL Benchmark Design}

SPARQL query benchmarks consist of three elements: RDF graphs (or datasets), SPARQL queries, and performance measures. We first discuss some design considerations regarding each of these elements.

\paragraph{Datasets} The RDF graphs and datasets proposed for use in SPARQL benchmarks are of two types: \emph{real-world} and \emph{synthetic}. Both have strengths and weaknesses.

Real-world graphs reflect the types of graphs that one wishes to query in practice. Graphs such as DBpedia, Wikidata, YAGO, etc., tend to be highly complex and diverse; for example, they can contain hundreds, thousands or tens of thousands of properties and classes. Presenting query performance over real-world graphs is thus a relevant test of how a store will perform over RDF graphs found in practice. Certain benchmarks may also include a number of real-world graphs for the purposes of distributed, federated or even decentralized (web-based) querying~\cite{FedBench2011}.

Synthetic graphs are produced using specific generators that are typically parameterized, such that graphs can be produced at different scales, or with different graph-theoretic properties. Thus synthetic graphs can be used to test performance at scales exceeding real-world graphs, or to understand how particular graph-theoretic properties (e.g., number of properties, distributions of degrees, cyclicity, etc.) affect performance. Synthetic graphs can also be constructed to emulate certain properties of real-world graphs~\cite{appleoranges2011}.

A number of measures have been proposed in order to understand different properties of benchmark graphs. Obvious ones include basic statistics, such as number of nodes, number of triples, number of properties and classes, node degrees, etc.~\cite{largerdfbench2018,appleoranges2011}. Other (less obvious) proposals of measures include \textit{structuredness}~\cite{appleoranges2011}, which measures the degree to which entities of the same class tend to have similar characteristic sets; \textit{relationship specialty}~\cite{rbench2015}, which indicates the degree to which the multiplicity of individual properties varies for different nodes, etc. Observations indicate that the real-world and synthetic graphs that have been used in benchmarks tend to vary on such measures, with more uniformity seen in synthetic graphs~\cite{appleoranges2011,rbench2015,triplestoreBench2019}. This may affect performance in different ways; e.g., property tables will work better over graphs with higher structuredness and (arguably) lower relationship specialty.

\paragraph{SPARQL Queries} The second key element of the benchmark is the queries proposed. There are three ways in which the queries for a benchmark may be defined:

\begin{itemize}
\item \textit{Manually-generated}: The benchmark designer may manually craft queries against the RDF graph, trying to balance certain criteria such as query features, complexity, diversity, number of results, etc.

\item \textit{Induced from the graph}: The queries may be induced from the RDF graph by extracting sub-graphs (e.g., using some variation on random walks), with constants in the sub-graphs replaced by variables to generate basic graph patterns.

\item \textit{Extracted from logs}: The queries to be used may be extracted from real-world SPARQL logs reflecting realistic workloads; since logs may contain millions of queries, a selection process is often needed to identify an interesting subset of queries in the log.
\end{itemize}

\noindent Aside from concrete queries, benchmarks may also define query templates, which are queries where a subset of variables are marked as placeholders. These placeholders are replaced by constants in the data, typically so that the resulting partially-evaluated query still returns results over the RDF graph. In this way, each template may yield multiple concrete queries for use in the benchmark, thus smoothing variance for performance that may occur for individual queries.

Queries can vary in terms of the language considered (SPARQL 1.0 vs. SPARQL 1.1) and the algebraic features used (e.g., projection, filters, paths, distinct, etc.), but also in terms of various measures of the complexity and diversity of the queries -- and in particular, the basic graph patterns -- considered. Some basic measures to characterize the complexity and diversity of queries in a benchmark include the number of queries using different features, measures for the complexity of the graph patterns considered (e.g., number of triple patterns, number of variables, number of joins variables, number of cyclic queries, mean degree of variables, etc.), etc. Calculating such measures across the queries of the benchmark, a high-level \textit{diversity score} can be computed for a set of queries~\cite{triplestoreBench2019}, based on the average coefficient of variation (dividing the mean by the standard deviation) across the measures. 

\paragraph{Performance Measures} The third key element of a benchmark is the performance measures used. Some benchmarks may be provided without a recommended set of measures, but at the moment in which a benchmark is run, the measures to be used must be selected. Such measures can be divided into four categories~\cite{triplestoreBench2019}:

\begin{itemize}
    \item \textit{Query Processing Related}: The most important dimension relating to query processing relates to runtimes. A benchmark usually contains many queries, and thus reporting the runtime for each and every query is often too fine-grained. Combined results can rather be presented with measures like Query Mix per Hour (QMpH), Queries per Second (QpS), or measures over the distributions of runtimes (max, mean, percentile values, standard deviation, etc.). Other statistics like the number of intermediate results generated, disk/memory reads, resource usage, etc., can be used to understand lower-level performance issues during query processing load~\cite{schmidt2009sp}.
    \item \textit{Data Storage Related:} This category includes measures like data loading time, storage space, index sizes, etc. Often there is a space--time trade-off inherent in different approaches, where more aggressive indexing can help to improve query runtimes but at the cost of space and more expensive updates. Hence these measures help to contextualize query-processing related measures. 
    \item \textit{Result Related:} Some systems may produce partial results for a query based on fixed thresholds or timeouts. An important consideration for a fair comparison between two RDF engines relates to the results produced in terms of correctness and completeness. This can often be approximately captured in terms of the number of results returned, the number of queries returning empty results (due to timeouts), the recall of queries, etc.
    \item \textit{Update Related:} In real-world scenarios, queries are often executed while the underlying data are being updated in parallel. While the previous categories consider a read-only scenario, benchmarks may also record measures relating to updates~\cite{DBLP:conf/sigmod/ErlingALCGPPB15,iguana2017}. Measures may include the number of insertions or deletions per second, the number of read/write transactions processed, etc.
\end{itemize}

Often a mix of complementary measures will be presented in order to summarize different aspects of the performance of the tested systems.

\subsection{Synthetic Benchmarks}

We now briefly survey the SPARQL benchmarks that have been proposed and used in the literature, and that are available for download and use. We start with benchmarks based on synthetic data.

\paragraph{LUBM (Lehigh)}~\cite{lubm2005} (2005) creates synthetic RDF graphs that describe universities, including students, courses, professors, etc. The number of universities described by the graph is a parameter that can be changed to increase scale. The benchmark includes 14 hand-crafted queries. LUBM further includes an OWL ontology to benchmark reasoning, though often the benchmark is run without reasoning.

\paragraph{BSBM (Berlin)}~\cite{bsbm2009} (2009) is based on an e-commerce use-case describing entities in eight classes relating to products. The number of products can be varied to produce RDF graphs of different scales. A total of 12 query templates are defined with a mix of SPARQL features. The benchmark is also given in SQL format, allowing to compare RDF stores with RDBMS engines.

\paragraph{SP$^2$Bench}~\cite{schmidt2009sp} (2009) creates synthetic RDF graphs that emulate an RDF version of the DBLP bibliographic database. Various distributions and parameters from the DBLP data are extracted and defined in the generator. A total of 17 queries are then defined for the benchmark in both SPARQL and SQL formats.

\paragraph{BowlognaBench}~\cite{bowlogna2012} (2012) creates synthetic RDF graphs inspired by the Bologna process of reform for European universities. The dataset describes entities such as students, professors, theses, degrees, etc. A total of 13 queries are defined that are useful to derive analytics for the reform process.

\paragraph{WatDiv}~\cite{alucc2014diversified} (2014) provides a data generator that produces synthetic RDF graphs with an adjustable value of structuredness, and a query template generator that generates a specified number of query templates according to specified constraints. The overall goal is to be able to generate diverse graphs and queries. 

\paragraph{LDBC-SNB}~\cite{DBLP:conf/sigmod/ErlingALCGPPB15} (2015) is a benchmark based on synthetically generated social networking graphs. Three workloads are defined: \textit{interactive} considers both queries and updates in parallel; \textit{business intelligence} considers analytics that may touch a large percentage of the graph; \textit{algorithms} considers the application of graph algorithms. 

\paragraph{TrainBench}~\cite{DBLP:journals/sosym/SzarnyasIRV18} (2018) is a synthetic benchmark inspired by the use-case of validating a railway network model. The graph describes entities such as trains, swi\-tches, routes, sensors, and their relations. Six queries are defined that reflect validation constraints. TrainBench is expressed in a number of data models and query languages, including RDF/SPARQL and SQL.

\subsection{Real-World Benchmarks}

Next we survey benchmarks that are based on real-world datasets and/or queries from real-world logs. 

\paragraph{DBPSB (DBpedia)}~\cite{dbpsb2011} (2011) clusters queries from the DBpedia logs, generating 25 query templates representative of common queries found. These queries can then be evaluated over DBpedia, where a dataset of 153 million triples is used for testing, though smaller samples are also provided.

\paragraph{FishMark}~\cite{fishmark2012} (2012) is based on the FishBase dataset and is provided in RDF and SQL formats. The full RDF graph uses 1.38 billion triples, but a smaller graph of 20 million triples is used for testing. In total, 22 queries from a log of real-world (SQL) queries are converted to SPARQL. 

\paragraph{BioBenchmark}~\cite{biobenchmark2014} (2014) is based on queries over five real-world RDF graphs relating to bioinformatics -- Allie, Cell, DDBJ, PDBJ and UniProt -- with the largest dataset (DDBJ) containing 8 billion triples. A total of 48 queries are defined for the five datasets based on queries generated by real-world applications.

\paragraph{FEASIBLE}~\cite{feasible2015} (2015) generates SPARQL benchmarks from real-world query logs based on clustering and feature selection techniques. The framework is applied to DBpedia and Semantic Web Dog Food (SWDF) query logs and used to extract 15--175 benchmark queries from each log. The DBpedia and SWDF datasets used contain 232 million and 295 thousand triples, respectively. 

\paragraph{WGPB}~\cite{aidan_hogan_2019_4035223} (2019) is a benchmark of basic graph patterns over Wikidata. The queries are based on 17 abstract patterns, corresponding to binary joins, paths, stars, triangles, squares, etc. The benchmark contains 850 queries, with 50 instances of each abstract pattern mined from Wikidata using guided random walks. Two Wikidata graphs are given: a smaller one with 81 million triples, and a larger one with 958 million triples. 

\subsection{Benchmark Comparison and Results} For a quantitative comparison of (most of) the benchmarks mentioned here, we refer to the work by Saleem et al.~\cite{triplestoreBench2019}, which provides a detailed comparison of various measures for SPARQL benchmarks. For benchmarks with results comparing different RDF stores, we refer to the discussion for (italicizing non-RDF/SPARQL engines):

\begin{itemize} 
\item BSBM~\cite{bsbm2009} (2009) with results for Jena, RDF4J, Virtuoso and \textit{MySQL};
\item SP$^2$Bench~\cite{schmidt2009sp} (2009) including results for Kowari, Jena, RDF4J, Redland and Virtuoso.
\item DBPSB~\cite{dbpsb2011} (2011) with results for GraphDB, Jena, RDF4J and Virtuoso;
\item BowlognaBench~\cite{bowlogna2012} (2012) with results for 4store, dip\-LODocus, RDF-3X and Virtuoso.
\item FishMark~\cite{fishmark2012} (2012) with results for Virtuoso, \textit{MySQL} and \textit{Quest};
\item BioBench~\cite{biobenchmark2014} (2014) with results for 4store, Blazegraph, GraphDB, Kowari and Virtuoso.
\item WatDiv~\cite{alucc2014diversified} (2014) with results for 4store, gSt\-ore, RDF-3X, Virtuoso and \textit{MonetDB};
\item FEASIBLE~\cite{feasible2015} (2015) with results for  Graph\-DB, Je\-na, RDF4J and Virtuoso;
\item LDBC-SB~\cite{DBLP:conf/sigmod/ErlingALCGPPB15} (2015), with results for SparkSee and Virtuoso;
\item TrainBench~\cite{DBLP:journals/sosym/SzarnyasIRV18} (2018) with results for Jena, RDF4J, \textit{Neo4j} and \textit{SQLite}, among others.
\end{itemize}

For a performance comparison of eleven distributed RDF stores (SHARD, H2RDF+, CliqueSquare, S2X, S2RDF, AdPart, TriAD, H-RDF-3x, SHAPE, gStore-D and DREAM) and two local RDF stores (gStore and RDF-3X) over various benchmarks (including LUBM and WatDiv), we refer to the experimental comparison by Abdelaziz et al.~\cite{AbdelazizHKK17}.}
\end{document}